\documentclass[hyper]{JHEP}
\preprint{Cavendish-HEP-99/03\\CERN-TH/2000-284\\RAL-TR-2000-048}

\usepackage{amssymb,cite}
\usepackage{longtable}

\newcommand{\vv}[1]{\begingroup\obeyspaces\tt#1\endgroup}

\newcommand{\IQ}{\vv{IQ}}
\newcommand{\IL}{\vv{IL}}
\newcommand{\ID}{\vv{ID}}
\newcommand{\IN}{\vv{IN}}
\newcommand{\IC}{\vv{IC}}
\newcommand{\IP}{\vv{IP}}
\newcommand{\ISQ}{\vv{ISQ}}
\newcommand{\ino}{\widetilde}
\newcommand{\gluino}{\ino{g}}

\newcommand{\squark}{\ino{q}} 
\newcommand{\slepton}{\ino{\ell}}

\newcommand{\snu}{\ino{\nu}}
\newcommand{\gaugino}{\ino{\chi}}
\newcommand{\ntlino}[1]{\ino{\chi}^0_{#1}}
\newcommand{\sss}[1]{{#1}}
\newcommand{\ra}{\rightarrow}
\newcommand{\PD}{\sss{PDG}}
\newcommand{\AL}{{\sf ALPGEN}}
\newcommand{\CI}{{\sf CIRCE}}
\newcommand{\HD}{{\sf HDECAY}}
\newcommand{\HP}{{\sf HERWIG++}}
\newcommand{\HW}{{\sf HERWIG}}
\newcommand{\Fo}{{\sf Fortran}}
\newcommand{\Fos}{{\sf Fortran~77}}
\newcommand{\JS}{{\sf JETSET}}
\newcommand{\PDF}{{\sf PDFLIB}}
\newcommand{\PY}{{\sf PYTHIA}}
\newcommand{\IS}{{\sf ISAJET}}
\newcommand{\IW}{{\sf ISAWIG}}
\newcommand{\MN}{{\sf MC@NLO}}
\newcommand{\TA}{{\sf TAUOLA}}
\newcommand{\SM}{\sss{SM}}
\newcommand{\MS}{\sss{MSSM}}
\newcommand{\SY}{\sss{SUSY}}
\newcommand{\QCD}{\sss{QCD}}

\newcommand{\DIS}{\sss{DIS}}
\newcommand{\LEP}{\sss{LEP}}
\newcommand{\LHC}{\sss{LHC}}

\newcommand{\CERN}{\sss{CERN}}
\newcommand{\RPV}{\rlap{/}{R}$_{\mbox{\scriptsize p}}$}
\newcommand{\BNV}{\rlap{/}{B}}

\newcommand{\gh}{\Gamma_{\rm H}}
\newcommand{\ycut}{$y_{{\rm cut}}$}
\newcommand{\mh}{m_{\rm H}}

\newcommand{\lms}{\Lambda_{\overline{\rm MS}}}
\newcommand{\as}{\alpha_{{\rm S}}}
\newcommand{\ee}{e^+e^-}
\newcommand{\MC}{Monte Carlo}
\newcommand{\VEV}[1]{\langle{#1}\rangle}

\newcommand{\qbar}{\bar{q}}
\newcommand{\Qbar}{\bar{Q}}
\newcommand{\dbar}{\bar{d}}
\newcommand{\ubar}{\bar{u}}
\newcommand{\sbar}{\bar{s}}
\newcommand{\cbar}{\bar{c}}
\newcommand{\bbar}{\bar{b}}
\newcommand{\tbar}{\bar{t}}
\newcommand{\pbar}{\bar{p}}

\newcommand{\lbar}{\bar{\l}}
\renewcommand{\l}{\ell}

\newcommand{\shat}{\hat{s}}
\newcommand{\that}{\hat{t}}
\newcommand{\uhat}{\hat{u}}

\newcommand{\SMH}{H^0_{{\rm SM}}}

\title{{\sf HERWIG 6.5}: an event generator for Hadron Emission Reactions With
Interfering Gluons (including supersymmetric processes)\footnote{ Work
supported in part by the U.K.\ Particle Physics and Astronomy Research
Council, by the Italian Ministero della Universit\`a e Ricerca
Scientifica, and by the E.U. Fourth Framework Programme `Training and
Mobility of Researchers', Network `Quantum Chromodynamics and the Deep
Structure of Elementary Particles', contract FMRX-CT98-0194 (DG 12 -
MIHT).  Research of G.C.\ supported by grant number DE-FG02-91ER40685
from the U.S. Department of Energy.}}   
 
\author{Gennaro Corcella\\
	Max-Planck-Institut f\"ur Physik, Werner-Heisenberg-Institut, Munich\\
	E-mail: \email{corcella@mppmu.mpg.de}} 
 
\author{Ian G.\ Knowles\\ 
	Department of Physics and Astronomy, University of Edinburgh, UK\\ 
	E-mail: \email{knowles@supanet.com}}
 
\author{Giuseppe Marchesini\\ 
	Dipartimento di Fisica, Universit\`a di Milano-Bicocca, and\\ 
	I.N.F.N., Sezione di Milano, Italy\\ 
	E-mail:  \email{Giuseppe.Marchesini@mi.infn.it}} 
 
\author{Stefano Moretti\\ 
	Theory Division, CERN, and IPPP, University of Durham, UK\\ 
	E-mail: \email{Stefano.Moretti@cern.ch}} 
 
\author{Kosuke Odagiri\\
	Theory Group, KEK, Japan\\ 
	E-mail: \email{odagirik@post.kek.jp}} 
 
\author{Peter Richardson\\
	Department of Applied Mathematics and Theoretical Physics and\\
	Cavendish Laboratory, University of Cambridge, UK\\ 
	E-mail: \email{richardn@hep.phy.cam.ac.uk}} 
 
\author{Michael H.\ Seymour\\ 
	 Department of Physics and Astronomy, University of Manchester, UK\\ 
	E-mail: \email{M.Seymour@rl.ac.uk}}
 
\author{Bryan R.\ Webber\\ 
	Cavendish Laboratory, University of Cambridge, UK\\
	E-mail: \email{webber@hep.phy.cam.ac.uk}}

\abstract{ \HW\ is a general-purpose Monte Carlo event generator,
which includes the simulation of hard lepton-lepton, lepton-hadron and
hadron-hadron scattering and soft hadron-hadron collisions in one
package.  It uses the parton-shower approach for initial- and
final-state \QCD\ radiation, including colour coherence effects and
azimuthal correlations both within and between jets.  This article
updates the description of \HW\ published in 1992, emphasising the new
features incorporated since then. These include, in particular, the
matching of first-order matrix elements with parton showers, a more
correct treatment of spin correlations and
heavy quark decays, and a wide range of new
processes, including many predicted by the Minimal Supersymmetric
Standard Model, with the option of R-parity violation.  At the same
time we offer a brief review of the physics underlying \HW, together
with details of the input and control parameters and the output data,
to provide a self-contained guide for prospective users of the
program.  This version of the manual (version 3) is updated to
\HW\ version 6.5, which is expected to be the last major
release of Fortran \HW. Future developments will be implemented
in a new C++ event generator, \HP.}


\begin{document}

\section{Introduction}\label{introduction}

\HW\ is a general-purpose event generator for high-energy
processes, with particular emphasis on the detailed simulation of
\QCD\ parton showers. The program provides a full simulation of
hard lepton-lepton, lepton-hadron and hadron-hadron scattering and
soft hadron-hadron collisions in a single package, and has the
following special features:
\begin{itemize}
\item
Initial- and final-state \QCD\ jet evolution with soft gluon
interference taken into account via angular ordering;
\item
Colour coherence of (initial and final) partons in all hard
subprocesses, including the production and decay of heavy quarks and
supersymmetric particles;
\item
Azimuthal correlations within and between jets due to gluon
interference and polarization;
\item
A cluster model for jet hadronization based on non-perturbative
gluon splitting, and a similar  cluster model for soft and underlying
hadronic  events;
\item
A space-time picture of event development, from parton showers to
hadronic decays, with an optional colour rearrangement model based on
space-time structure.
\end{itemize}
Several of these features were already present in \HW\ version 5.1 and
were described accordingly in some detail in ref.~\cite{hw51}.  This
was in turn based on the work of refs.~\cite{Bass,March2,Webb1,Webb2,EMW,March1,March3,Knowl1,Knowl3,March4,March5,Catani1,Abbi}. In
the present article we concentrate therefore on the new features
incorporated into \HW\ since 1992.  The most important of these are
the matching of first-order matrix elements with parton
showers~\cite{Seym0,Seym,Seym1,Seym2,CorSey,Corcella:2000gs}, a more correct treatment
of spin correlations \cite{Richardson:2001df} and
heavy quark decays, again including matrix element
matching~\cite{CorSey}, and a wide range of new hard processes. In
particular \HW\ now includes the production and decay of
supersymmetric particles, with or without the assumption of R-parity
conservation~\cite{Moretti:2002eu}.

The precise \HW\ version described here is 6.500, which we shall
generally refer to as ``version 6'' in the following.

Let us recall briefly the main features of a generic hard, i.e.\ high
momentum transfer, process of the type simulated by \HW.  It can be
divided notionally into four components, corresponding roughly to
increasing scales of distance and time:
\begin{enumerate}
\item \emph{Elementary hard subprocess.} A pair of incoming beam
particles or their constituents interact to produce one or more
primary outgoing fundamental objects.  This can be computed exactly to
lowest order in perturbation theory.  The hard momentum transfer scale
$Q$ together with the colour flow of the subprocess set the boundary
conditions for the initial- and final-state parton showers, if there
are any.
\item \emph{Initial- and final-state parton showers.}  A parton
constituent of an incident beam hadron with low spacelike virtuality
can radiate timelike partons.  In the process it decreases its energy
to a fraction $x$ of that of the beam and increases its spacelike
virtuality, which is bounded in absolute value by the scale $Q$ of the
hard subprocess. This initial-state emission process leads to the
evolution of the structure function $F(x,Q)$ of the incident hadron.
On a similar time-scale, an outgoing parton with large timelike
virtuality can generate a shower of partons with lower virtuality.
The amount of emission depends on the upper limit on the virtuality of
the initiating parton, which is again controlled by the momentum
transfer scale $Q$ of the hard subprocess. Timelike partons from the
initial-state emission may in turn initiate parton showering.  The
coherence of soft gluon emission from different parton showers is
controlled by the colour flow of the subprocess. Within the showers,
it is simulated by angular ordering of successive emissions.
\item \emph{Heavy object decays.} Massive produced objects such as top
quarks, electroweak gauge and Higgs bosons, and possibly non-Standard
Model (e.g.\ \SY) particles, can decay on time-scales that may be
shorter than or comparable to that of the \QCD\ parton showers.
Depending on their nature and the decay mode, they may also initiate
parton showers before and/or after decaying.
\item \emph{Hadronization process.} In order to construct a realistic
simulation one needs to combine the partons into hadrons. This applies
to the constituent partons of incoming hadronic beams as well as to
the outgoing products of parton showering, which give rise to hadronic
jets.  This hadronization process takes place at a low momentum
transfer scale, for which the strong coupling $\as$ is large and
perturbation theory is not applicable. In the absence of a firm
theoretical understanding of non-perturbative processes, it must be
described by a phenomenological model. The model adopted in \HW\ is
intended to disrupt as little as possible the event structure
established in the parton showering phase. Showering is terminated at
a low scale, $Q_0<1$~GeV, and the \emph{preconfinement} property of
perturbative \QCD~\cite{AV,LPHD} is exploited to form colour-neutral
clusters~\cite{Webb1} which decay into the observed
hadrons. Initial-state partons are incorporated into the incoming
hadron beams through a soft, non-perturbative ``forced branching''
phase of spacelike showering.  The remnants of incoming hadron, i.e.\
those constituent partons which do not participate in the hard
subprocess, undergo a soft ``underlying event'' interaction modelled
on soft minimum bias hadron-hadron collisions.
\end{enumerate}

After a brief, practical section on the use of \HW, in the following
sections we discuss each of these components in turn, concentrating on
the new features since version 5.1.

\section{Using \HW}\label{using}
   
The latest version of the program, together with all relevant
information, is always available via the official \HW\ web page:

\begin{center}
\href{http://hepwww.rl.ac.uk/theory/seymour/herwig/}{\tt
http://hepwww.rl.ac.uk/theory/seymour/herwig/}
\end{center}
which is also mirrored at CERN:
\begin{center}
\href{http://home.cern.ch/seymour/herwig/}{\tt
http://home.cern.ch/seymour/herwig/}
\end{center}

The program is written in \Fo\ and the user has to modify the main
program \vv{HWIGPR} to generate the type and number of events
required.  See section~\ref{program} for a sample main program.  The
program operates by setting up parameters in common blocks and then
calling a sequence of subroutines to generate an event. Parameters
not set explicitly in the block data \vv{HWUDAT} or in \vv{HWIGPR} are
set to default values in the main initialisation routine
\vv{HWIGIN}. Output data are delivered in the LEP standard common
block \vv{HEPEVT} ~\cite{LEP1,LEP2}.  Note that all real variables
accessible to the user, including those in \vv{HEPEVT}, are of type
\vv{DOUBLE PRECISION}.

Since version~6.3, to take account of the increased energy and
complexity of interactions at
LHC and future colliders, the default value of the parameter \vv{NMXHEP},
which sets the array sizes in the standard \vv{/HEPEVT/} common block,
has been increased to 4000.

To generate events the user must first set up the beam particle names
\vv{PART1}, \vv{PART2} (type \vv{CHARACTER*8})
and momenta \vv{PBEAM1}, \vv{PBEAM2} (in GeV/$c$),
a process code \vv{IPROC} and the
number of events required \vv{MAXEV}.
   
See section~\ref{processes} for beams and processes available.

All  analysis of  generated  events  (histogramming, etc.)  should be
performed  by  the  user-provided  routines \vv{HWABEG} (to initialise
the run),
\vv{HWANAL} (to analyse an event) and \vv{HWAEND} (to terminate the run).

A detailed  event  summary is printed out for the first \vv{MAXPR} events
(default $\vv{MAXPR}=1$). Setting $\vv{IPRINT}=2$ lists the particle identity
codes, properties and decay schemes used in the program.

The  programming language is  standard \Fos\ as far as possible.
However,  the  following  may  require  modification  for  running on
some computers
\begin{itemize}
\item
Most common blocks are inserted by \vv{INCLUDE 'HERWIG65.INC'}
statements --- the file \vv{HERWIG65.INC} is part of the standard
program package.
\item
Many common blocks are initialised by \vv{BLOCK DATA HWUDAT}. Although
\vv{BLOCK DATA} is standard \Fos, it can cause linkage problems
for some systems.
\item
Variables of type \vv{DOUBLE COMPLEX} are used, which may be called
\vv{COMPLEX*16} on some systems.
\end{itemize}
In the \vv{INCLUDE} file, common blocks have been regrouped so that
all commons that were present in version 6.1 are unchanged.
All new variables since version 6.1 are in separate common blocks,
which are frozen after each version.

It is anticipated that version 6.5, with minor modifications in the
series 6.5xx, will be the {\em last Fortran version} of \HW.  Substantial
physics improvements will be included in the new C++ event
generator \HP\ \cite{HW++}, to be released in 2003. Watch for progress
on the \HP\ web page:
\begin{center}
\href{http://www.hep.phy.cam.ac.uk/~gieseke/Herwig++/}{\tt
http://www.hep.phy.cam.ac.uk/}$\sim${\tt gieseke/Herwig++/}
\end{center}

\section{Physics simulated by \HW}\label{physics}

\subsection{Elementary hard subprocess}\label{sec:elem}

In \HW\ version~6 there is a fairly large library of \QCD,
electroweak and supersymmetric elementary subprocesses. These are
listed and discussed in section~\ref{processes}.

The hard subprocess plays an important r\^ole in defining the phase
space of any associated initial- and final-state parton showers.  As
discussed in ref.~\cite{hw51} and references therein, the parton
showers are branching processes in which the branchings are ordered in
angle from a maximum to a minimum value determined by the
cutoff~$Q_0$. The maximum value is determined by the elementary
subprocess and is due to interference among soft gluons.  The general
result~\cite{March1,EMW,DKTM} is that the initial and final branchings
are approximately confined within cones around the incoming and
outgoing partons from the elementary subprocess.  For the branching of
parton~$i$, the aperture of the cone is defined by the direction of
the other parton $j$ that is colour-connected to~$i$.  The relation
between soft gluon interference and the colour connection structure of
the elementary subprocess leads to many detectable effects in hadronic
final states. For recent examples, see e.g.\
refs.~\cite{CDFcolcoh,D0colcoh}.

For a general process there are various contributions with different
colour connections. The \HW\ library of elementary subprocesses
includes the separate colour connection contributions.  In general
there is some ambiguity in the separation of contributions which are
suppressed by inverse powers of the number of colours $N_c$.  In
earlier versions of \HW, these sub-leading terms were divided amongst
the various colour connection contributions according to the recipe in
ref.~\cite{March1}.  In version~6 the following improved
prescription~\cite{ko_colour} has been followed, for both the \QCD\
and \SY\ \QCD\ subprocesses.  The matrix element squared $|{\cal
M}_i|^2$ for each colour connection is computed in the limit
$N_c\to\infty$ and the corresponding contribution is defined as
$$ 
\frac{|{\cal M}_i|^2}{\sum_j|{\cal M}_j|^2} |{\cal M}|^2\,,
$$
where $|{\cal M}|^2$ represents the sum over all colour connections
\emph{including terms sub-leading in $N_c$}. This ensures that each
colour connection contribution is positive-definite and has the
correct pole structure, while the sum of contributions yields the
exact (Born-level) cross section.

Parton emission into phase space regions which are outside the
above-mentioned angular cones, called in the following ``dead zones'',
do not contribute to leading order and often not even to
next-to-leading order. However, for a more complete description of the
event it is also necessary to take into account emission into these
``dead zones'', which may be done using exact fixed-order matrix
elements (see the following sections).

Another important function of the elementary subprocess is to set up
the polarizations of any electroweak bosons, heavy (e.g.\ \SY)
particles or gluon jets that may be
involved.  These polarizations give rise to angular asymmetries and
correlations in boson decays and jet fragmentation. They are included
in \HW\ for many of the subprocesses provided, using the approach of
refs.~\cite{Knowl1,Knowl3,Richardson:2001df} to generate all correlations
in decays, and in jet fragmentation to leading-logarithmic accuracy.

\subsection{Parton showers}\label{sec:showers}

\subsubsection{Final-state showers}

Final-state parton showering in \HW\ is generated by a \emph{coherent
branching algorithm} with the following properties:
\begin{enumerate}
\item
The energy fractions are distributed according to the leading-order
Dokshitzer-Gribov-Lipatov-Altarelli-Parisi (DGLAP) splitting functions.
\item
The full available phase space is restricted to an angular-ordered 
region. Such a restriction is the result of interference and
correctly takes important infrared singularities into account.
At each branching, the angle between the two emitted partons
is smaller than that of the previous branching.
\item 
The emission angles are distributed according to the Sudakov form 
factors, which sum the virtual corrections and unresolved real emissions.
The Sudakov form factor normalizes the branching distributions to 
give the probabilistic interpretation needed for a \MC\ simulation. 
This fact is a consequence of unitarity and of the infrared finiteness
of inclusive quantities.
\item 
The azimuthal angular distribution in each branching is determined 
by two effects: 
\begin{itemize}
\item[($a$)] for a soft emitted gluon the azimuth is 
distributed according to the eikonal dipole distribution~\cite{March1};
\item[($b$)] for non-soft emission one finds azimuthal correlations due to
spin effects. See~\cite{Knowl1,Knowl3} for the method used to
implement these correlations in full, to leading collinear logarithmic
accuracy, in \HW.
\end{itemize}
\item 
In each branching the scale of $\as$ is the relative transverse
momentum of the two daughter partons.
\par\noindent
\item 
In the case of heavy flavour production the mass of the quark modifies
the angular-ordered phase space.  The most important effect is that
the soft radiation in the direction of the heavy quark is
depleted. One finds that emission within an angle of order $M/E$ is
suppressed, $M$ and $E$ being the mass and energy of the heavy quark:
this is known as the ``dead cone''~\cite{deadcone}.
\end{enumerate}
Specifically, the \HW\ parton shower evolution is done in terms of the
parton energy fraction $z$ and an angular variable $\xi$.  In the
parton splitting $i\to jk$, $z_j = E_j/E_i$ and $\xi_{jk} = (p_j\cdot
p_k)/(E_j E_k)$.  Thus $\xi_{jk}\simeq{1\over 2}\theta_{jk}^2$ for
massless partons at small angles.  The values of $z$ are chosen
according to the DGLAP splitting functions and the distribution of
$\xi$ values is determined by the Sudakov form factors. Angular
ordering implies that each $\xi$ value must be smaller than the $\xi$
value for the previous branching of the parent parton.

\looseness=-1 The parton showers are terminated as follows. For partons there is a
cutoff of the form $Q_i = m_i + Q_0$, where $m_i$ ($i=1,\dots,6$ for
$d$, $u$, $s$, $c$, $b$, $t$) is set by the relevant mass parameter
$\vv{RMASS}(i)$ and $Q_0$ is set by the quark and gluon virtuality
cutoff parameters \vv{VQCUT} and \vv{VGCUT} (see
section~\ref{sec:params}).  Showering from any parton stops when a value
of $\xi$ below $Q_i^2/E_i^2$ is selected for the next branching. For
heavy quarks, the condition $\xi > Q_i^2/E_i^2$ corresponds to the
``dead cone" mentioned above.  At this point the parton is put on
mass-shell or given a small non-zero effective mass in the case of
gluons.\footnote{The quark mass parameters should also be thought of
as effective or constituent masses rather than current quark masses.}
Working backwards from these on-shell partons, one can now construct
the virtual masses of all the internal lines of the shower and the
overall jet mass, from the energies and opening angles of the
branchings. Finally one can assign the azimuthal angles of the
branchings, including EPR-type correlations (from
Einstein-Podolski-Rosen \cite{EPR}), and deduce completely all the
4-momenta in the shower.

\pagebreak[3]

Next the parton showers are used to replace the (on mass-shell)
partons that were generated in the original hard process. This is done
in such a way that the jet 3-momenta have the same directions as the
original partons in the c.m.\ frame of the hard process, but they are
boosted to conserve 4-momentum taking into account their extra masses.

The main improvements in the final-state emission algorithm of
\HW\ version~6, relative to version 5.1, are as follows.

The Sudakov form factors can be calculated using the one-loop or
two-loop $\as$, according to the variable \vv{SUDORD} (default = 1).  The
parton showering  still incorporates  the two-loop $\as$ in either
case but if \vv{SUDORD} = 1 this is done using a veto algorithm,  whereas
if \vv{SUDORD} = 2 no vetoes  are used in the  final-state evolution.
The usefulness of this option is discussed briefly in section~\ref{form}.

Matrix element corrections have been introduced into final-state
parton showers in $\ee$ and deep inelastic processes~\cite{Seym0,Seym},
in heavy flavour decays~\cite{CorSey}
and in Drell-Yan processes~\cite{Corcella:2000gs}~(see
section~\ref{mecorr}).

In \HW, the angular-ordering constraint, which is derived for soft
gluon emission, is applied to all parton shower vertices, including
$g\to q\qbar$.  In versions before 6.1, this resulted in a severe
suppression (an absence in fact) of configurations in which the gluon
energy is very unevenly shared between the quarks.  For light quarks
this is irrelevant, because in this region one is dominated by gluon
emissions, which are correctly treated.  However, for heavy quarks
this energy sharing (or equivalently the quarks' angular distribution
in their rest frame) is a directly measurable quantity and was badly
described.  Related to this was an inconsistency in the calculation of
the Sudakov form factor for $g\to q\qbar$. This was calculated using
the entire allowed kinematic range (with massless kinematics) for the
energy fraction, $0\le z\le 1$, while the $z$ distribution generated
was actually confined to the angular-ordered region, $z,1\!-\!z \ge
m/E\sqrt{\xi}$.

In version~6, these defects are corrected as follows.  We generate the
$E,\xi$ and $z$ values for the shower as before.  We then apply an
\emph{a posteriori} adjustment to the kinematics of the $g\to q\qbar$
vertex during the kinematic reconstruction.  At this stage, the masses
of the $q$ and $\qbar$ showers are known.  We can therefore guarantee
to stay within the kinematically allowed region.  In fact, the
adjustment we perform is purely of the angular distribution of the $q$
and $\qbar$ showers in the $g$ rest frame, preserving all the masses
and the gluon four-momentum.  Therefore we do not disturb the
kinematics of the rest of the shower at all.

Although this cures the inconsistency above, it actually introduces a
new one: the upper limit for subsequent emission is calculated from
the generated $E,\xi$ and $x$ values, rather than from the
finally-used kinematics.  This correlation is of NNLL importance, so
we can formally neglect it.  It would be manifested as an incorrect
correlation between the masses and directions of the produced $q$ and
$\qbar$ jets.  This is, in principle, physically measurable, but it
seems less important than getting the angular distribution itself
right.  In fact the solution we propose maps the old angular
distribution smoothly onto the new, so the sign of the correlation
will still be preserved, even if the magnitude is wrong.
Even with this modification, the \HW\  kinematic reconstruction can
only cope with particles that are emitted into the forward hemisphere in
the showering frame.  Thus one cannot populate the whole of
kinematically-allowed phase space.  Nevertheless, we find that this is
usually a rather weak condition and that most of phase space is
actually populated.

Using this procedure, we find that the predicted angular distribution
for secondary $b$ quarks at LEP energies is well-behaved, i.e.\ it
looks reasonably similar to the leading-order result
($1+\cos^2\theta^*$), and has relatively small hadronization
corrections.

Real photon emission is included in timelike parton showering.  The
infrared photon cutoff is \vv{VPCUT}, which defaults to 0.4\,GeV.
Agreement with LEP data is satisfactory if showering is used together
with the matrix element correction to produce photons in the
back-to-back region.  The results are insensitive to \vv{VPCUT}
variations in the range 0.1--1.0\,GeV. Setting \vv{VPCUT} greater than
the total c.m.\ energy switches off such emission.  As an expedient
way of improving the efficiency of photon final-state radiation
studies, the electromagnetic coupling \vv{ALPHEM} can be multiplied by
a factor \vv{ALPFAC} (default = 1) for all quark-photon vertices in jets,
and in the `dead zone' in $\ee$.  Results at small photon-jet
separation become sensitive to \vv{ALPFAC} above about 5.

\subsubsection{Initial-state showers}

The theoretical analysis of initial-state showering is more complex
than the final-state case. The most relevant parameters of the hard
subprocess are the hard scale $Q$ and the energy fraction $x$ of the
incoming parton after the emission of initial state radiation.  For
lepton-hadron processes $x$ corresponds to the Bjorken variable, while
for hadron-hadron processes $x$ is related to $Q^2/s$ where $s$ is the
c.m.\ energy squared.

The main result is that for any value of $x$, even for $x$
small~\cite{GMcoh}, the initial-state emission process factorizes and
can be described as a \emph{coherent branching} process suitable for
\MC\ simulations.  The properties which characterize this process
include all those discussed above for the final-state
emission. However, in the initial-state case the angular-ordering
restriction on the phase space applies to the angles $\theta_i$
between the directions of the incoming hadron and the emitted partons
$i$.

For large $x$, the coherent branching algorithm sums
correctly~\cite{Catani1} not only the leading but also the
next-to-leading contributions.  This accuracy allows us to identify
the relation between the \QCD\ scale used in the \MC\ program and the
fundamental parameter $\lms$.  This is achieved by using the one-loop
Altarelli-Parisi splitting functions and the two-loop expression for
$\as$ with the following universal relation between the scale
parameter $\Lambda_{\rm phys}$ ~\cite{Catani1} used in the simulation
and $\lms$ (here, $N_f$ is the number of flavours)
$$
\Lambda_{\rm phys} = \exp\left(\frac{67 - 3 \pi^2 - 10 N_f/3}
{2 (33 - 2 N_f)}\right) \lms \;\;
\simeq 1.569\,\lms \qquad \hbox{for $N_f=5$}\,.
$$
Therefore a \MC\ simulation with next-to-leading accuracy can be used
to determine $\lms$ from semi-inclusive data at large momentum
fractions.\footnote{This applies also to final-state emission, i.e.\
to jet fragmentation at large values of the jet momentum fraction.}

In the case of small values of $x$, the initial state branching
process has additional properties, which are not yet included fully in
\HW.  This was discussed in ref.~\cite{hw51} and the situation remains
unchanged since version 5.1.

The initial-state branching algorithm in \HW\ is of the \emph{backward
evolution} type. It proceeds from the elementary subprocess, at a hard
scale set by colour coherence (see section~\ref{sec:elem}), back to the
hadron scale, set here by the spacelike cutoff parameter
\vv{QSPAC}.  At this point there is a forced non-perturbative stage of
branching which ensures that the emitting parton fits smoothly with
the valence parton distributions of the incoming hadron.

Matrix-element corrections have been introduced into initial-state
parton showers in deep inelastic~\cite{Seym} and Drell-Yan
processes~\cite{Corcella:2000gs}, as discussed in the following
subsection.

To avoid double-counting of hard parton emission, all radiation at
transverse momenta greater than the hard process scale \vv{EMSCA} is
vetoed.  In the case of initial-state radiation, this affects all
events, while for final-state radiation it only affects those events
in which the two jets have a rapidity difference of more than about
3.4.

In the backward evolution of initial-state radiation for photons the
``anomalous'' branching $q\qbar\leftarrow\gamma$ is included.  Variables
\vv{ANOMSC}(1,\vv{IBEAM}) and \vv{ANOMSC}(2,\vv{IBEAM}) record the
evolution scale and transverse momentum, respectively, at which an
anomalous splitting was generated in the backward evolution of beam
\vv{IBEAM}.  If zero, then no such splitting was generated.

The treatment of forced branching of gluons and sea (anti-)quarks in
backward evolution has been improved, by allowing it to occur at a
random scale between the spacelike cutoff \vv{QSPAC} and the infrared
cutoff, instead of exactly at \vv{QSPAC} as before.  A new option
\vv{ISPAC} = 2 allows the freezing of structure functions at the scale
\vv{QSPAC}, while evolution continues to the infrared cutoff.  The
default, \vv{ISPAC} = 0 is equivalent to previous versions, in which
perturbative evolution stops at \vv{QSPAC}.

The width of the (gaussian) intrinsic transverse momentum distribution
of valence partons in incoming hadrons at scale \vv{QSPAC} is set by
the parameter \vv{PTRMS} (default value zero).  The intrinsic
transverse momentum is chosen before the initial state cascade is
performed and is held fixed even if the generated cascade is rejected.
This is done to avoid correlation between the amount of perturbative
and non-perturbative transverse momentum generated.

It is possible to completely switch off initial-state emission, by
setting \vv{NOSPAC} = \vv{.TRUE.}, in which case only the forced
splitting of non-valence partons is generated.

\subsubsection{Matrix-element corrections}\label{mecorr}

One of the new features of \HW\ 6 is the matching of first-order
matrix elements with parton showers.

The \HW\ parton showers are performed in the soft or collinear
approximation and emission is allowed only in regions of the phase
space satisfying the condition $\xi<1$ or $\xi<z^2$, for the final-
(timelike branching) and initial-state (spacelike branching) radiation
respectively, where $\xi$ and $z$ are the showering variables defined
above.

The emission is entirely suppressed inside the so-called dead zones
($\xi >1$ or $\xi>z^2$), corresponding to hard and/or large-angle
parton radiation.  According to the exact matrix elements, the
radiation in the dead zones is suppressed, since it is not soft or
collinear logarithmically enhanced, but it is not completely absent as
happens in the \HW\ standard shower algorithm. The \HW\ parton
cascades need to be supplemented by matrix-element corrections for a
full description of the physical phase space.

The method of matrix-element corrections to the \HW\ parton showers is
discussed in~\cite{Seym,Seym1}.  The radiation in the dead zones is
generated according to the exact first-order matrix element (`hard
correction'); the shower in the already-populated region of the phase
space is corrected by the use of the exact ${\cal O}(\alpha_S)$
amplitude any time an emission is capable of being the `hardest so
far' (`soft correction'\footnote{We point out that in the expression
`soft correction', `soft' refers to the phase space where such
corrections are applied and not to the amplitude, since we still use
the `hard' exact matrix element for the soft correction as well.}).
 
By `hardest-so-far', we mean the radiation of a parton
whose transverse momentum relative to the splitting one is larger than
all those previously emitted. This is not always the first emission, 
as angular ordering does not necessarily imply ordering in transverse
momentum. As shown in~\cite{Seym}, if we corrected only the first
emission, we would have problems in the implementation of the Sudakov
form factor~whenever~a subsequent harder emission occurs, as we would
find that the probability of hard radiation would depend on the
infrared cutoff, which is clearly unphysical. Using the ${\cal O}
(\alpha_S)$ result for the hardest-so-far emission in the
already-filled phase space as well as in the dead zone allows one to
have matching over the boundary of the dead zone itself.

Since the fraction of events which receive a hard correction is
typically small, we neglect multiple hard emissions in the dead zones
and rely on the first-order result plus showering in those regions.

\looseness=-2 Our method is quite different from the one used to
implement matrix-element cor-\linebreak rections in \JS\,\cite{jetset}, where the
parton shower probability is applied over~the~whole phase space and
the first-order amplitude is used only to correct the first emission.

\pagebreak[3]

Following these general prescriptions, matrix-element corrections have
been implemented in some $\ee$ processes~\cite{Seym0} (including
$\ee\to WW/ZZ$), deep inelastic lepton scattering~\cite{Seym2}, top
quark decay~\cite{CorSey}, and Drell-Yan
processes~\cite{Corcella:2000gs}.
Those for the process $gg\to$~Higgs are now in progress \cite{ggME}.

The variables \vv{HARDME} (default = \vv{.TRUE.}) and \vv{SOFTME}
(default = \vv{.TRUE.})  allow respectively the application of hard and
soft matrix-element corrections to the \HW\ parton cascades.

\subsection{Heavy flavour production and decay}\label{heavy}
  
Heavy quark decays are treated as  secondary hard subprocesses. 
Top quarks and any hypothetical  heavier quarks  always decay before
hadronization. Heavy-flavoured hadrons are split into 
collinear heavy quark and spectator and the former decays
independently.  After decay, parton showers may be  generated  from 
coloured  decay products, in the usual way. See ref.~\cite{March4} for
details  of the  treatment of colour coherence in these showers.

In \HW\ version~6 matrix-element corrections to the simulation of top
quark decays are available.  The routine \vv{HWBTOP} implements the
hard corrections; \vv{HWBRAN} has been modified to implement the soft
corrections.  Since the dead zone includes part of the soft
singularity, a cutoff is required: only gluons with energy above
\vv{GCUTME} (default value 2\,GeV) in the top rest frame are
corrected. Physical quantities are not strongly dependent on
\vv{GCUTME} in the range 1 to 5\,GeV, after the typical experimental
cuts are applied. For more details see ref.~\cite{CorSey}.

The structure of the program has been altered so that the secondary
hard subprocess and subsequent fragmentation associated with each
partonic heavy hadron decay appear separately in the event
record. Thus top quark decays are treated individually as are any
subsequent bottom hadron partonic decays. Note that the statement
\vv{CALL HWDHOB}, which deals with the decays of all heavy objects
(including \SY\ particles), must appear in the main program between
the calls to \vv{HWBGEN} and \vv{HWCFOR}, in order to carry out any
decays before hadronization.

The partonic decay fractions of heavy quarks are specified in the
decay tables like the decay modes of other particles. This permits
different decays to be given to individual heavy hadrons.  Changes to
the decay table entries can be made on an event by event basis if
desired. Partonic decays of charm hadrons and quarkonium states are
also now supported.  The order of the products in a partonic decay
mode is significant.  For example, if the decay is $Q \to W+q \to
(f+\bar f')+q$ occurring inside a $Q\sbar$ hadron, the required
orderings are:
\begin{eqnarray*}
Q+\sbar &\to& (f+\bar f')+(q+\sbar) \\
        &\mbox{or}&(q+\bar f')+(f+\sbar)
\qquad\hbox{(`colour rearranged')}\,.
\end{eqnarray*}
In both cases the $(V-A)^2$ matrix element-squared is proportional to
$(p_Q\cdot p_2)(p_1\cdot p_3)$, where $p_1$ etc.\ correspond to the
ordering given.  Decays of heavy-flavoured hadrons to exclusive
non-partonic final states are also supported. No check is made against
double counting from partonic modes. However this is not expected to
be a major problem for the semi-leptonic and two-body hadronic modes
supplied.

The default masses of the $c$ and $b$ quarks have been lowered to 1.55
and 4.95 respectively: this corresponds to the mass of the lightest
meson minus the $u$ or $d$ quark mass. This increases the number of
heavy mesons, and hence total multiplicities, and slightly softens
their momentum spectrum.  The rate of photoproduced charm states
increases and B-$\pi$ momentum correlations become smoother.  The
default top quark mass is 174.3\,GeV/$c^2$.  The same value is used in
the production and decay matrix elements and for all kinematics. Note
that higher-order corrections are not fully included, and so the \HW\
top mass does not necessarily correspond to that defined in any
particular rigorous scheme (e.g.\ the pole mass or the
$\overline{\mbox{MS}}$ running mass).  However, since it is probably
the decay kinematics that are most sensitive to this parameter, it
should be close to the pole mass.  See subsection~\ref{quarkmass} for
notes on the treatment of quark masses in various processes.

\subsection{Gauge and Higgs boson decays}\label{sec:gaugebosons}

\TABULAR{|c|c|c|}{\hline
$\vv{MODBOS(}i)$&$W^\pm$ Decay&$Z^0$ Decay\\
\hline
0&all&all\\
1&$q\qbar$&$q\qbar$\\
2&$e\nu$&$\ee$\\
3&$\mu\nu$&$\mu^+\mu^-$ \\
4&$\tau\nu$&$\tau^+\tau^-$\\
5&$e\nu + \mu\nu$&$\ee + \mu^+\mu^-$\\
6&all&$\nu\bar\nu$ \\
7&all&$b\bbar$\\
$>7$&all&all\\
\hline}{$W,Z$ decays.\label{tab.1}}

{\sloppy The total decay widths of the electroweak gauge bosons $V=W,Z$ are
specified by the input parameters \vv{GAMW} and \vv{GAMZ}.  Their
branching fractions to various final states are computed automatically
from the other \SM\ input parameters.  Which decays actually occur is
controlled as follows. The variable $\vv{MODBOS}(i)$ controls
the decay of the $i$th gauge boson per event (table~\ref{tab.1}).

}
All entries of \vv{MODBOS} default to 0.  Bosons which are produced in
pairs (i.e.\ from $VV$ pair production, or Higgs decay) are
symmetrized in $\vv{MODBOS}(i)$ and $\vv{MODBOS}(i+1)$.  For processes
which directly produce gauge bosons, the event weight includes the
branching fraction to the requested decay, but this is only true for
Higgs production if decay to $W^+W^-/Z^0Z^0$ is forced (\vv{IPROC} = 310,
311 but not 399, etc.).  Users can thus force $Z \to b\bbar$ decays,
with $\vv{MODBOS}(i)=7$.  For example, \vv{IPROC} = 250, \vv{MODBOS}(1) = 7,
\vv{MODBOS}(2) = 0 gives $Z^0Z^0$ production with one $Z^0$ decaying to
$b\bbar$.

The spin correlations in the decays are handled in one of two ways:
\begin{enumerate}
\item The diagonal members of the spin density matrix are stored in
$\vv{RHOHEP}(i,\vv{IHEP})$, where $i=1,2,3$ for helicity$=i-2$ in the
centre-of-mass frame of their production, for processes where this
matrix is diagonal (i.e.\ there is no interference between spin
states).
\item The correlations in the decay are handled directly by the
production routine where (1) is not possible.
\end{enumerate}

The processing of the parton showers in hadronic $W$ and $Z$ decays is
handled in the rest frame of the vector boson if \vv{WZRFR} is
\vv{.TRUE.}  (the default), otherwise in the lab frame.  In the latter
case, which was the default in earlier versions, the initial cone
angles of the showers depend on the velocity of the boson, which leads
to a slight Lorentz non-invariance of decay distributions.

The total decay width of the \SM\ Higgs boson is computed from its
input mass \vv{RMASS(201)} and stored as \vv{GAMH}. Its decay branching
fractions are also computed and stored in \vv{BRHIG(I)}: \vv{I} = 1--6 for
$d\dbar$,\dots, $t\tbar$; \vv{I} = 7--9 for $\ee$,\dots, $\tau^+\tau^-$;
\vv{I} = 10,11,12 for $W^+W^-$, $Z^0Z^0$, $\gamma\gamma$.  Non-\SM\ Higgs
bosons, on the other hand, such as those in supersymmetric models,
have to have their widths and decay tables provided as input data (see
section~\ref{sec:SUSYdata}).  To avoid any ambiguity, the \SM\ Higgs
boson has a distinct identity code in \HW\ and is represented by the
special symbol $\SMH$.

There are two choices for the treatment of the \SM\ Higgs width, both
controlled by the variable \vv{IOPHIG}:
$$      
\vv{IOPHIG} = 2I + J\,,
$$
where $I$ and $J$ are both zero or one.  Whenever a Higgs boson is
generated, its mass is chosen from a distribution that, for heavy \SM\
Higgs bosons, can be rather broad.  The choice of $I$ makes a
significant difference to the physical meaning of the distribution
generated: for $I=0$, the cross section corresponds to the tree level
process containing an $s$-channel Breit-Wigner resonance for the Higgs
boson with a running Higgs width.  As discussed in~\cite{MHS}, this
neglects important contributions from interference with non-resonant
diagrams and can violate unitarity at high energy, so $I=1$ (the
default) uses the improved prescription of~\cite{MHS}.  This replaces
the $s$-channel propagator by an effective propagator that sums the
interference terms to all orders.  This increases the cross section
below resonance and decreases it above, causing an overall increase in
cross section.  More details can be found in~\cite{MHS}.

The variable $J$ is a more technical parameter that does not affect
the physical results, only the method by which they are generated:
$J=1$ (the default) generates the mass according to a fixed-width
Breit-Wigner resonance, while $J=0$ biases the distribution more
towards higher masses.  In either case, the appropriate jacobian
factor is included in the event weight, so that the physical cross
section is independent of $J$.

In all the above cases, the \SM\ Higgs mass distribution is restricted
to the range $[\mh-\vv{GAMMAX}\times\gh , \mh+\vv{GAMMAX}\times\gh].$
\vv{GAMMAX} defaults to 10, but in the non-perturbative region
$\mh\gtrsim 1$\,TeV should be reduced to protect against poor weight
distributions.  These considerations do not affect the distribution
noticeably for $\mh\lesssim~500$\,GeV, and \vv{GAMMAX} can safely be
increased if necessary.

For a \SY\ Higgs, the width is never large enough for unitarity to be
violated and these issues are unimportant.  In this case, the mass
distribution is chosen according to a fixed-width Breit-Wigner
resonance, like that of any other \SY\ particle.

The \SM\ Higgs decays that can occur are normally controlled by the
process code \vv{IPROC}, as in $\vv{IPROC}=300+\ID$ for example:
\ID = 1--6 for quarks, 7--9 for leptons, 10/11 for $W^+W^-/Z^0Z^0$
pairs, and 12 for photons.  In addition \ID = 0 gives quarks of all
flavours, and \ID = 99 gives all decays.  For each process, the average
event weight is the cross section in nb times the branching fraction
to the requested decay.  The branching ratios to quarks use the
next-to-leading logarithm corrections, those to $W^+W^-/Z^0Z^0$ pairs
allow for one or both bosons being off mass-shell.

All Higgs vertices include an optional enhancement factor to account
for non-\SM\ and non-\MS\ couplings.  The amplitudes for all
Higgs vertices are multiplied by the factor \vv{ENHANC(ID)} where
\ID\ is the same as in $\vv{IPROC}=300+\ID$ except the
$\gamma\gamma H$ `vertex' which is calculated from \vv{ENHANC(6)}
and \vv{ENHANC(10)} for the top and $W^\pm$ loops.  This allows the
simulation of the production of any chargeless scalar Higgs-like
particle.  Note however that pseudoscalar and charged Higgs
bosons, and processes involving more than one Higgs particle
(e.g.\ the decay $H^0\to h^0Z$) are not included this way (see
section~\ref{sec:SUSYproc}).

The array \vv{ENHANC(ID)} is initialised as usual in \vv{HWIGIN}.  Note,
however, that it will be overwritten if \MS\ Higgs production is
required by \vv{IPROC}. In that case, as mentioned earlier, the Higgs
widths and decay modes are simply read from an input particle data
file (see section~\ref{sec:SUSYdata}).

\subsection{Supersymmetry}\label{sec:SUSY}

\TABULAR{|lcclcc|}{
\hline
Particle & & Spin & Particle & & Spin \\
\hline
quark  & $q$ & 1/2 & squarks & $\ino{q}_{L,R}$ & 0 \\
charged lepton & $\ell$ & 1/2 & charged sleptons & $\ino{\ell}_{L,R}$ & 0 \\
neutrino  & $\nu$ & 1/2 &  sneutrino & $\ino{\nu}$ & 0 \\
gluon & $g$ & 1 & gluino & $\ino{g}$ & 1/2 \\
photon & $\gamma$ & 1 & photino & $\ino{\gamma}$ & 1/2 \\
neutral gauge boson &$Z^0$ & 1 & zino & $\ino{Z}$ & 1/2 \\
neutral Higgs bosons & $h^0,H^0,A^0$ & 0 & neutral Higgsinos &
$\ino{H}^{0}_{1,2}$ & 1/2 \\ 
charged gauge boson & $W^\pm$ & 1 & wino & $\ino{W}^\pm$ & 1/2 \\
charged Higgs boson & $H^\pm$ & 0 & charged Higgsino &
$\ino{H}^{\pm}$ & 1/2 \\ 
graviton & $G$ & 2 & gravitino & $\ino{G}$ & 3/2 \\
\hline
\multicolumn{6}{|c|}{\rule{0pt}{15pt}
$\ino{W}^\pm, \ino{H}^\pm$ mix to form 2 chargino 
mass eigenstates $\gaugino^{\pm}_{1}, \gaugino^{\pm}_{2}$}\\
\multicolumn{6}{|c|}{
$\ino{\gamma}$, $\ino{Z}$, $\ino{H}^0_{1,2}$ mix to
form 4
neutralino mass eigenstates $\ntlino{1},\ntlino{2},\ntlino{3},\ntlino{4}$}\\ 
\multicolumn{6}{|c|}{${\ino t}_L,{\ino t}_R$ (and similarly
${\ino b}, {\ino\tau}$) mix to form the mass eigenstates
${\ino t}_1, {\ino t}_2$} \\ 
\hline}{\SY\ particle content.\label{tab.2}}

\HW\ now includes the production and decay of superparticles, as given
by the Minimal Supersymmetric Standard Model (\MS)~\cite{Moretti:2002eu}.
The mass spectrum and decay modes, being read from input files (see
below), are completely general. The particle content, listed in the
following table (table~\ref{tab.2}), includes the gravitino/goldstino.
For sparticles that mix, the subscripts label the mass eigenstates in
the ascending order of mass.  The two Higgs Doublet Model (2HDM) Higgs
sector, intrinsic to the \MS, is also included. The three neutral
Higgs bosons are denoted by $h^0$, $H^0$ and~$A^0$.

The \SY\ particle names in \HW\ are as shown in table~\ref{tab.3}.
\vv{IDHW} is the \HW\ identity code and \vv{IDPDG} is the
corresponding Particle Data Group code~\cite{PDG}.

\TABULAR{|rccr|rccr|}{
\hline
 \vv{IDHW}& &  \vv{NAME}    &\vv{IDPDG}&
 \vv{IDHW}& &  \vv{NAME}    &\vv{IDPDG}\\\hline
\rule{0pt}{15pt}
401 & $\ino{d}_L$ & \vv{'SSDL    '} & 1000001 &
413 & $\ino{d}_R$ & \vv{'SSDR    '} & 2000001 \\ 
402 & $\ino{u}_L$ & \vv{'SSUL    '} & 1000002 &
414 & $\ino{u}_R$ & \vv{'SSUR    '} & 2000002 \\
403 & $\ino{s}_L$ & \vv{'SSSL    '} & 1000003 &
415 & $\ino{s}_R$ & \vv{'SSSR    '} & 2000003 \\
404 & $\ino{c}_L$ & \vv{'SSCL    '} & 1000004 &
416 & $\ino{c}_R$ & \vv{'SSCR    '} & 2000004 \\
405 & $\ino{b}_1$ & \vv{'SSB1    '} & 1000005 &
417 & $\ino{b}_2$ & \vv{'SSB2    '} & 2000005 \\
406 & $\ino{t}_1$ & \vv{'SST1    '} & 1000006 &
418 & $\ino{t}_2$ & \vv{'SST2    '} & 2000006 \\
& & & & & & & \\
425 & $\ino{e}_L$ & \vv{'SSEL-   '} & 1000011 &
437 & $\ino{e}_R$ & \vv{'SSER-   '} & 2000011 \\
426 & $\ino{\nu}_e$ & \vv{'SSNUEL  '} & 1000012 & & & & \\
427 & $\ino{\mu}_L$ & \vv{'SSMUL-  '} & 1000013 &
439 & $\ino{\mu}_R$ & \vv{'SSMUR-  '} & 2000013 \\
428 & $\ino{\nu}_\mu$ & \vv{'SSNUMUL '} & 1000014 & & & & \\
429 & $\ino{\tau}_1$ & \vv{'SSTAU1- '} & 1000015 &
441 & $\ino{\tau}_2$ & \vv{'SSTAU2- '} & 2000015 \\
430 & $\ino{\nu}_\tau$ & \vv{'SSNUTL  '} & 1000016 & & & & \\
& & & & & & & \\
449 & $\ino{g}$   & \vv{'GLUINO  '} & 1000021 &
458 & $\ino{G}$ & \vv{'GRAVTINO'} & 1000039 \\
450 & $\ntlino{1}$ & \vv{'NTLINO1 '} & 1000022 &
451 & $\ntlino{2}$ & \vv{'NTLINO2 '} & 1000023 \\
452 & $\ntlino{3}$ & \vv{'NTLINO3 '} & 1000025 &
453 & $\ntlino{4}$ & \vv{'NTLINO4 '} & 1000035 \\
454 & $\ino{\chi}^+_1$ & \vv{'CHGINO1+'} & 1000024 &
455 & $\ino{\chi}^+_2$ & \vv{'CHGINO2+'} & 1000037 \\
203 & $h^0$ & \vv{'HIGGSL0 '} &      26 &
204 & $H^0$ & \vv{'HIGGSH0 '} &      35 \\
205 & $A^0$ & \vv{'HIGGSA0 '} &      36 &
206 & $H^+$ & \vv{'HIGGS+  '} &      37 \\
\hline}{\SY\ particle names.\label{tab.3}}

Antiparticles generally appear in sequence after the corresponding
particles, e.g.\ antisquarks $\ino{d}_L^*-\ino{t}_1^*$ at
\vv{IDHW} = 407--412, $\ino{d}_R^*-\ino{t}_2^*$ at 419--424. They have
'\vv{BR}' added to the name, e.g.\ \vv{'SSDLBR '}, or opposite charge,
and negative \PD\ codes. A full list can be obtained using the print
option \vv{IPRINT} = 2 (see section~\ref{control}).

Note that the \HW\ particle labelling of the lightest \MS\ Higgs
boson departs from the \PD\ recommendation: it is given \PD\ code 26,
to avoid confusion with the \SM\ Higgs boson (\PD\ code 25) in our
implementation (specifically, in our use of the array \vv{ENHANC} for
the \MS\ processes: see the relevant Higgs sections for more
details).

\HW\ does not contain any built-in models for \SY\ scenarios beyond
the \MS, such as, Supergravity (\sss{SUGRA}) or Gauge Mediated
Symmetry Breaking (\sss{GMSB}). In all cases the \SY\ particle spectrum
and decay tables must be provided just like those for any other
particles.  The subroutine \vv{HWISSP}, if called, reads these from an
input file. The production subprocesses are then generated by
\vv{HWHESP}, in lepton-antilepton collisions, \vv{HWHSSP}, in
hadron-hadron collisions, or by one of the \RPV\ production
routines. The decays of the sparticles produced, as well as any top
quarks or Higgs bosons, are then performed by \vv{HWDHOB}.

\subsubsection{Data input}\label{sec:SUSYdata}

A package \IW, see section~\ref{sect:ISAWIG},
 has been created to work with \IS~\cite{ISA} to produce
a file containing the \SY\ particle masses, lifetimes and decay modes.
This package takes the outputs of the \IS\ SUGRA or general MSSM
programs and produces a data file in a format that can be read into
\HW\ by the subroutine \vv{HWISSP}. In principle the user can produce a
similar file provided that the correct format is used, as explained
below.

For the mixing terms of the \MS\ lagrangian we follow the
Haber-Kane~\cite{HabKan,HHG} conventions, so that we differ from \IS\
on the sign for gaugino masses, the ordering and signs of the gaugino
current eigenstates, the interchange of the rows and columns of the
gaugino mixing matrices, and the sign of the neutral Higgs mixing
angle $\alpha$.

In addition to the decay modes included in the \IS\ package \IW\
allows for the possibility of violating R-parity and includes the
calculation of all 2-body squark and slepton, and 3-body gaugino and
gluino R-parity violating (\RPV) decay modes.

It can happen that some of the \SY\ particle decay modes generated by
\IS\ are found to be kinematically forbidden in \HW, owing to the
slightly different values assumed for the light quark masses. In this
case a warning message is printed by \HW\ and these modes are deleted,
the other branching ratios being rescaled accordingly.  Such modes
normally have negligible \IS\ branching ratios anyway, because of
their tiny phase space.

The input file organisation expected by \vv{HWISSP} is as follows.
First the \SY\ particle and top quark masses and lifetimes (in seconds)
are given according to their \HW\ identity codes \vv{IDHW}, for example:
\small\begin{verbatim}
         65
         401 927.3980    0.74510E-25
         402 925.3307    0.74009E-25
         ....etc.
\end{verbatim}\normalsize
That is,
\begin{tabbing}
That is,m\=   \kill
         \>\vv{NSUSY} = Number of SUSY + top particles\\
         \>\vv{IDHW}, \vv{RMASS(IDHW)}, \vv{RLTIM(IDHW)}\\
         \>repeated \vv{NSUSY} times.
\end{tabbing}
Next each particle's decay modes together with their branching ratios
and matrix element codes are given as, for example:
\small\begin{verbatim}
         6
         401  0.18842796E-01     0   450     1     0     0     0
          :         :            :    :      :     :     :     :
         401  0.32755006E-02     0   457     2     0     0     0
         6
         402  0.94147678E-02     0   450     2     0     0     0
         ....etc.
\end{verbatim}\normalsize
That is,
\begin{tabbing}
That is,m\=   \kill
         \>Number of decay modes for a given particle \vv{IDK}\\
         \>\vv{IDK(IM)}, \vv{BRFRAC(IM)}, \vv{NME(IM)}, \vv{IDKPRD(1-5,IM)}\\
         \>repeated for each mode \vv{IM}\\
         \> all repeated for each particle (NSUSY times).
\end{tabbing}
The order in which the decay products appear is important in order to
obtain appropriate showering and hadronization.  The correct orderings
are indicated in the table below (table~\ref{tab.4}).

{\small
\TABULAR{@{}|c|c|c|c|c|c|@{}}{\hline
Decaying & No.\ of &Type of mode &
\multicolumn{3}{c|}{Order of decay products}\\\cline{4-6}
Particle & products & & 1st & 2nd & 3rd \\
\hline
top     &  2&  two-body to Higgs      & Higgs & bottom & \\
\cline{2-6}
& 3&  three-body via Higgs/W  &
\multicolumn{2}{c|}{quarks or leptons} & bottom \\
& &                   &\multicolumn{2}{c|}{from W/Higgs} & \\
\hline
gluinos  & 2 & without gluon & \multicolumn{2}{c|}{any order} &\\
\cline{3-6}
&  & with gluon    & gluon & colour  & \\
& &             &       & neutral & \\
\cline{2-6}
& 3 & R-parity conserved& colour  &
\multicolumn{2}{c|}{quark or antiquark} \\ 
&  &                & neutral & \multicolumn{2}{c|}{  } \\
\hline
squark & 2 &gaugino/gluino   & gaugino & quark  & \\ 
or slepton & & quark/lepton     & gluino  & lepton & \\
\cline{2-6}
&3 &  weak  & sparticle & \multicolumn{2}{c|}{particles from } \\
&  &  & & \multicolumn{2}{c|}{W decay } \\
\hline
squarks & 2 & lepton number violated & quark & lepton & \\
\cline{3-6}
&  & baryon number violated & quark & quark  & \\
\hline
sleptons & 2 & lepton number violated &
\multicolumn{2}{c|}{quark or antiquark} & \\
\hline
Higgs & 2 & (s)quark-anti(s)quark  &
\multicolumn{2}{c|}{(s)quark or anti(s)quark} & \\ 
\cline{3-6}
&  & (s)lepton-anti(s)lepton&
\multicolumn{2}{c|}{(s)lepton or anti(s)lepton} & \\
\cline{2-6}
& 3 & all three-body &  colour & \multicolumn{2}{c|}{quark or antiquark}\\
&   &           & neutral & \multicolumn{2}{c|}{lepton or antilepton} \\
\hline
gaugino & 2 & squark-quark & \multicolumn{2}{c|}{quark or squark} & \\
\cline{3-6}
&   & slepton-lepton & \multicolumn{2}{c|}{lepton or slepton} & \\
\cline{2-6}
& 3 & R-parity conserved &   colour &
\multicolumn{2}{c|}{quark or antiquark}\\
&   &                    &  neutral &
\multicolumn{2}{c|}{lepton or antilepton} \\
\hline
gaugino  & 3 & R-parity violating &
\multicolumn{3}{c|}{particles in the order $i,j,k$ as} \\
 or gluino&   &          &\multicolumn{3}{c|}{in the superpotential} \\\hline}{\SY\ decay product ordering.\label{tab.4}}}

New matrix element codes have been added for \SY\ and Higgs decays:
\begin{itemize}
\item \vv{NME}=200, describing the $1\to 3$ body heavy-quark decays
via a virtual $H^\pm$ boson. A  new function,
\vv{HWDHWT}, was introduced to this end. This can be used
to emulate e.g.\ $t\to b H^+(\to f\bar f')$ decays.
\item\vv{NME} = 300 for three-body \RPV\ gaugino and gluino decays.
\end{itemize}

The indices $i,j,k$ in \RPV\ gaugino/gluino decays refer to the
ordering of the indices in the \RPV\ couplings in the
superpotential. The convention is as in ref.~\cite{Rtheory}.

Next a number of parameters derived from the \SY\ lagrangian must be
given. These are: the ratio of Higgs VEVs, $\tan\beta$, and the scalar
Higgs mixing angle, $\alpha$; the mixing parameters for the Higgses,
gauginos and the sleptons; the trilinear couplings; and the Higgsino
mass parameter $\mu$.

Finally the logical variable \vv{RPARTY} must be set \vv{.FALSE.}  if
R-parity is violated, and the \RPV\ couplings must also be supplied;
otherwise not.

Details of the \vv{FORMAT} statements employed can be found by
examining the subroutine \vv{HWISSP}.

\vv{HWISSP} reads the data from \vv{UNIT = LRSUSY} (default
\vv{LRSUSY} = 66).  If the data are stored in a \vv{fort.LRSUSY} file on a
UNIX system\footnote{Or the equivalent, e.g.\ \vv{fortLRSUSY.dat} on a
VAX system.} no further action is required, but if the data are to be
read from a file named \vv{fname.dat} then appropriate \vv{OPEN} and
\vv{CLOSE} statements must be added by hand: 
\small\begin{verbatim}
OPEN(UNIT=LRSUSY,FORM='FORMATTED',STATUS='UNKNOWN',FILE='fname.dat')
CALL HWISSP CLOSE(UNIT=LRSUSY)
\end{verbatim}\normalsize
A number of sets of \SY\ parameter files, produced using \IS, for
the standard LHC SUGRA and GMSB points are available from the \IW\
home page:
\href{http://www.hep.phy.cam.ac.uk/~richardn/HERWIG/ISAWIG/}{\tt
http://www.hep.phy.cam.ac.uk/}$\sim${\tt richardn/HERWIG/ISAWIG/}

\subsubsection{Processes}\label{SUSYprocesses}

The implementation of supersymmetric particle production processes in
lepton-anti\-lepton and hadron-hadron collisions is described in
sections~\ref{sec_susy_lept} and~\ref{sec:SUSYproc}, respectively.  As
in \SM\ processes, we do not include any higher-order \QCD\
corrections to the relevant matrix elements, even though these are now
known in many cases. This is in order to avoid double counting of
corrections generated approximately by parton showering. 

The procedure by which the \MS\ matrix elements describing the
elementary hard subprocesses are interfaced to the initial- and
final-state parton showers is similar to that described in
sections~\ref{sec:elem} and~\ref{sec:showers} for the \SM\
case. However, at present showers are generated from partons but not
from spartons, whose short lifetimes should make this a reasonable
approximation.

We include matrix elements and 
spin correlations in all SUSY decays as described
in~\cite{Richardson:2001df}.  Breit-Wigner mass distributions
have been implemented for all unstable \SY\ particles, according to
their widths as given in the input file. For a list of the most
relevant decay modes of \MS\ particles, see ref.~\cite{Moretti:2002eu}.

One difference between the \SM\ and \MS\ implementations of the
Higgs decay channels should be mentioned. Whereas for the \SM\ Higgs
boson all decay rates are calculated by \HW\ itself, in terms of the
other (known) parameters of the \SM, in the case of the \MS\ scalars
these are passed to the generator through the data files.

A more detailed description of the new \SY\ reactions in \HW, along
with the relevant formulae for the hard scattering processes involved,
can be found in~\cite{Moretti:2002eu}.

We include the possibility of violating R-parity both in the decays of
the sparticles and in the initial hard subprocess.  The \RPV\ model we
consider is specified by a spectrum which can be given in either the
\sss{SUGRA}, \sss{GMSB} or a more general \MS\ scenario, and a set of
\RPV\ couplings at the weak scale. We include only the tri-linear
couplings and not the bi-linear terms which mix the leptons and
gauginos; a recent review of these models can be found
in~\cite{herbi}. However we do include the possibility that more than
one of the \RPV\ couplings can be non-zero.  All the two-body squark
and slepton and three-body gaugino and gluino decay modes, and
resonant production processes in hadron-hadron collisions are
included, as well as a range of production processes in $e^+e^-$
collisions.  A decay matrix element is implemented for the three-body
decays. The colour structure of these events is very different from
that of the \MS\ \cite{Rtheory}, due to the presence of the
baryon-number violating (\BNV) vertex. This means that a new
subroutine \vv{HWBRCN} is required to handle the colour connections
between jets in this case. A full discussion of the colour connections
in these processes and the matrix elements for the cross sections and
decays can be found in~\cite{Rtheory}.

\subsubsection{Related changes}

A large number of changes have been made in \HW\ to enable \SY\
processes to be included in hadron-hadron collisions. The main changes
are:
\begin{itemize}
\item
The subroutine \vv{HWDHQK} has been replaced by \vv{HWDHOB} which does
both heavy quark and \SY\ particle decays.
\item
The subroutines \vv{HWBCON}, \vv{HWCGSP} and \vv{HWCFOR} have been
adapted to handle the colour connections found in normal \SY\ decays.
\item
The subroutine \vv{HWBRCN} has been included to deal with the inter-jet
colour connections arising in \RPV\ \SY\@. Also \vv{HWCBVI}, \vv{HWCBVT}
and \vv{HWCBCT} have been added to handle the hadronization of baryon
number violating (\BNV) \SY\ decays and processes. If the variable
\vv{RPARTY=.TRUE.} (default) then the old \vv{HWBCON} colour connection
code is used, else the new \vv{HWBRCN}.
\end{itemize}

\subsection{Spin correlations}
    In addition to the spin correlations in parton showers and vector
    boson decays discussed above, in versions 6.4 and higher
    spin correlation  effects are included in  processes where top
    quarks, $\tau$ leptons and \SY\ particles are produced, as described in
    \cite{Richardson:2001df}. At the moment the effects  are only calculated
    for the production of these particles  in the  following  processes:
    \vv{IPROC}=100-199, 700-799, 800-899, 1300-1399, 1400-1499, 1500-1599, 1700-1799,
    2000-2099, 2800-2825, 3000-3030, 4000-4199.
    However, if these particles  are  produced in other
    processes, the  spin  correlation  algorithm will still  be  used to
    perform  their  decays. The correlations  are also  included for the
    decay of the \MS\ Higgs bosons, regardless of how they are produced.

    The spin correlations are controlled by the logical
    variable\footnote{Default values for input variables are
      shown in square brackets.} \vv{SYSPIN} [\vv{.TRUE.}]
    which  switches  the  correlations  on. If  required the 
    correlations are initialised by the new routine \vv{HWISPN}. This routine
    initialises the two, three and four body matrix elements.

    The three and four  body matrix  elements can be  used separately to
    generate  the decay  distributions without spin correlation effects.
    These are switched on by the switches \vv{THREEB} [\vv{.TRUE.}]
for three body
    decay and \vv{FOURB} [\vv{.FALSE.}] for four body decays.
The four body decays
    are only important  in \SY\ Higgs studies,  and have small branching
    ratios. However, they take some time to initialise and are therefore
    switched off by default.

   The  initialisation  of the spin  correlations  and/or  decay matrix
    elements  can be  time  consuming  and we have therefore included an
    option to read/write the information.  The information is written to
    unit \vv{LWDEC} [88] and read from \vv{LRDEC} [0]. If either are zero
the data
    is not  written/read.  If \vv{IPRINT}=2 then information on the branching
    ratios  for the decay modes  and the maximum  weights for the matrix
    elements is outputted.

    If the spin correlation (\vv{SYSPIN}) or matrix element 
    switches  (\vv{THREEB}, \vv{FOURB}) are \vv{.TRUE.}, then the matrix element
    codes (\vv{NME} entries) for the decays concerned are not used;
    the calculated matrix elements are used instead.

    When we included spin correlations in \HW6.4 \cite{Richardson:2001df}
    we did not  include
    either R-parity violating  decays or decays producing  gravitinos in
    the  algorithm. This  led to \HW\  stopping when  such decays were 
    included. This of  course could be  stopped by  switching  the  spin
    correlations  off, i.e.\  \vv{SYSPIN}=\vv{.FALSE.}. In version 6.5,
    we have included the 
    relevant  matrix  elements for R-parity  violating  decays and  hard 
    processes and decays producing gravitinos. At the same  time we have
    made changes so that at both the initialisation and event generation
    stages many of the terminal warnings  which were caused  by the code
    not having the correct matrix elements are now information-only
    warnings. If you still  get terminal  error messages from any of the
    spin correlation  routines please let us know.

    The effect of polarization for incoming leptonic beams in
\MS\ and \RPV\ \SY\ processes has also been included. These effects are included both
    in the  production  of \SY\  particles  and via  the  spin  correlation
    algorithm in their decays.
\subsection{Hadronization}

For a general hard process in hadron-hadron collisions, we have to
consider:
($a$)~the representation of the incoming partons as constituents
of the incident hadrons;
($b$)~the conversion of the emitted partons into outgoing hadrons;
($c$)~the `underlying soft event' associated with the presence of
spectator partons.

The first of these is dealt with through the use of non-perturbative
parton distribution functions, which are discussed below in
section~\ref{sec_pdfs}, and by the remnant hadronization model.  The
cluster model for hadron formation, remnant hadronization and the
underlying event is as follows.

\subsubsection{Cluster model}\label{sec:cluster}

The preconfinement property mentioned in section~\ref{introduction} is
used by \HW\ as the basis for a simple hadronization model which is
local in colour and independent of the hard process and the
energy~\cite{Webb1,March1}.

After the perturbative parton showering, all outgoing gluons are split
non-pertur\-batively, into light quark-antiquark or
diquark-antidiquark pairs (the default option is to disallow diquark
splitting). At this point, each jet consists of a set of outgoing
quarks and antiquarks (also possibly some diquarks and antidiquarks)
and, in the case of spacelike jets, a single incoming valence quark or
antiquark.  The latter is replaced by an outgoing spectator carrying
the opposite colour and the residual flavour and momentum of the
corresponding beam hadron.

In the limit of a large number of colours, each final-state colour
line can now be followed from a quark/anti-diquark to an
antiquark/diquark with which it can form a colour-singlet
cluster.\footnote{The situation when baryon number is violated is more
complicated and is discussed in~\cite{Gibbs} for the Standard Model
and in~\cite{Rtheory} for R-parity violating SUSY models.}  By virtue
of pre-confinement, these clusters have a distribution of mass and
spatial size that peaks at low values, falls rapidly for large cluster
masses and sizes, and is asymptotically independent of the hard
subprocess type and scale.

The clusters thus formed are fragmented into hadrons. If a cluster is
too light to decay into two hadrons, it is taken to represent the
lightest single hadron of its flavour. Its mass is shifted to the
appropriate value by an exchange of 4-momentum with a neighbouring
cluster in the jet.  Similarly, any diquark-antidiquark clusters with
masses below threshold for decay into a baryon-antibaryon pair are
shifted to the threshold via a transfer of 4-momentum to a
neighbouring cluster.

Those clusters massive enough to decay into two hadrons, but below a
fission threshold to be specified below, decay isotropically
\footnote{Except for those containing a `perturbative' quark when
\vv{CLDIR} = 1 --- see below.}  into pairs of hadrons selected in the
following way. A flavour $f$ is chosen at random from among $u$, $d$,
$s$, the six corresponding diquark flavour combinations, and $c$. For
a cluster of flavour $f_1 \bar{f_2}$, this specifies the flavours $f_1
\bar{f}$ and $f \bar{f_2}$ of the decay products, which are then
selected at random from tables of hadrons of those flavours. See
section~\ref{pdata} for details of the hadrons included.  The selected
choice of decay products is accepted in proportion to the density of
states (phase space times spin degeneracy) for that
channel. Otherwise, $f$ is rejected and the procedure is repeated.

The above method of selection for cluster decays is simple and fast
but does not automatically satisfy constraints such as strong isospin
symmetry. The decay rate into hadrons of a certain flavour depends on
the average phase space for channels involving that flavour.  Thus,
for example, the existence of the $\eta$ or $\eta'$, with the same
quark content as the $\pi^0$, leads to a slight reduction of direct
$\pi^0$ production relative to $\pi^+$ and $\pi^-$. Quantitatively,
the effect is too small to be observed even with the high statistics
of the LEP1 data.  However, the method can give rise to strange
effects if the particle data tables are extended, and modifications to
avoid this have been proposed~\cite{Kupco}.

In the decays of clusters to $\eta$ or $\eta'$, the parameter
\vv{ETAMIX} gives the $\eta_8/\eta_0$ mixing angle in degrees (default
= --20). Rates are not very sensitive to its exact value, as the
$\eta'/\eta$ suppression is dominated by mass effects in the cluster
model. See section~\ref{pdata} for more details.

A fraction of clusters have masses too high for isotropic two-body
decay to be a reasonable ansatz, even though the cluster mass spectrum
falls rapidly (faster than any power) at high masses. These are
fragmented using an iterative fission model until the masses of the
fission products fall below the fission threshold. In the fission
model the produced flavour $f$ is limited to $u$, $d$ or $s$ and the
product clusters $f_1 \bar{f}$ and $f \bar{f_2}$ move in the
directions of the original constituents $f_1$ and $\bar{f_2}$ in their
c.m.\ frame. Thus the fission mechanism is not unlike string
fragmentation~\cite{Lund}.

In \HW\ there are three main fission parameters, \vv{CLMAX}, \vv{CLPOW}
and \vv{PSPLT}.  The maximum cluster mass parameter \vv{CLMAX} and
\vv{CLPOW} specify the fission threshold $M_f$ according to the formula
$$
M_f^{\vv{CLPOW}} = \vv{CLMAX}^{\vv{CLPOW}}+ (m_1 + m_2)^{\vv{CLPOW}}\,,
$$
where $m_1$ and $m_2$ are the quark mass parameters $\vv{RMASS}(i)$ for
flavours $f_1$ and $f_2$ (see section~\ref{sec:showers}).  The parameter
\vv{PSPLT} specifies the mass spectrum of the produced clusters, which
is taken to be $M^{\vv{PSPLT}}$ within the allowed phase
space. Provided the parameter \vv{CLMAX} is not chosen too small (the
default value is 3.35\,GeV), the gross features of events are
insensitive to the details of the fission model, since only a small
fraction of clusters undergo fission.  However, the production rates
of high-$p_t$ or heavy particles (especially baryons) are affected,
because they are sensitive to the tail of the cluster mass
distribution.
The default value of the power \vv{CLPOW} is 2.
Smaller values will  increase the  yield of  heavier clusters  (and hence of
baryons) for heavy quarks, without affecting light quarks much.  For
example, the default value gives no $b$-baryons (for the default value
of \vv{CLMAX}) whereas \vv{CLPOW} = 1.0 makes the ratio of $b$-baryons to
$b$-hadrons about 1/4.

There is also a switch \vv{CLDIR} for cluster decays. If \vv{CLDIR} = 1
(the default) then a cluster that contains a `perturbative' quark,
i.e.\ one coming from the perturbative stage of the event (the hard
process or perturbative gluon splitting) `remembers' its
direction. Thus when the cluster decays, the hadron carrying its
flavour continues in the same direction (in the cluster c.m.\ frame)
as the quark. This considerably hardens the spectrum of heavy hadrons,
particularly of $c$- and $b$-flavoured hadrons. It also introduces a
tendency for baryon-antibaryon pairs preferentially to align
themselves with the event axis (the `TPC/$2\gamma$ string
effect'~\cite{tpc}). \vv{CLDIR} = 0 turns off this option, treating
clusters containing quarks of perturbative and non-perturbative origin
equivalently.  In the \vv{CLDIR} = 1 option, the parameter \vv{CLSMR}
(default = 0.0) allows for a gaussian smearing of the direction of the
perturbative quark's momentum.  The smearing is actually exponential
in $1-\cos\theta$ with mean \vv{CLSMR}.  Thus increasing \vv{CLSMR}
decorrelates the cluster decay from the initial quark direction.

\looseness=-1 The process of $b$-quark hadronization requires special treatment and
the results obtained using \HW\ are still not fully satisfactory.
Generally speaking, it is difficult to obtain a sufficiently hard
B-hadron spectrum and the observed $b$-meson/$b$-baryon ratio. These
depend not only on the perturbative subprocess and parton shower but
also on non-perturbative issues such as the fraction of $b$-flavoured
clusters that become a single B meson, the fractions that decay into a
B meson and another meson, or into a $b$-baryon and an antibaryon, and
the fraction that are split into more clusters. Thus the properties of
$b$-jets depend on the parameters \vv{RMASS(5)}, \vv{CLMAX}, \vv{CLPOW}
and \vv{PSPLT} in a rather complicated way. In practice these
parameters are tuned to global final-state properties and one needs
extra parameters to describe $b$-jets.

A parameter \vv{B1LIM} has therefore been introduced to allow clusters
somewhat above the B$\pi$ threshold mass $M_{\rm th}$ to form a single B
meson if
$$
M < M_{\rm lim} = (1+\vv{B1LIM}) M_{\rm th}\,.
$$
The probability of such single-meson clustering is assumed to decrease
linearly for $M_{\rm th} < M < M_{\rm lim}$.  This has the effect of hardening
the B spectrum if \vv{B1LIM} is increased from the default value of
zero.  In addition, in version~6, the parameters \vv{PSPLT}, \vv{CLDIR}
and \vv{CLSMR} have been converted into two-dimensional arrays, with
the first element controlling clusters that do not contain a $b$-quark
and the second those that do. Thus tuning of $b$-fragmentation can now
be performed separately from other flavours, by setting \vv{CLDIR(2)} = 1
and varying \vv{PSPLT(2)} and \vv{CLSMR(2)}.  By reducing the value of
\vv{PSPLT(2)}, further hardening of the B-hadron spectrum can be
achieved.

\subsubsection{Underlying soft event}\label{sec:sue}

In hadron-hadron and lepton-hadron collisions there are `beam
clusters' containing the spectators from the incoming hadrons. In the
formation of beam clusters, the colour connection between the
spectators and the initial-state parton showers is cut by the forced
emission of a soft quark-antiquark pair.  The underlying soft event in
a hard hadron-hadron collision is then assumed to be a soft collision
between these two beam clusters. In a lepton-hadron collision the
corresponding `soft hadronic remnant' is represented by a soft
collision between the beam cluster and the adjacent cluster, i.e.\ the
one produced by the forced emission mentioned above.

The model used for the underlying event is based on the minimum-bias
$p\pbar$ event generator of the UA5 Collaboration~\cite{Soft},
modified to make use of our cluster fragmentation algorithm. This
model is explained in the following subsection.

Adding 10000 to the \HW\ process code \vv{IPROC} suppresses the
underlying event, in which case the beam clusters are simply
fragmented like other clusters, without any soft collision. The
parameter \vv{PRSOF} enables one to produce an underlying event in only
a fraction \vv{PRSOF} of events (default = 1.0). Adding 10000 to
\vv{IPROC} is thus equivalent to setting \vv{PRSOF} = 0.

A parameter \vv{BTCLM} is available to users to adjust the mass
parameter equivalent to \vv{PMBM1} (see below) in remnant cluster
formation.  Its default value, 1.0, is identical to previous versions.
There is also an option for the special treatment of the splitting of
clusters containing hadron (or photon) remnants. \vv{IOPREM} = 0 gives
the fragments a gaussian mass spectrum typical of soft processes.
When \vv{IOPREM} = 1 (default), the child containing the remnant is
treated as before but the other cluster, containing a perturbative
parton, is treated as a normal cluster, with mass spectrum
$M^{\vv{PSPLT}}$.

Two special remnant `particles' have been defined: \vv{'REMG '} with
\vv{IDHW} = 71, \vv{IDHEP} = 98 and \vv{'REMN '} with \vv{IDHW} = 72,
\vv{IDHEP} = 99.  These are remnant photons and nucleons respectively.
They are identical to photons and nucleons, except that gluons are
labelled as valence partons and, for the nucleon, valence quark
distributions are set to zero. They are used by an external package
for simulating multi-parton interactions, called
{\sf JIMMY}~\cite{jimmy}. See section~\ref{sect:JIMMY} for further details.

\subsubsection{Minimum bias processes}\label{minbiproc}

The minimum-bias event generator of the UA5 Collaboration~\cite{Soft}
starts from a parametrization of the $p\pbar$ inelastic charged
multiplicity distribution as a negative binomial distribution.  In
\HW\ version~6, the relevant parameters are made available to the user
for modification, the default values being the UA5 ones as used in
previous versions. These parameters are given in table~\ref{tab.5}.

\TABULAR[t]{|c|l|r|}{
\hline
Name&Description&Default\\
\hline
& & \\
\vv{PMBN1}  & $a$ in $\bar n =as^b+c$  & 9.110 \\
\vv{PMBN2}  & $b$ in $\bar n =as^b+c$  & 0.115 \\
\vv{PMBN3}  & $c$ in $\bar n =as^b+c$  & $-9.500$ \\
& & \\
\vv{PMBK1}  & $a$ in $1/k =a\ln s+b$  & 0.029  \\
\vv{PMBK2}  & $b$ in $1/k =a\ln s+b$  & $-0.104$ \\
& & \\
\vv{PMBM1}  & $a$ in $(M-m_1-m_2-a)e^{-bM}$ & 0.4  \\
\vv{PMBM2}  & $b$ in $(M-m_1-m_2-a)e^{-bM}$ & 2.0 \\
& & \\
\vv{PMBP1}  & $p_t$ slope for $d,u$ & 5.2  \\
\vv{PMBP2}  & $p_t$ slope for $s,c$ & 3.0  \\
\vv{PMBP3}  & $p_t$ slope for $qq$  & 5.2  \\
& & \\
\hline}{Soft/min.bias parameters.\label{tab.5}}

The first three parameters control the mean charged multiplicity $\bar
n$ at c.m.\ energy $\sqrt{s}$ as indicated. The next two specify the
parameter $k$ in the negative binomial charged multiplicity
distribution,
$$ 
P(n) = \frac{\Gamma(n+k)}{n!\,\Gamma(k)}
          \frac{(\bar n/k)^n}{(1+\bar n/k)^{n+k}}\,. 
$$
The parameters \vv{PMBM1} and \vv{PMBM2} describe the distribution of
cluster masses $M$ in the soft collision. These soft clusters are
generated using a flat rapidity distribution with gaussian
shoulders. The transverse momentum distribution of soft clusters has
the form
$$
P(p_t)\propto p_t\exp\left(-b\sqrt{p_t^2+M^2}\right)\,,
$$
where the slope parameter $b$ depends as indicated on the flavour of
the quark or diquark pair created when the cluster was produced.  As
an option, for underlying events the value of $\sqrt{s}$ used to
choose the multiplicity $n$ may be increased by a factor \vv{ENSOF} to
allow for an enhanced underlying activity in hard events. The actual
charged multiplicity is taken to be $n$ plus the sum of the moduli of
the charges of the colliding hadrons or clusters.

There is now also an interface to the multiple-interaction model
{\sf JIMMY}~\cite{jimmy}. For this purpose, several routines have been
added or modified.  New are \vv{HWHREM} for identifying and cleaning up
the beam remnants and \vv{HWHSCT} to administer the extra scatters.
Minor modifications to \vv{HWBGEN} and \vv{HWSBRN} suppress energy
conservation errors when \vv{ISLENT} $= -1$; \vv{HWSSPC} has an improved
approximation for remnant mass at high energies; and \vv{HWUPCM}
improves safety against negative square roots.

\subsection{Spacetime structure}\label{spacetime}

The space-time structure of events is available for all types of
subprocess. The production vertex of each parton, cluster, unstable
resonance and final-state particle is supplied in the \vv{VHEP} array
of \vv{/HEPEVT/}. Set $\vv{PRVTX} = \vv{.TRUE.}$ to include this
information when printing the event record (120 column format).  The
units are: $x,y,z$ in mm and $t$ in mm/$c$. In the case of partons and
clusters the production points are always given in a local coordinate
system with its origin at the relevant hard subprocess. This helps to
separate the fermi-scale partonic showers from millimetre-scale
distances possible in particle decays, for example the partonic decays
of heavy ($c,b$) hadrons. The vertices of hadrons produced in cluster
decays are always corrected back into the laboratory coordinate
system.

It is possible to vary the principal interaction point, assigned to
the c.m.\ frame entry in \vv{/HEPEVT/} (with $\vv{ISTHEP}=103$), by
setting $\vv{PIPSMR}=\vv{.TRUE.}$ The smearing is generated by the
routine \vv{HWRPIP} according to a triple gaussian given by parameters
\vv{VIPWID(I)} (\vv{I} = 1,2,3 for $x,y,z$ widths): the default values
correspond to LEP1.

It is also possible to veto particle decays that would occur outside a
specified volume by setting $\vv{MAXDKL}=\vv{.TRUE.}$ Each putative
decay is tested in \vv{HWDXLM} and if the particle would have decayed
outside the chosen volume it is frozen and labelled as final state.
Using $\vv{IOPDKL} = 1,2$ selects a cylindrical or spherical allowed
region (centred about the origin): then parameters \vv{DXRCYL},
\vv{DXZMAX} or \vv{DXRSPH} specify the dimensions of the region.

\subsubsection{Particle decays}

Lepton and hadron lifetimes (in seconds) are supplied in the array
\vv{RLTIM}.  In the case of \MS\ (s)particles, including Higgs
states, \vv{RLTIM} values are entered through the input files (see
discussion in section~\ref{SUSYprocesses}).  The lifetimes of heavy
quarks (top and any hypothetical extra generations) and weak bosons
(including the \SM\ Higgs) are derived from their calculated or
specified widths in \vv{HWUDKS}, whilst light quarks and gluons are
given an effective minimum width that acts as a lifetime cutoff --- see
below.  All particles whose lifetimes are larger than \vv{PLTCUT} are
set stable.

The proper (i.e.\ rest-frame) time $t^*$ at which an unstable lepton
or hadron decays is generated according to the exponential decay law
with mean lifetime $\VEV{t^*}=\tau\equiv\vv{RLTIM}$:
$$ 
\hbox{Prob}(\hbox{proper time}> t^*) =  \exp(-t^*/\tau)\,.
$$
The laboratory-frame decay time $t$ and distance travelled $d$ are
obtained by applying a boost: $t=\gamma t^*$, $d=\beta\gamma t^*$
where $\beta=v/c$ and $\gamma = 1/\sqrt{1-\beta^2}$.  The production
vertices of the daughter particles are then calculated by adding the
distance travelled by the mother particle as given above to its
production vertex.
The mean lifetime $\tau$ of a particle is set, taking into account its width
and virtuality, by:
$$
\tau(q^2)=\frac{\hbar\sqrt{q^2}}{\sqrt{(q^2-M^2)^2 + (\Gamma q^2/M)^2}}\,.
$$
This formula is used for all particles: light partons; heavy quarks
and weak bosons, which have appreciable widths; resonances; and
unstable leptons. It interpolates between $\tau=\hbar/\Gamma$ for a
particle that is on mass-shell and $\tau=\hbar\sqrt{q^2}/(q^2-M^2)$
for one that is far off mass-shell.

\subsubsection{Parton showers}

The above prescription, based on an exponential proper lifetime
distribution, is also used to describe parton showers. For light
quarks and gluons, whose natural widths are small, this could lead to
unreasonably large distances being generated in the final, low
virtuality steps of showering. To avoid this they are given a width
$\Gamma=\vv{VMIN2}/M$; the parameter \vv{VMIN2} (default value
0.1\,GeV$^2$) acts effectively as a lower limit on a parton's
virtuality. This is particularly important for the forced splitting of
gluons (see section~\ref{sec:cluster}), which uses
$\tau=\hbar\,\vv{RMASS(13)}/\vv{VMIN2}$.

\subsubsection{Hadronization}

In the case of a cluster its initial production vertex is taken as the
midpoint of a line perpendicular to the cluster's direction of travel
and with its two ends on the trajectories of the constituent
quark-antiquark pair. If such a cluster undergoes a forced splitting
to two clusters the string picture is adopted. The vertex of the light
quark pair is positioned so that the masses of the two daughter
clusters would be the same as those for two equivalent string
fragments. The production vertices of the daughter clusters are given
by the first crossing of their constituent $q\qbar$ pairs.  The
production positions of primary hadrons from cluster decays are
smeared, relative to the cluster position, according to a gaussian
distribution of width 1/(cluster mass).

\subsubsection{Colour rearrangement}

\HW\ version~6 contains a colour rearrangement model based on the
space-time structure of an event at the end of the parton shower. This
is illustrated in the simple example shown below where showering
results in a colour-neutral $qgg\qbar$ final state.  In the
conventional \HW\ hadronization model (corresponding to the default
value of the reconnection parameter, $\vv{CLRECO}=\vv{.FALSE.}$), after
a non-perturbative splitting of the final-state gluons, colour singlet
clusters are formed from neighbouring $q\qbar$ pairs: $(ij)(pq)(kl)$.
However when $\vv{CLRECO}=\vv{.TRUE.}$ the program first creates colour
singlet clusters as normal but then checks all (non-neighbouring)
pairs of clusters to test if a colour rearrangement lowers the sum of
the clusters' spatial sizes added in quadrature. A cluster's size
$d_{ij}$ is defined to be the Lorentz-invariant space-time distance
between the production points of its constituent quark $q_i$ and
antiquark $\qbar_j$. If an allowed alternative is found, that is,
$(ij)(kl)\to(il)(jk)$ such that $|d_{ij}|^2+|d_{kl}|^2 >
|d_{il}|^2+|d_{kl}|^2$, then it is accepted with a probability given
by the parameter \vv{PRECO} (default value $1/9$).

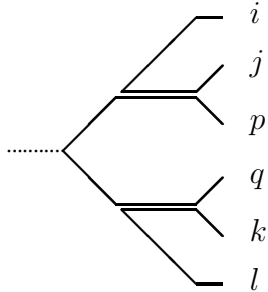
\setlength{\unitlength}{.71mm}
\begin{center}
{\begin{picture}(50,50)
\thicklines
\put(10,25){\line(1,1){10}}
\put(20,35){\line(1,0){15}}
\put(21,36){\line(1,1){14}}
\put(21,36){\line(1,0){14}}
\put(35,35){\line(1,-1){5}}
\put(35,36){\line(1, 1){5}}
\put(35,50){\line(1, 0){5}}
\put(10,25){\line(1,-1){10}}
\put(20,15){\line(1,0){15}}
\put(21,14){\line(1,-1){14}}
\put(21,14){\line(1,0){14}}
\put(35,15){\line(1, 1){5}}
\put(35,14){\line(1,-1){5}}
\put(35, 0){\line(1, 0){5}}
\put(45,49){$i$}
\put(45,40){$j$}
\put(45,29){$p$}
\put(45,19){$q$}
\put(45,8){$k$}
\put(45,-1){$l$}
\multiput(0,25)(1,0){10}{\circle*{.3}}
\end{picture}}
\end{center}

\noindent
Note that not all colour rearrangements are allowed, for instance in
the example shown $(ij)(pq)\to (iq)(jp)$ is forbidden since the
cluster $(jp)$ is a colour octet --- it contains both products from a
non-perturbative gluon splitting.

Multiple colour rearrangements are considered by the program, as are
those between clusters in jets arising from a single, colour neutral
source, for example $Z^0$ decay (as shown above), or due to more than
one source, for example $\ee\to W^+W^- \to 4$ jets. In the latter case
a new parameter, \vv{EXAG}, is available to exaggerate the lifetime of
the $W^\pm$ or any other weak boson, so that any dependence of
rearrangement effects on source separation can be investigated.

The \vv{CLRECO} option can be used for all the processes available in
\HW.  Note, however, that before using the program with
$\vv{CLRECO}=\vv{.TRUE.}$ for detailed physics analyses the default
parameters should be retuned to \LEP\ data with this option switched
on.

\subsubsection[B-$\overline{\mbox{B}}$ mixing]{
B-$\overline{\mbox{B}}$ mixing}

When $\vv{MIXING}=\vv{.TRUE.}$, particle-antiparticle mixing for
B$^0_{d,s}$ mesons is implemented. The probability that a meson is
mixed when it decays is given in terms of its lab-frame decay time $t$
by:
$$
P_{\rm mix}(t) =
\frac 1 2 -\frac{\cos(Xtm/c\tau E)}{2\cosh(Ytm/c\tau E)}\,,
$$
where $X=\Delta M/\Gamma$, $Y=\Delta\Gamma/2\Gamma$ and $m,\;\tau,\;E$
are the B$^0$ mass, lifetime and energy.  The ratios $X$ and $Y$ are
stored in \vv{XMIX(I)} and \vv{YMIX(I)}, $\vv{I}=1,2$ for $q=s,d$.
Whenever a neutral B meson occurs in an event, a copy of the original
entry is always added to the event record, with $\vv{ISTHEP}=200$,
which gives the particle's flavour at the production (or cluster
decay) time.  This is in addition to the usual decaying particle entry
with $\vv{ISTHEP}=199$.

\section{Processes}\label{processes}

\subsection{Beams}\label{beams}

\TABULAR[t]{|c|l|c|}{
\hline
Name&Description&Default    \\
\hline
\vv{PART1}    & Type of particle in beam 1       & \vv{'PBAR    '}\\
\vv{PART2}    & Type of particle in beam 2       & \vv{'P       '}\\
\vv{PBEAM1}   & Momentum of beam 1               & 900.0   \\
\vv{PBEAM2}   & Momentum of beam 2               & 900.0   \\
\vv{IPROC}    & Type of process to generate      & $\:1500$      \\
\vv{MAXEV}    & Number of events to generate     & $\;\;\,100$       \\
\hline}{Main program variables.\label{tab.6}}

As indicated in table~\ref{tab.6}, a number of variables must be set in
the main program \vv{HWIGPR} to specify what is to be simulated.  The
beam particle types \vv{PART1}, \vv{PART2} can take any of the values
\vv{NAME} listed in table~\ref{tab.7}.

\TABULAR{|l|c|l|c|}{
\hline
&\vv{NAME}& &\vv{NAME} \\
\hline
$e^+$      & \vv{'E+      '}& $e^-$         &\vv{'E-      '}\\
$\mu^+$    & \vv{'MU+     '}& $\mu^-$       &\vv{'MU-     '}\\
$\nu_e$    & \vv{'NU\_E    '}& $\bar\nu_e$   &\vv{'NU\_EBAR '}\\
$\nu_\mu$  & \vv{'NU\_MU   '}& $\bar\nu_\mu$ &\vv{'NU\_MUBAR'}\\
$\nu_\tau$ & \vv{'NU\_TAU  '}& $\bar\nu_\tau$&\vv{'NU\_TAUBR'}\\
$p$        & \vv{'P       '}& $\bar p$      &\vv{'PBAR    '}\\
$n$        & \vv{'N       '}& $\bar n$      &\vv{'NBAR    '}\\
$\pi^+$    & \vv{'PI+     '}& $\pi^-$       &\vv{'PI-     '}\\
$\gamma$   & \vv{'GAMMA   '} &  & \\
\hline}{Beam particles.\label{tab.7}}

\noindent
{\sloppy\looseness=-1 In the case of point-like photon/\QCD\ processes,
\vv{IPROC} = 5000--5999, the first particle must be the photon or a
lepton.  In addition, beams \vv{'K+ '} and \vv{'K- '} are supported for
minimum bias non-diffractive soft hadronic events ($\vv{IPROC}=8000$)
only.  In the case that the beam momenta $\vv{PBEAM1}$ and $\vv{PBEAM2}$
are not equal, the default procedure ($\vv{USECMF}=\vv{.TRUE.}$) is to
generate events in the beam-beam centre-of-mass frame and boost them
back to the laboratory frame afterwards.

} 

In hadronic processes with lepton beams (e.g.\ photoproduction in
$ep$), the lepton $\to$ lepton + photon vertex uses the full
transverse-momentum dependent splitting function, with exact
light-cone kinematics, i.e\ the Equivalent Photon Approximation
\pagebreak[3](EPA).  This means that the photon-hadron collision has a transverse
momentum in the lepton-hadron frame and must be boosted to a frame
where it has no transverse momentum. Thus the c.m.f.\ boost described
above is always used in these processes, regardless of the value of
\vv{USECMF}.  The correct lower energy cutoff appropriate to the
hadronic process is applied to the photon. The $Q^2$ of the photon is
generated within the kinematically allowed limits, or the user-defined
limits \vv{Q2WWMN} and \vv{Q2WWMX} (defaults 0 and 4) whichever is more
restrictive.\footnote{The \vv{WW} in parameter names is a relic from
earlier versions that used the less accurate Weizsacker-Williams
approximation.}  Similarly for the photon's light-cone momentum
fraction, with user-defined limits \vv{YWWMIN} and \vv{YWWMAX} (default
0 and 1).  Together with the Bjorken $y$-variable limits \vv{YBMIN} and
\vv{YBMAX}, this allows different ranges for the tagged and untagged
photons in two-photon \DIS.

\subsubsection{Parton distributions}\label{sec_pdfs}

The parton momentum fraction distributions of the beam particles are
used in the generation of initial-state parton showers and also in the
non-perturbative process of linking the shower with the beam hadron
and its remnant.  Since the parton showering is done in
leading-logarithmic order, there is no strong motivation to use
next-to-leading order parton distributions, although this has become
customary since the most up-to-date distributions are deduced from
next-to-leading order fits to (inclusive) data.  Thus the most common
option is to use the interface to the \PDF\ parton distribution
library~\cite{PDF}.

\looseness=-1 The \HW\ interface is compatible with \PDF\ version~4. \vv{AUTPDF}
should be set to the author group as listed in the \PDF\ manual, e.g.\
\vv{'MRS'}, and \vv{MODPDF} to the set number in the new convention.  It
is permissible to choose the \PDF\ set independently for each of the
two beams.  For example, to use MRS D-- for the proton and
Gordon-Storrow set 1 for the photon in $\gamma$-hadron or
lepton-hadron collisions, one sets:
\small
\begin{verbatim}
          AUTPDF(2)='MRS'
          MODPDF(2)=28
          AUTPDF(1)='GS'
          MODPDF(1)=2
\end{verbatim}
\normalsize

\noindent
\looseness=-1 If the \PDF\ interface is not used, the parton distributions
are chosen from the \HW\ internal sets according to the value of the
parameter \vv{NSTRU}.
  The default parton distributions in \HW\ versions prior to 6.3
  were very old and did not include fits to any of the HERA data.
  Therefore several new PDFs have been
  included in versions 6.3 and higher. These are shown in table~\ref{tab:MRST}.

  It should be noted that we have only added leading-order fits because
  the evolution algorithms in \HW, in particular the backward-evolution
  algorithm for initial-state parton showering,
  are only leading-order and therefore inconsistencies could occur
  with next-to-leading-order distributions.

  The new default structure function set \vv{NSTRU}=8 is the average of 
  two of the published fits \cite{Martin:1998np}, because this has been
  found \cite{Thorne} to be closer to 
  the central value of more recent next-to-leading-order fits. The
  other fits can then be used to assess the effects of varying the high-$x$
  gluon.

  These new \HW\ parton distributions are only available for nucleons.
  For pion beams, either the old \vv{NSTRU}=1,2 pion sets or \PDF\
  should be used.

For photons, the default is to use the Drees-Grassie parton
distributions \cite{Drees}.  The heavy quark content of the photon
uses the corrections to the Drees-Grassie distribution functions for
light quarks, calculated by Drees and Kim~\cite{DK}.  There is also an
interface to the Schuler-Sj\"ostrand~\cite{SchSjo} parton distribution
functions for the photon, version~2.  These appear as \PDF\ sets with
author group \vv{`SaSph'}, but are actually implemented via a call to
their \vv{SASGAM} code.  The value in \vv{MODPDF} specifies the set (1-4
for 1D [recommended set],1M, 2D,2M),
whether the Bethe-Heitler process is used for heavy flavours (add
10), whether the $P^2$-dependence is included (add 20), and which of
their $P^2$ models is used (add 100 times their \vv{IP2} parameter).

\begin{table}[!t]
\begin{center}
\begin{tabular}{|c|l|}
\hline
\vv{NSTRU} & Description\\
\hline
6 & Central $\alpha_S$ and gluon leading-order fit of \cite{Martin:1998np}\\
7 & Higher gluon leading-order fit of \cite{Martin:1998np}\\
8 & Average of central and higher gluon leading-order fits of \cite{Martin:1998np}\\
\hline
\end{tabular}\caption{New internal MRST parton distributions.}\label{tab:MRST}
\end{center}
\end{table}

An option to damp the parton distributions of off mass-shell photons
relative to on-shell photons, according to the scheme of Drees and
Godbole~\cite{DrGod} has been introduced.  The adjustable parameter
\vv{PHOMAS} defines the crossover from the non-suppressed to suppressed
regimes.  Recommended values lie in the range from \vv{QCDLAM} to
1\,GeV.  The default value \vv{PHOMAS} = 0 corresponds to no suppression,
as in previous versions.

\subsection{Summary of subprocesses}

We give in table~\ref{tab.8} a list of the currently available hard
subprocesses \vv{IPROC}.  More detailed descriptions are given in
sections~\ref{sec_lept}--\ref{sec_DIS}, and then in
section~\ref{including} there are instructions to users on how to add
a new process.
\setlongtables
\begin{longtable}[t]{|c|l|}
\hline
\vv{\vv{IPROC}} & Process \\
\hline
\endhead
\caption*{\small {\bf Table~\ref{tab.8}:} Process codes. (Continues)}
\endfoot
\caption*{\small {\bf Table~\ref{tab.8}:} Process codes.}
\endlastfoot
\label{tab.8}
   100     & $\l^+ \l^- \to q \qbar(g)$ (all $q$ flavours) \\
   100+\IQ & $\l^+ \l^- \to q \qbar(g)$ ($\IQ=1,2,3,4,5,6$ for
$q=d,u,s,c,b,t$)\\
   107     & $\l^+ \l^- \to g g (g)$ (fictitious process)\\
   110     & $\l^+ \l^- \to q \qbar g$ (all flavours)  \\
   110+\IQ & $\l^+ \l^- \to q \qbar g$ (\IQ\ as above) \\
   120     & $\l^+ \l^- \to q \qbar$ (all flavours, no hard gluon
correction)\\
   120+\IQ & $\l^+ \l^- \to q \qbar$ (\IQ\ as above, no hard gluon
correction)\\
   127     & $\l^+ \l^- \to g g$ (fictitious process, no hard gluon
correction)\\
   150+\IL & $\l^+ \l^- \to \l' \lbar'$ ($\IL=1,2,3$ for $\l'=e,\mu,\tau$,
N.B. $\l\neq\l'$) \\
\hline
   200     & $\l^+ \l^- \to  W^+ W^-$ (see sect.~\ref{electroweak} on
control of $W/Z$ decays) \\
   250     & $\l^+ \l^- \to  Z^0 Z^0$ (see sect.~\ref{electroweak} on
control of $W/Z$ decays) \\
\hline
   300     & $\l^+ \l^- \to Z^0\SMH \to Z^0 q \qbar$ (all flavours)  \\
   300+\IQ & $\l^+ \l^- \to Z^0\SMH \to Z^0 q \qbar$ (\IQ\ as above) \\
   306+\IL & $\l^+ \l^- \to Z^0\SMH \to Z^0 \l \lbar$ (\IL\ as above) \\
   310, 311& $\l^+ \l^- \to Z^0\SMH \to Z^0 W^+ W^-$, $Z^0 Z^0 Z^0$ \\
   312     & $\l^+ \l^- \to Z^0\SMH \to Z^0 \gamma \gamma $ \\
   399     & $\l^+ \l^- \to Z^0\SMH \to Z^0 $ anything \\
\hline
   400+\ID & $\l^+ \l^- \to \nu \bar\nu\SMH + \l^+ \l^-\SMH$ (\ID\ as in
$\vv{IPROC}=300+\ID$)\\
\hline
   500+\ID & $\begin{array}[t]{r}
	     \l^+ \l^- \to \l^+\l^-\gamma\gamma\to
	     \l^+\l^-q\qbar/\l\lbar/W^+W^- \\
	     \hbox{(\ID=0--10 as in $\vv{IPROC}=300+\ID$)}
	     \end{array}$ \\
   550+\ID & $\l^+ \l^- \to\l\nu_\l \gamma W\to
\l\nu_\l q\qbar'/\l\lbar'$ (\ID=0--9 as in $\vv{IPROC}=300+\ID$)\\
\hline
   600     & $\l^+ \l^-\to q\qbar gg, q\qbar q'\qbar'$ (all $q$ flavours) \\
   600+\IQ & $\l^+ \l^-\to q\qbar gg, q\qbar q'\qbar'$ (\IQ\ as above)\\
           & After generation, \vv{IHPRO} is subprocess (see
sect.~\ref{fourjet}) \\
\hline
   700-99  & Minimal Supersymmetric Standard Model (\MS) processes\\
   700     & $\l^+ \l^- \to$~2-sparticle processes (sum of 710, 730, 740
             and 760)\\
   710     & $\l^+ \l^- \to$~neutralino pairs (all neutralinos) \\
706+4\vv{IN1}+\vv{IN2} &$\l^+ \l^- \to \gaugino^0_{\mbox{\scriptsize IN1}}
			 \gaugino^0_{\mbox{\scriptsize IN2}}$
		      (\vv{IN1,2}=neutralino mass eigenstate)\\
   730     & $\l^+ \l^- \to$~chargino pairs (all charginos) \\
728+2\vv{IC1}+\vv{IC2} &$\l^+ \l^- \to \gaugino^+_{\mbox{\scriptsize IC1}}
			 \gaugino^-_{\mbox{\scriptsize IC2}}$
		      (\vv{IC1,2}=chargino mass eigenstate) \\
   740     & $\l^+ \l^- \to$~slepton pairs (all flavours) \\
   736+5\IL& $\l^+ \l^- \to \slepton_{L,R} \slepton_{L,R}^*$
	     ($\IL=1,2,3$ for $\slepton=\tilde{e},\tilde{\mu},\tilde{\tau}$) \\
   737+5\IL& $\l^+ \l^- \to \slepton_{L} \slepton_{L}^*$ (\IL\ as above) \\
   738+5\IL& $\l^+ \l^- \to \slepton_{L} \slepton_{R}^*$ (\IL\ as above)\\
   739+5\IL& $\l^+ \l^- \to \slepton_{R} \slepton_{R}^*$ (\IL\ as above)\\
   740+5\IL& $\l^+ \l^- \to \snu_{L} \snu_{L}^*$ 
             ($\IL=1,2,3$ for $\snu_e, \snu_\mu, \snu_\tau$) \\
   760      & $\l^+ \l^- \to$~squark pairs (all flavours) \\
   757+4\IQ & $\l^+ \l^- \to \squark_{L,R} \squark^*_{L,R}$
	     ($\IQ=1,2,3,4,5,6$ for $\squark=\tilde{d},\tilde{u},\tilde{s},
					     \tilde{c},\tilde{b},\tilde{t}$)\\
   758+4\IQ & $\l^+ \l^- \to \squark_{L} \squark^*_{L}$
		(\IQ\ as above)\\
   759+4\IQ & $\l^+ \l^- \to \squark_{L} \squark^*_{R}$
		(\IQ\ as above)\\
   760+4\IQ & $\l^+ \l^- \to \squark_{R} \squark^*_{R}$
		(\IQ\ as above)\\
\hline
   800-99  & R-parity violating supersymmetric processes \\
   800     & Single sparticle production, sum of 810--840 \\
   810     & $\l^+ \l^- \to \gaugino^0 \nu_i$, (all neutralinos)\\
   810+\IN & $\l^+ \l^- \to \gaugino^0_{\mbox{\scriptsize IN}} \nu_i$,
	     (\IN=neutralino mass state)\\
   820     & $\l^+ \l^- \to \gaugino^- e^+_i$ (all charginos) \\
   820+\IC & $\l^+ \l^- \to \gaugino^-_{\mbox{\scriptsize IC}} e^+_i$,
	     (\IC=chargino mass state) \\
   830     & $\l^+ \l^- \to \snu_i Z^0$ and 
             $\l^+ \l^- \to \slepton^+_i W^-$  \\
   840     & $\l^+ \l^- \to \snu_i h^0/H^0/A^0$ and 
             $\l^+ \l^- \to \slepton^+_i H^-$  \\
   850     & $\l^+ \l^- \to \snu_i \gamma$ \\
   860     & Sum of 870 and 880 \\
   870     & $\l^+ \l^- \to \l^+ \l^-$, via LLE only \\
   867+3\vv{IL1}+\vv{IL2} & 
$\l^+ \l^- \to \l^+_{\mbox{\scriptsize IL1}} \l^-_{\mbox{\scriptsize IL2}}$
          (\vv{IL1,2}=1,2,3 for $e,\mu,\tau$) \\
   880     & $\l^+ \l^- \to \dbar d$, via LLE and LQD \\
   877+3\vv{IQ1}+\vv{IQ2} & 
$\l^+ \l^- \to d_{\mbox{\scriptsize IL1}} \dbar_{\mbox{\scriptsize IL2}}$
          (\vv{IQ1,2}=1,2,3 for $d,s,b$) \\
\hline
    910    &      $\ell^+ \ell^- \to \nu_e \bar\nu_e h^0 + e^+ e^- h^0$\\
    920    &      $\ell^+ \ell^- \to \nu_e \bar\nu_e H^0 + e^+ e^- H^0$\\
    960    &      $\ell^+ \ell^- \to Z^0 h^0$\\ 
    970    &      $\ell^+ \ell^- \to Z^0 H^0$\\ 
    955    &      $\ell^+ \ell^- \to H^+ H^-$\\
    965    &      $\ell^+ \ell^- \to A^0 h^0$\\
    965    &      $\ell^+ \ell^- \to A^0 H^0$\\
\hline
1000+\ID & $\l^+\l^-\to t\,\bar t\,\SMH$ (\ID\ as in \vv{IPROC}=300+\ID)\\
\hline
1110+\IQ & $\l^+\l^-\to q\,\qbar\, h^0$  (\IQ\ as in \vv{IPROC}=100+\IQ)\\
1116+\IL & $\l^+\l^-\to \l^+\l^- h^0$ (\IL=1,2,3 for $e,\mu,\tau$)\\
1120+\IQ & $\l^+\l^-\to q\,\qbar\, H^0$ (\IQ\ as in \vv{IPROC}=100+\IQ)\\
1126+\IL & $\l^+\l^-\to \l^+\l^- H^0$ (\IL=1,2,3 for $e,\mu,\tau$)\\
1130+\IQ & $\l^+\l^-\to q\,\qbar\, A^0$ (\IQ\ as in \vv{IPROC}=100+\IQ)\\
1136+\IL & $\l^+\l^-\to \l^+\l^- A^0$ (\IL=1,2,3 for $e,\mu,\tau$)\\
1140    &  $\l^+\l^-\to d\,\bar u\, H^+ +$ ch. conj.\\
1141    &  $\l^+\l^-\to s\,\bar c\, H^+ +$ ch. conj.\\
1142    &  $\l^+\l^-\to b\,\bar t\, H^+ +$ ch. conj.\\
1143    &  $\l^+\l^-\to e\,\bar\nu_e H^+ +$ ch. conj.\\
1144    &  $\l^+\l^-\to \mu\,\bar\nu_\mu H^+ +$ ch. conj.\\
1145    &  $\l^+\l^-\to \tau\,\bar\nu_\tau H^+ +$ ch. conj.\\
\hline
   1200--99 & Reserved for other $\l^+\l^-$ processes \\
\hline
   1300    & $q \qbar \to Z^0/\gamma \to q' \qbar'$ (all flavours)\\
   1300+\IQ& $q \qbar \to Z^0/\gamma \to q' \qbar'$ ($\IQ=1,2,3,4,5,6$ for
$q=d,u,s,c,b,t$) \\
   1350    & $q \qbar \to Z^0/\gamma \to \l \lbar$ (all lepton species)\\
   1350+\IL& $q \qbar \to Z^0/\gamma \to \l \lbar$
($\IL=1-6$ for $\l=e,\nu_e,\mu,\nu_\mu$, etc.) \\
   1399    & $q \qbar \to Z^0/\gamma \to $ anything\\
\hline
   1400    & $q \qbar \to W^\pm \to q' \qbar''$ (all flavours) \\
   1400+\IQ& $q \qbar \to W^\pm \to q' \qbar''$ ($q'$ or $q''$ as above)\\
   1450    & $q \qbar \to W^\pm \to \l \nu_\l$ (all lepton species) \\
   1450+\IL& $q \qbar \to W^\pm \to \l \nu_\l$ ($\IL=1,2,3$ for
$\l=e,\mu,\tau$)     \\
   1499    & $q \qbar \to W^\pm \to$ anything           \\
\hline
   1500    & \QCD\ $2 \to 2$ hard parton scattering      \\
           & After generation, \vv{IHPRO} is subprocess (see
sect.~\ref{qcd}) \\
\hline
   1600+\ID& $g g/q\qbar \to\SMH$ (\ID\ as in $\vv{IPROC}=300+\ID$)  \\
\hline
   1700+\IQ& \QCD\ heavy quark production (\IQ\ as above) \\
           & After generation, \vv{IHPRO} is subprocess (see
sect.~\ref{qcd}) \\
\hline
   1800    & \QCD\ direct photon + jet production \\
           & After generation, \vv{IHPRO} is subprocess (see
sect.~\ref{direct}) \\
\hline
   1900+\ID&  $q\qbar\to q'\qbar'W^+W^-/Z^0Z^0\to q'\qbar'H^0_{\rm{SM}}$
(\ID\ as in $\vv{IPROC}=300+\ID$) \\
\hline
   2000    &    $t$ production via $W^\pm$ exchange (sum of 2001--2008) \\
   2001--4 &    $\ubar \bbar \to \dbar \tbar\;,\;\;\; d \bbar \to u
\tbar\;,\;\;\;\dbar \bbar \to \ubar \tbar\;,\;\;\; u b\to d t$ \\
   2005--8 &    $\cbar \bbar \to \sbar \tbar\;,\;\;\; s\bbar \to c\tbar
\;,\;\;\;\sbar b \to \cbar t\;,\;\;\; c b\to s t$ \\
\hline
   2100   & $W^\pm$ + jet production \\
   2110   & $W^\pm$ + jet production (Compton only: $g q \to W q$) \\
   2120   & $W^\pm$ + jet production (annihilation only: $q \qbar \to W g$)\\
   2150   & $Z^0$ + jet production \\
   2160   & $Z^0$ + jet production (Compton only: $g q \to Z q$) \\
   2170   & $Z^0$ + jet production (annihilation only: $q \qbar \to Z g$)\\
\hline
   2200   & \QCD\ direct photon pair production \\
           & After generation, \vv{IHPRO} is subprocess (see
sect.~\ref{direct}) \\
\hline
   2300+\ID & \QCD\ SM Higgs + jet production\ (\ID\ as in \vv{IPROC}=300+\ID)\\
           & After generation, \vv{IHPRO} is subprocess (see
sect.~\ref{qcd_higgs}) \\
\hline
   2400    & Mueller-Tang colour singlet exchange \\
   2450    & Quark scattering via photon exchange \\
\hline
   2500+\ID& $gg/q\qbar\to t\tbar\SMH$ (\ID\ as in \vv{IPROC}=300+\ID)\\
\hline
   2600+\ID& $q\qbar' \to W^\pm\SMH$ (\ID\ as in \vv{IPROC}=300+\ID)\\
\hline
   2700+\ID& $q\qbar \to Z^0\SMH$ (\ID\ as in \vv{IPROC}=300+\ID)\\
\hline
2800 & $W^+W^-$ production in hadron-hadron collisions\\
2810 & $Z^0Z^0$ production in hadron-hadron collisions (including photon terms)\\
2815 & $Z^0Z^0$ production in hadron-hadron collisions ($Z^0$ only)\\
2820 & $W^\pm Z^0$ production in hadron-hadron collisions (including photon terms)\\
2825 & $W^\pm Z^0$ production in hadron-hadron collisions ($Z^0$ only)\\
    2850    &  hadron-hadron $\to W^+ W^- X$ using \MN \\
    2860    &  hadron-hadron $\to Z^0 Z^0 X$ using \MN \\
    2870    &  hadron-hadron $\to W^+ Z^0 X$ using \MN \\
    2880    &  hadron-hadron $\to W^- Z^0 X$ using \MN \\
\hline
2900+\IQ & $gg+q\qbar\to Q\Qbar Z^0$ for massless $Q$ and $\Qbar$ (\IQ=1\ldots6
for $Q=d\ldots t$)\\
2910+\IQ & $gg+q\qbar\to Q\Qbar Z^0$, for massive $Q$ and $\Qbar$ (\IQ=1\ldots6
for $Q=d\ldots t$)\\
\hline
   3000-3999& Minimal Supersymmetric Standard Model (\MS) processes\\
   3000    & 2-parton $\to$ 2-sparticle processes (sum of those below)\\
   3010    & 2-parton $\to$ 2-sparton processes \\
   3020    & 2-parton $\to$ 2-gaugino processes \\
   3030    & 2-parton $\to$ 2-slepton processes \\
\hline
   3100+\ISQ& $gg/q\qbar\to {\tilde q}{\tilde q}^{'*} {H^\pm}$
   (\ISQ=\vv{IPROC}$-3100$ as from table~\ref{tab:sqch}) \\
\hline
   3200+\ISQ& $gg/q\qbar\to {\tilde q}{\tilde q}^{'*} {h,H,A}$
   (\ISQ=\vv{IPROC}$-3200$ as from table~\ref{tab:sqne}) \\
\hline
   3310,3315    & $\begin{array}[t]{l}
		  q\qbar' \to W^\pm h^0,H^\pm h^0 \hbox{(all $q,q'$ flavours
                  -- gauge bosons mediated only)} 
                  \end{array}$ \\
   3320,3325    & $q\qbar' \to W^\pm H^0,H^\pm H^0$ ('') \\
 ~~~~~~~3335    & $q\qbar' \to H^\pm A^0$ ('') \\
 3350~~~~~~~    & $q\qbar  \to W^\pm H^\mp$ (Higgstrahlung and
                  Higgs mediated) \\
 ~~~~~~~3355    & $q\qbar  \to H^\pm H^\mp $ (all $q$ flavours ---
                  gauge boson mediated only) \\
   3360,3365    & $q\qbar  \to Z^0 h^0,A^0 h^0$ ('') \\
   3370,3375    & $q\qbar  \to Z^0 H^0,A^0 H^0$ ('') \\
\hline
   3410      & $bg \to b~h^0$ + ch.\ conj.\\
   3420      & $bg \to b~H^0$ + ch.\ conj.\\
   3430      & $bg \to b~A^0$ + ch.\ conj.\\
   3450      & $bg \to t~H^-$ + ch.\ conj.\\
\hline
   3500     & $b q \to b q' H^\pm$ + ch.\ conj. \\
\hline
   3610& $q\qbar/gg \to h^0\:$ (light scalar Higgs) \\
   3620& $q\qbar/gg \to H^0\;$ (heavy scalar Higgs) \\
   3630& $q\qbar/gg \to A^0\,$ (pseudoscalar Higgs) \\
\hline
   3710&  $q\qbar\to q'\qbar'W^+W^-/Z^0Z^0\to q'\qbar'h^0$  \\
   3720&  $q\qbar\to q'\qbar'W^+W^-/Z^0Z^0\to q'\qbar'H^0$  \\
\hline
   3810+\IQ& $gg+q\qbar\to Q\Qbar h^0$ (all $q$ flavours in $s$-channel,
                                    \IQ\ as usual for $Q$ flavour) \\
   3820+\IQ& $gg+q\qbar\to Q\Qbar H^0$ ('') \\
   3830+\IQ& $gg+q\qbar\to Q\Qbar A^0$ ('') \\
3839~~~~~~~& $gg+q\qbar\to b\bar t H^+$ + ch. conjg. (all $q$ flavours in 
$s$-channel) \\
   3840+\IQ& $gg       \to Q\Qbar h^0$ (\IQ\ as above) \\
   3850+\IQ& $gg       \to Q\Qbar H^0$ ('') \\
   3860+\IQ& $gg       \to Q\Qbar A^0$ ('') \\
3869~~~~~~~& $gg       \to b\bar t H^+$ + ch. conjg.  \\
   3870+\IQ& $q\qbar   \to Q\Qbar h^0$ (all $q$ flavours in $s$-channel,
\IQ\ as above) \\
   3880+\IQ& $q\qbar   \to Q\Qbar H^0$ ('') \\
   3890+\IQ& $q\qbar   \to Q\Qbar A^0$ ('') \\
3899~~~~~~~& $q\qbar   \to b\bar t H^+$ + ch. conjg. (all $q$ flavours in 
$s$-channel) \\
\hline
   3900--99  & Reserved for other hadron-hadron \MS\ processes \\ 
\hline
   4000--99  &  R-parity violating supersymmetric processes via LQD\\
   4000      & single sparticle production, sum of 4010--4050 \\
   4010      & $\ubar_j d_k \to \gaugino^0 l^-_i$,
               $\dbar_j d_k \to \gaugino^0 \nu_i$ (all neutralinos)\\
  4010+\IN  & $\ubar_j d_k \to \gaugino^0_{\mbox{\scriptsize IN}} l^-_i$,
               $\dbar_j d_k \to \gaugino^0_{\mbox{\scriptsize IN}} \nu_i$
(\IN=neutralino mass state)\\
   4020      & $\ubar_j d_k \to \gaugino^- \nu_i$, 
               $\dbar_j d_k \to \gaugino^- e^+_i$ (all charginos) \\
   4020+\IC & $\ubar_j d_k \to \gaugino^-_{\mbox{\scriptsize IC}} \nu_i$,
               $\dbar_j d_k \to \gaugino^-_{\mbox{\scriptsize IC}} e^+_i$ (\IC=chargino mass state) \\
   4040      & $u_j \dbar_k \to \tilde{\tau}^+_i Z^0$,
               $u_j \dbar_k \to \snu_i W^+$ and 
               $d_j \dbar_k \to \slepton^+_i W^-$  \\
   4050      & $u_j \dbar_k \to \slepton^+_i h^0/H^0/A^0$,
               $u_j \dbar_k \to \snu_i H^+$ and 
               $d_j \dbar_k \to \slepton^+_i H^-$  \\
   4060      & Sum of 4070 and 4080 \\
   4070      & $\ubar_j d_k \to \ubar_l d_m $ and 
               $\dbar_j d_k \to \dbar_l d_m $, via LQD only \\
   4080      & $\ubar_j d_k \to \nu_j l^-_k $ and 
               $\dbar_j d_k \to l^+_j l^-_k $, via LQD and LLE \\
\hline
   4100-99 & R-parity violating supersymmetric processes via UDD\\
   4100    & single sparticle production, sum of 4110--4150\\
   4110      & $u_i d_j \to \gaugino^0 \dbar_k$,  
               $d_j d_k \to \gaugino^0 \bar{u_i}$ (all neutralinos)\\
  4110 +\IN  & $u_i d_j \to \gaugino^0_{\mbox{\scriptsize IN}} \dbar_k$,  
               $d_j d_k \to \gaugino^0_{\mbox{\scriptsize IN}}
 \bar{u_i}$(\IN\ as above)\\
  4120       & $u_i d_j \to \gaugino^+ \ubar_k$,  
               $d_j d_k \to \gaugino^- \bar{d_i}$  (all charginos) \\
  4120 +\IC  & $u_i d_j \to \gaugino^+_{\mbox{\scriptsize IC}} \ubar_k$,  
               $d_j d_k \to \gaugino^-_{\mbox{\scriptsize IC}}
 \bar{d_i}$  (\IC\ as above) \\
  4130       & $u_i d_j \to \gluino \dbar_k$,  
               $d_j d_k \to \gluino \bar{u_i}$ \\
  4140       & $u_i d_j \to \tilde{b}^*_1 Z^0$,
$d_j d_k \to \tilde{t}^*_1 Z^0$, 
               $u_i d_j \to \tilde{t}^*_i W^+$
and $d_j d_k \to \tilde{b}^*_i W^-$  \\
  4150       & $\begin{array}[t]{rcl}
	       u_i d_j \to \tilde{d}^*_{k1} h^0/H^0/A^0, 
               d_j d_k \to \tilde{u}^*_{i1} h^0/H^0/A^0,
               u_i d_j \to \tilde{u}^*_{k\alpha} H^+, \\
               d_j d_k \to \tilde{d}^*_{i\alpha} H^-  
	       \end{array} $ \\
  4160       & $u_i d_j \to u_l d_m$, $d_j d_k \to d_l d_m$ via UDD. \\
\hline
   4200-99  & Graviton resonance production\\
   4200     & Sum of 4210, 4250 and 4270\\
   4210     & $gg/q\qbar\to G \to gg/q\qbar$ (all partons)\\
   4210+\IQ & $gg/q\qbar\to G \to q\qbar$ (\IQ\  as above)\\
   4220     & $gg/q\qbar\to G \to gg$\\
   4250     & $gg/q\qbar\to G \to \l\lbar$ (all leptons)\\
   4250+\IL & $gg/q\qbar\to G \to \l\lbar$ 
	      ($\IL=1-6$ for $\l=e,\nu_e,\mu,\nu_\mu$, etc.) \\
   4260     & $gg/q\qbar\to G \to \gamma \gamma$ \\
   4270     & $gg/q\qbar\to G \to W^+W^-/Z^0Z^0/\SMH\SMH$ \\
   4271     & $gg/q\qbar\to G \to W^+W^-$\\
   4272     & $gg/q\qbar\to G \to Z^0Z^0$\\
   4273     & $gg/q\qbar\to G \to \SMH\SMH$\\
\hline
   5000    & Pointlike photon-hadron jet production (all flavours) \\
   5100+\IQ& Pointlike photon heavy flavour pair production (\IQ\ as above)\\
   5200+\IQ& Pointlike photon heavy flavour single excitation
(\IQ\ as above)\\
           & After generation, \vv{IHPRO} is subprocess (see
sect.~\ref{direct}) \\
   5300    & Quark-photon Compton scattering \\
   5500    & Pointlike photon production of light ($u,d,s$) L=0 mesons\\
   5510,20 & S=0 mesons only, S=1 mesons only \\
 & After generation, \vv{IHPRO} is subprocess (see sect.~\ref{direct}) \\
\hline
   6000    & $\gamma\gamma\to q\qbar$ (all flavours)\\
   6000+\IQ& $\gamma\gamma\to q\qbar$ (\IQ\ as above)\\
   6006+\IL& $\gamma\gamma\to \l\lbar$ ($\IL=1,2,3$ for $\l=e,\mu,\tau$)\\
   6010    & $\gamma\gamma\to W^+W^-$ \\
\hline
   7000 $-$ & Baryon-number violating and other multi-$W^\pm$ processes \\
   7999    & generated by \sf{HERBVI} package \\
\hline
   8000    & Minimum bias soft hadron-hadron event \\
\hline
   9000   & Deep inelastic lepton scattering (all neutral current) \\
  9000+\IQ& Deep inelastic lepton scattering (NC on flavour \IQ) \\
   9010   & Deep inelastic lepton scattering (all charged current) \\
  9010+\IQ& Deep inelastic lepton scattering (CC on flavour \IQ) \\
\hline
  9100    & Boson-gluon fusion in neutral current \DIS\ (all flavours) \\
  9100+\IQ& Boson-gluon fusion in neutral current \DIS\ (\IQ\ as above) \\
  9107    & $J/\psi$ + gluon production by boson-gluon fusion \\
  9110    & \QCD\ Compton process in neutral current \DIS\ (all flavours) \\
  9110+\IP& \QCD\ Compton process in NC \DIS\ (\IP=1--12 for
$d-t,\dbar-\tbar$)\\
  9130    & All ${\cal O}(\as)$ NC processes (i.e.\ 9100+9110) \\
  9140+\IP& Heavy quark production by charged-current boson-gluon fusion \\
          & \IP: $1 = s \cbar, 2 = b \cbar, 3 = s \tbar, 4
= b \tbar$ (+ ch.\ conj.) \\
\hline
  9500+\ID& $W^+W^-/Z^0Z^0\to\SMH$ in \DIS\ (\ID\ as in $\vv{IPROC}=300+\ID$)\\
\hline
 10000+\IP& as $\vv{IPROC}=\IP$ but with soft underlying event \\
          & (soft remnant fragmentation in lepton-hadron) suppressed\\
\hline
\end{longtable}

\subsubsection{Treatment of quark masses}\label{quarkmass}

The extent to which quark mass effects are included in the hard
process cross section is different in different processes.  In many
processes, they are always treated as massless: \vv{IPROC} = 1300,
1800, 1900, 2100, 2300, 2400, 5300, 9000.  In two processes they are
all treated as massless except the top quark, for which the mass is
correctly incorporated: 1400, 2000.  In the case of massless pair
production, only quark flavours that are kinematically allowed are
produced.  In all cases the event kinematics incorporate the quark
mass, even when it is not used to calculate the cross section.  In two
processes, quarks are always treated as massive: 500, 9100.  Finally,
in several processes, the behaviour is different depending on whether
a specific quark flavour is requested, in which case its mass is
included, or not, in which case all quarks are treated as
massless. These are: \vv{IPROC} = 100, 110, 120, \QCD\ $2\to 2$
scattering (1500 vs.\ 1700+\IQ), jets in direct photoproduction (5000
vs.\ 5100+\IQ\ and 5200+\IQ). In the case of \vv{IPROC} = 2900,2910 one has
the option of using the massless or massive matrix element.

These differences can cause inconsistencies between different ways of
generating the same process.  The most noticeable example is in direct
photoproduction, where one can use process 9130, which uses the exact
$2\to 3$ matrix element $e+g \to e+q+\qbar$, or process 5000, which
uses the Equivalent Photon Approximation (EPA) for $e \to e+\gamma$
and the $2\to 2$ matrix element for $\gamma+g \to q+\qbar$. For
typical \sss{HERA} kinematics, the EPA is valid to a few per cent, but
the difference between the two processes is much larger, about 20\%
for $\vv{PTMIN}=2$\,GeV.  This is entirely due to the difference in
quark mass treatments, as can be checked by comparing process 9130
with processes 5100+\IQ\ and 5200+\IQ\ summed over \IQ.

\subsubsection{Couplings}\label{couples}

The two-loop \QCD\ coupling at scale $Q$ is given by subroutine
\vv{HWUALF} with arguments $\vv{IOPT}=1$ and $\vv{SCALE}=Q$. Threshold
matching is performed at the quark mass scales
$Q=\vv{RMASS}(i)$. Setting $\vv{IOPT}=0$ initialises the coupling using
the 5-flavour value $\lms=\vv{QCDLAM}$.  Other values of \vv{IOPT} are
for internal use only.

The electromagnetic coupling is given by $\vv{HWUAEM}(Q^2)=e^2/(4\pi)$;
it runs according to the prescription in ref.~\cite{yrep} with the
hadronic term as given in ref.~\cite{burk}. The parameter
$\vv{ALPHEM}\equiv\vv{HWUAEM}(0)$, default value 0.0072993, provides the
normalization at the Thomson limit; it is used for all processes
involving real photons. Photon emission in parton showers and in the
`dead-zone' in $\ee$ processes can be enhanced by a factor of \vv{ALPFAC}
(default = 1). The normalised electric charges of the fundamental
fermions are stored in the array \vv{QFCH(I)}, where \vv{I} = 1--6 for the
quarks $d,u,s,c,b,t$ (e.g.\ $\vv{QFCH(4)}=2./3.$) and 11--16 for the
leptons $e,\nu_e,\mu,\nu_\mu, \tau,\nu_\tau$.

The weak neutral current is taken to be of the form
$e(v_f+a_f\gamma_5)\gamma_\mu$, where the electric charge is evaluated
at a scale appropriate to the process. The arrays \vv{VFCH(I,J)} and
\vv{AFCH(I,J)} store the couplings: \vv{I} as before, \vv{J} = 1 for the
minimal Standard Model and 2 for possible $Z'$ couplings (only used if
\vv{ZPRIME=.TRUE.}). Note that universality is not assumed --- couplings
can be arbitrarily set separately for each fermion species. The
default couplings are given in terms of of \vv{SWEIN}$=\sin^2\theta_W$,
default value 0.2319, as:
$$
       v_f=(T_3/2-Q\sin^2\theta_W)/(\cos\theta_W\sin\theta_W)\,,
\qquad  a_f=T_3/(2\cos\theta_W\sin\theta_W)\,.
$$

The weak charged current is given in terms of $g=e/\sin\theta_W$ and
the Cabbibo-Kobayashi-Maskawa mixing matrix, the elements squared of
which are stored in \vv{VCKM(K,L)}, $\vv{K}=1,2,3$ for $u,c,t$,
$\vv{L}=1,2,3$ for $d,s,b$. The variable
$\vv{SCABI}=\sin^2\theta_{\mbox{\scriptsize Cabibbo}}$ is however also
retained for the present. Note the Fermi constant
$G_{\mbox{\scriptsize Fermi}}$ is eliminated from all cross sections.

The overall scale for all cross sections, given in nanobarns, is set
by $\vv{GEV2NB}=(\hbar c/e)^2$, default value 389379.

We now give more detailed descriptions of the various subprocesses,
concentrating again on the new features since ref.~\cite{hw51}.

\subsection{Lepton-antilepton Standard Model processes}\label{sec_lept}

Lepton beam polarization effects are included in $\ee\to$ 2/3 jet
production and the Bjorken process ($ZH$ production).  Incoming lepton
and antilepton beam polarizations are specified by setting the two
3-vectors \vv{EPOLN} and \vv{PPOLN}: component 3 is longitudinal and 1,2
transverse.

Photon initial-state radiation (\sss{ISR}) in $\ee$ annihilation events
is allowed. The parameter \vv{TMNISR} sets the minimum $\shat/s$ value
(default = 10$^{-4}$), \vv{ZMXISR} sets the (arbitrary) separation
between unresolved and resolved emission (default = $1-10^{-6}$).
Setting \vv{ZMXISR} = 0 switches off photon \sss{ISR}.

Where processes are listed for $\ell^+ \ell^-$ they are available for
$\ee$ and $\mu^+\mu^-$ annihilation. Many of the processes listed for
$\ee$ will also work for $\mu^+\mu^-$, but we have not been systematic
in ensuring this. 

\subsubsection{\vv{IPROC}=100--127: hadron production}

A correction to hard gluon emission in $\ee$ events has been added and
is now the default process for \vv{IPROC}=100+\IQ.  The ${\cal O}(\as)$
matrix element is used to add events in the `dead zone' of phase-space
corresponding to a quark-antiquark pair recoiling from a hard
gluon~\cite{Seym}.  Although this is asymptotically negligible, and
cannot be produced within the shower itself, it has a significant
effect at LEP1 energies.  As a result, the default parameters have
been retuned, and show a marked improvement in agreement with $\ee$
data for event shapes sensitive to three-jet configurations.

The routine \vv{HWBDED} implements this hard correction while
\vv{HWBRAN} has been modified to include the soft matrix-element
corrections described in section~\ref{mecorr}.

When \vv{IPROC}=100+\IQ, hard gluons emitted into the dead zone are
assigned to the quark or antiquark shower and do not appear explicitly
in the event record.

The $q\qbar g$ process alone, generated according to the ${\cal
O}(\as)$ $q\qbar g$ matrix element with a maximum thrust cutoff
\vv{THMAX} (default 0.9), is given by \vv{IPROC}=110+\IQ.

The uncorrected $q\qbar$ process has been retained for comparative
purposes and is available as \vv{IPROC}=120+\IQ.

The fictional $\ee$ processes $\ee \to g+g(+g)$, \vv{IPROC}=107 and
127, is treated just like $\ee \to q\qbar$, summed over light quark
flavours, for direct comparisons between quark and gluon jets.

\subsubsection{\vv{IPROC}=150--250: lepton and electroweak boson production}\label{electroweak}

In \vv{IPROC}=150, only the $s$-channel process, mediated by a virtual
photon or $Z^0$, is included, so the final-state leptons must be
different from the initial ones.

The processes of $W^+W^-$ and $Z^0Z^0$ pair production, \vv{IPROC}=200
and 250, are based on a program kindly supplied by Zoltan Kunszt,
which fully includes decay correlations. The QCD ${\cal O}(\as)$
matrix element correction for hard gluon emission in hadronic $W$ and
$Z$ decays has also been implemented in these processes, according to
the method described in section~\ref{mecorr}. In contrast to
\vv{IPROC}=100+\IQ, any hard gluons emitted into the dead zone are
shown explicitly in the event record.

\subsubsection{\vv{IPROC}=300--499: Higgs boson production}\label{SMHiggs}

\HW\ generates \SM\ Higgs bosons in lepton-antilepton collisions
through the Bjorken process $Z^{(*)} \to Z^{(*)}\SMH$ with one or both
$Z^0$'s off-shell ($\vv{IPROC}=300+\ID$) and $W^+W^-/Z^0Z^0$ fusion
($\vv{IPROC}=400+\ID$).  See section~\ref{sec:gaugebosons} for
explanation of how the Higgs decay is controlled by the value of
\vv{ID}.

\subsubsection{\vv{IPROC}=500--559: two-photon/photon-boson processes}
\label{twophot}

In the $\ee$ two-photon processes, $\vv{IPROC} = 500+\vv{ID}$,
$\vv{ID}=0-10$ is the same as in Higgs processes for $q\qbar$,
$\l\lbar$ and $W^+W^-$.  The Equivalent Photon Approximation (EPA) is
used for the $e\to e\gamma$ vertices. The phase space is controlled by
\vv{EMMIN} and \vv{EMMAX} for the two-photon centre-of-mass frame
(\sss{CMF}) mass, \vv{PTMIN} and \vv{PTMAX} for the transverse momentum of
the \sss{CMF} in the lab, and \vv{CTMAX} for the c.m.\ angle of the
outgoing particles.  The additional phase-space variable \vv{WHMIN}
sets the minimum allowed hadronic mass and affects photoproduction
reactions ($\gamma$-hadron and $\gamma$-$\gamma$) and \DIS.

In photon-$W^\pm$ fusion, $\vv{IPROC} = 550+\ID$, $\vv{ID}=0-9$ is
also the same as in Higgs processes, except that \vv{ID}=1 or 2 both
give the sum of $d\ubar$ and $u\dbar$ etc.  The EPA is used for the
$e\to e\gamma$ vertex.  The phase space is controlled by \vv{EMMIN}
and \vv{EMMAX} only.  The full $2\to 3$ matrix elements for $\gamma
e\to f\bar f'\nu$ are used, so the cross section for real $W^\pm$
production is correctly included.  In the case of $\gamma\gamma\to W
W$ the decay correlations are not yet correctly included: the $W$'s
currently decay isotropically.

\subsubsection{\vv{IPROC}=600--656: four jet production}\label{fourjet}

Electron-positron annihilation to four jets is provided by
\vv{IPROC=600+IQ}, where a non-zero value for \IQ\ guarantees
production of quark flavour \IQ\ whilst \vv{IQ=0} corresponds to the
natural flavour mix.  \vv{IPROC=650+IQ} is as above but without those
terms in the matrix element which orient the event w.r.t.\ the lepton
beam direction. The matrix elements are based on those of Ellis, Ross
and Terrano~\cite{ERT} with orientation terms from Catani and
Seymour~\cite{CatSey}. The soft and collinear divergences are avoided
by imposing a minimum \ycut, \vv{Y4JT} (default 0.01), on the initial
four partons.  The interjet distance \ycut\ is calculated using either
the Durham or JADE metrics, as selected by the logical variable
\vv{DURHAM} (default \vv{.TRUE.}).  In order to improve efficiency
parameterizations of the volume of four-body phase space are used:
these are accurate up to a few percent for \ycut\ values less than
0.14.  Note also that the phase space is for massless partons, as are
the matrix elements, though a mass threshold cut is applied.

The argument of the strong coupling is set equal to \vv{EMSCA}, the
scale for the parton showers.  This is taken to be the smallest of the
\ycut\ values times the c.m.\ energy squared if \vv{FIX4JT=.FALSE.}
(default); otherwise the argument is fixed at \vv{Y4JT} times the c.m.\
energy squared.

\TABULAR{|c|c|c|}{
\hline
\vv{IHPRO} & $\gamma^\star\to$ 1+2+3+4 & c/f conn.\\
\hline
91 & $q+\qbar+g+g$       & 3 1 4 2 \\
92 & $q+\qbar+g+g$       & 4 1 2 3 \\
93 & $q+\qbar+q+\qbar$   & 4 1 2 3 \\
94 & $q+\qbar+q+\qbar$   & 2 1 4 3 \\
95 & $q+\qbar+q'+\qbar'$ & 4 1 2 3 \\
\hline
}{Four jet subprocesses.\label{tab.9}}

The matrix elements for the $q\qbar gg$ and $q\qbar q\qbar$ (same
flavour quark) final states receive contributions from two colour
flows each. The actual process and colour flow generated is indicated
by \vv{IHPRO} as shown in table~\ref{tab.9}.
The meaning of `c/f conn.' is discussed in section~\ref{qcd} below.  The
treatment of the interference terms between the two colour flows is
controlled by the array \vv{IOP4JT(1)} for $q\qbar gg$ and
\vv{IOP4JT(2)} for $q\qbar q\qbar$ (identical quark flavour):
$$
\vv{IOP4JT(1)}=
\left\{\begin{array}{ll}
0 & \mbox{neglect}  \\
1 & \mbox{extreme 3142} \\
2 & \mbox{extreme 4123} \\
\end{array}\right.
\>\>\>\>\>
\vv{IOP4JT(2)}=
\left\{\begin{array}{ll}
0 & \mbox{neglect}  \\
1 & \mbox{extreme 4123} \\
2 & \mbox{extreme 2143} \\
\end{array}\right.
$$
In both instances the default value is 0.

See ref.~\cite{4jstef} for some applications and discussions of the
new four-jet implementation in $e^+e^-$ annihilations.

\subsection[Lepton-antilepton supersymmetric processes (MSSM)]%
{Lepton-antilepton supersymmetric processes (MSSM)}\label{sec_susy_lept}

The R-parity conserving lepton-antilepton SUSY processes have
$\vv{IPROC}=700-799$ and \RPV\ processes have $\vv{IPROC}=800-899$.
Lepton beam polarization effects are included for all of the SUSY
production processes. The processes have all been implemented in such
a way as to allow either $e^+e^-$ or $\mu^+\mu^-$ as the initial
state.  As with the SM lepton-antilepton processes, ISR is allowed for
the SUSY production processes.

As, by probing the individual thresholds, it may be possible to study
the production of a given sparticle pair in lepton-antilepton
collisions, we have provided more control over the sparticles produced
than for the hadron-hadron SUSY production processes described in
section~\ref{sec:SUSYproc}.

We remind the reader here that all \SY\ particle data have to be read
from an input file before event generation (see
section~\ref{sec:SUSYdata}).

\subsubsection{\vv{IPROC}=700--799: gaugino, slepton and/or squark production}

\looseness=1 With $\vv{IPROC}=700$ one obtains the four processes
$\vv{IPROC}=710,730,740,760$ in the correct proportions. The matrix
elements have been derived independently and the cross sections are in
good agreement with those from {\sf SUSYGEN}~\cite{Katsanevas:1998fb}.
For all these processes the hard process scale \vv{EMSCA} has been set
to the centre-of-mass energy.

The gaugino and sfermion mixing conventions of Haber and Kane
\cite{HabKan} are adop\-ted in all cases. In addition, the neutralino
mixing matrix \vv{ZMIXSS} is defined internally in terms of the
photino, zino and current eigenstate neutral higgsino components,
instead of the bino, $W_3$-ino and higgsino components adopted for
\vv{ZMXNSS}, equivalent to $N_{ij}$ in~\cite{HabKan}.

$\vv{IPROC}=710$ gives lepton-antilepton$\to$neutralino pair
production. A number of additional \vv{IPROC} codes have been provided
to enable the user to produce given neutralino mass eigenstates. It
should be noted that in order to provide a compact code for these
processes there is more than one possible \vv{IPROC} number for some
processes.  For example the final state $\gaugino^0_1 \gaugino^0_3 $
can be produced using either the \vv{IPROC} codes 713 or 721.

$\vv{IPROC}=730$ gives lepton-antilepton$\to$chargino pair
production. As with the neutralino production there are codes to allow
a given pair of charginos to be produced.

\looseness=1 $\vv{IPROC}=740$ gives lepton-antilepton$\to$slepton pair
production. In these processes for the first two generations the
left/right eigenstates are produced, while for the third generation
the mass eigenstates are produced. In the processes producing given
slepton pairs for $\tilde{\tau}$ production processes the left
eigenstate is replaced by the lighter mass eigenstate and the right
eigenstate by the heavier mass eigenstate.

$\vv{IPROC}=760$ gives lepton-antilepton$\to$squark pair production. As
with the other processes additional codes are provided to allow the
production of given $\squark\squark^*$ pairs. For stop and sbottom
production the mass eigenstates are produced, while for the first two
generations the left/right eigenstates are generated. As with the
slepton production for the third generation when a given squark pair
is requested the left eigenstate is replaced by the lighter mass
eigenstate and the right eigenstate by the heavier mass eigenstate.

\subsection{Other lepton-antilepton non-Standard-Model processes}

\subsubsection{\vv{IPROC}=800--899: R-parity violating SUSY processes}

A range of possible \RPV\ production processes in lepton-antilepton
collisions is included. Unlike the case of \RPV\ production in
hadron-hadron collisions, section \ref{sec:RPVhadronproc}, we have
included processes for which there is either no $s$-channel resonance
or the resonance is not kinematically accessible.

All the possible single sparticle production mechanisms which occur
via the first term in the superpotential given in~\cite{Rtheory} are
included. This includes the process $\l^+\l^-\to\gamma\snu$ for which
there is no $s$-channel resonance. As the cross section for this
process diverges in the limit that the photon is collinear with the
incoming lepton beams a cut on the $p_T$ of the outgoing particles
$p_T>\vv{PTMIN}$ has been imposed for this process (\vv{IPROC}=850).
The ISR is switched off for this process as including it would lead to
a double counting of the photon radiation. For this reason this
process is not included in the code \vv{IPROC}=800 which generates all
the other single sparticle production mechanisms.

We have also included some processes for the production of SM
particles via $s$-channel sneutrino exchange. In addition to the
$s$-channel sneutrino diagrams the $t$-channel sparticle exchanges, SM
diagrams and all the interference terms are included. This uses a
generalization of the formulae of~\cite{Kalinowski:1997bc}.  A cut
$p_T>\vv{PTMIN}$ is used in the process $\l^+\l^-\to\l^+\l^-$
(\vv{IPROC}=870) to avoid the divergence as $t\rightarrow0$ in the
Bhabha scattering cross section.

Except where stated explicitly above, no \vv{PTMIN} cut is applied.

\subsubsection{\vv{IPROC}=900--999: MSSM Higgs production processes}

    For \MS\ Higgs pair production (\vv{IPROC}=955--965)
    a new subroutine  has been  introduced, \vv{HWHIHH},
    whereas for the other processes we make use of the 
    implementation of  their
    \SM\ counterparts, which are based on the subroutines
    \vv{HWHIGW} and \vv{HWHIGZ}.

\subsubsection{\vv{IPROC}=1000--1199: SM and MSSM Higgs associated production}
\SM\ and \MS\ Higgs production in association with fermion pairs
in lepton-lepton collisions is available in version 6.5.
These processes were introduced in \cite{Djouadi:1991tk} and their
phenomenological relevance was discussed in \cite{Djouadi:gp}.

Fermion masses are retained in the final state
according  to the \HW\ defaults. The same values appear in
the Yukawa couplings. Notice that in the case of charged Higgs boson
production the Cabibbo-Kobayashi-Maskawa mixing matrix has been assumed 
to be diagonal. Furthermore, due to the rather different phase space 
distribution of the final state products, all processes can only be produced 
separately, not collectively. Initial- and final-state radiation (both QED 
and QCD) and beamsstrahlung are included via the usual \HW\ algorithms. 
Finally notice that the use of the \vv{IPROC}\ series 1000 and 1100 for $\l^+\l^-$
processes required some internal modification to \HW, which was 
originally designed to generate leptonic processes only for $\vv{IPROC}< 1000$.
Now we assume an $\l^+\l^-$ process whenever $\vv{IPROC}< 1300$.
These modifications have no implications for the traditional user, but
may affect more knowledgeable ones who have edited previous versions
of the main \HW\ code.

\subsection{Hadron-hadron Standard Model processes}

\subsubsection{\vv{IPROC}=1300--1499: Drell-Yan processes}

The Drell-Yan code is extended to the production of all fermion
pairs; 1300 gives all quark flavours; 1300+\IQ\ a specific quark
flavour, 1350 all leptons (including neutrinos) 1350+\vv{IL} a specific
lepton flavour.  The $s$-channel component of the interference with
like-flavour $q\qbar$ scattering is included here.

The initial-state parton showers in Drell-Yan processes are matched
to the exact ${\cal O} (\alpha_S)$ matrix-element result as discussed
in section~\ref{mecorr}.  The routine \vv{HWBDYP} implements the hard
corrections whilst \vv{HWSBRN} includes the soft corrections to the
initial-state radiation. For further details see
ref.~\cite{Corcella:2000gs}.

\subsubsection{\vv{IPROC}=1500: QCD $2\to 2$ processes}\label{qcd}

{\renewcommand\belowcaptionskip{-.5em}
\TABLE[t]{\centerline{\begin{tabular}{|c|rcl|c|}\hline
\vv{IHPRO} &$ 1 + 2         $&$\to$&$ 3 + 4         $& c/f conn.\\
\hline
1  &$ q + q         $&$\to$&$ q + q         $& 3 4 2 1 \\
2  &$ q + q         $&$\to$&$ q + q         $& 4 3 1 2 \\
3  &$ q + q'        $&$\to$&$ q + q'        $& 3 4 2 1 \\
4  &$ q + \qbar     $&$\to$&$ q'+ \qbar'    $& 2 4 1 3 \\
5  &$ q + \qbar     $&$\to$&$ q + \qbar     $& 3 1 4 2 \\
6  &$ q + \qbar     $&$\to$&$ q + \qbar     $& 2 4 1 3 \\
7  &$ q + \qbar     $&$\to$&$ g + g         $& 2 4 1 3 \\
8  &$ q + \qbar     $&$\to$&$ g + g         $& 2 3 4 1 \\
9  &$ q + \qbar'    $&$\to$&$ q + \qbar'    $& 3 1 4 2 \\
10  &$ q + g         $&$\to$&$ q + g         $& 3 1 4 2 \\
11  &$ q + g         $&$\to$&$ q + g         $& 3 4 2 1 \\
12  &$ \qbar + q     $&$\to$&$ \qbar' +q'    $& 3 1 4 2 \\
13  &$ \qbar + q     $&$\to$&$ \qbar + q     $& 2 4 1 3 \\
14  &$ \qbar + q     $&$\to$&$ \qbar + q     $& 3 1 4 2 \\
15  &$ \qbar + q     $&$\to$&$ g + g         $& 3 1 4 2 \\
16  &$ \qbar + q     $&$\to$&$ g + g         $& 4 1 2 3 \\
17  &$ \qbar + q'    $&$\to$&$ \qbar + q'    $& 2 4 1 3 \\
18  &$ \qbar + \qbar $&$\to$&$ \qbar + \qbar $& 4 3 1 2 \\
19  &$ \qbar + \qbar $&$\to$&$ \qbar + \qbar $& 3 4 2 1 \\
20  &$ \qbar + \qbar'$&$\to$&$ \qbar + \qbar'$& 4 3 1 2 \\
21  &$ \qbar + g     $&$\to$&$ \qbar + g     $& 2 4 1 3 \\
22  &$ \qbar + g     $&$\to$&$ \qbar + g     $& 4 3 1 2 \\
23  &$ g + q         $&$\to$&$ g + q         $& 2 4 1 3 \\
24  &$ g + q         $&$\to$&$ g + q         $& 3 4 2 1 \\
25  &$ g + \qbar     $&$\to$&$ g + \qbar     $& 3 1 4 2 \\
26  &$ g + \qbar     $&$\to$&$ g + \qbar     $& 4 3 1 2 \\
27  &$ g + g         $&$\to$&$ q + \qbar     $& 2 4 1 3 \\
28  &$ g + g         $&$\to$&$ q + \qbar     $& 4 1 2 3 \\
29  &$ g + g         $&$\to$&$ g + g         $& 4 1 2 3 \\
30  &$ g + g         $&$\to$&$ g + g         $& 4 3 1 2 \\
31  &$ g + g         $&$\to$&$ g + g         $& 2 4 1 3 \\\hline
       \end{tabular}}%
\caption{\QCD\ subprocesses.\label{tab.10}}}}

At present only $2\to2$ subprocesses are implemented. They are
classified in table~\ref{tab.10}. Here and in other subprocess
tables, `c/f conn.' refers to the colour/flavour connections between
the partons: `$i\;j\;k\;l$' means that the colour of parton 1 comes
from parton $i$, that of 2 from $j$, etc.  For antiquarks, which have
no colour (only anticolour), the label shows instead to which parton
the flavour is connected.  For this colour/flavour labelling all
partons are defined as outgoing.  Thus, for example, process 10 has
colour connections 3 1 4 2, corresponding to the colour flow diagram:
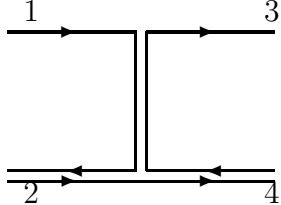
\begin{center}
{\begin{picture}(65,30)
\thicklines
\put( 7,0){\vector(1,0){13}}
\put(20,0){\vector(1,0){26}}
\put(46,0){\line(1,0){11}}
\put( 7,28){\vector(1,0){13}}
\put(20,28){\line(1,0){11}}
\put(31,28){\line(0,-1){26}}
\put(31,2){\vector(-1,0){13}}
\put(18,2){\line(-1,0){11}}
\put(57,2){\vector(-1,0){13}}
\put(44,2){\line(-1,0){11}}
\put(33,2){\line(0,1){26}}
\put(33,28){\vector(1,0){13}}
\put(46,28){\line(1,0){11}}
\put(10,-4){$2$}
\put(10,30){$1$}
\put(55,-4){$4$}
\put(55,30){$3$}
\end{picture}}
\end{center}

When different colour flows are possible, they are listed as separate
subprocesses.  This separation is not exact but is normally a good
approximation~\cite{EMW,March1}. The sepa\-ration is now performed using
the improved method proposed in~\cite{ko_colour}, as outlined in
section~\ref{sec:elem}.  The sum of the colour flows is the exact
lowest-order cross section.

\subsubsection[\vv{IPROC}=1600--1699: Higgs boson production by parton fusion]
{\vv{IPROC}=1600--1699: Higgs boson production by parton fusion}

\noindent $\vv{IPROC} = 1600+\vv{ID}$  gives the sum of $gg$ and $q\qbar$ fusion.
The lowest-order formulae that we have used can be found in~\cite{HHG}.
The hard processes are implemented in the subroutine \vv{HWHIGS}. 

\subsubsection{\vv{IPROC}=1700--1706: heavy quark production}
The separation of colour flows~is~now performed using the improved
me\-thod proposed in~\cite{ko_colour}, as outlined in
section~\ref{sec:elem}.  The classification of subprocesses according to
\vv{IHPRO} is as for \vv{IPROC}=1500.

\subsubsection{\vv{IPROC}=1800: QCD direct photon plus jet production}
\label{direct}

The relevant \vv{IHPRO} codes are 41--47 in table~\ref{tab.11}.  For
future referen\-ce we also collect here the codes for other processes
that involve outgoing direct photons or incoming pointlike photons
(\vv{IPROC}=2200, 5000--5520).
Note that the photon is colour/flavour-connected to itself. In the
cases \vv{IHPRO}=71--76, $M$ represents an $L=0$ meson (see
\vv{IPROC}=5500).

{\renewcommand\belowcaptionskip{-.5em}
\TABULAR{@{}|@{\,}c@{\,}|r@{\,\,}c@{\,\,}l|@{\,}c@{\,}|@{}}{\hline
\vv{IHPRO} &$ 1 + 2 $&$\to$&$ 3 + 4         $& c/f conn.\\\hline
41 &$ q + \qbar     $&$\to$&$ g + \gamma    $& 2 3 1 4\\
42 &$ q + g         $&$\to$&$ q + \gamma    $& 3 1 2 4\\
43 &$ \qbar + q     $&$\to$&$ g + \gamma    $& 3 1 2 4\\
44 &$ \qbar + g     $&$\to$&$ \qbar + \gamma$& 2 3 1 4\\
45 &$ g + q         $&$\to$&$ q + \gamma    $& 2 3 1 4\\
46 &$ g + \qbar     $&$\to$&$ \qbar + \gamma$& 3 1 2 4\\
47 &$ g + g         $&$\to$&$ g + \gamma    $& 2 3 1 4\\\hline
51 &$ \gamma+  q    $&$\to$&$ g+ q          $& 1 4 2 3\\
52 &$ \gamma+ \qbar $&$\to$&$ g+     \qbar  $& 1 3 4 2\\
53 &$ \gamma+ g     $&$\to$&$ q    + \qbar  $& 1 4 2 3\\\hline
61 &$ q + \qbar     $&$\to$&$ \gamma+\gamma $& 2 1 3 4\\
62 &$ \qbar +  q    $&$\to$&$ \gamma+\gamma $& 2 1 3 4\\
63 &$ g + g         $&$\to$&$ \gamma+\gamma $& 2 1 3 4\\\hline
71 &$ \gamma+ q     $&$\to$&$ M(S=0) +q'    $& 1 4 3 2\\
72 &$ \gamma+ q     $&$\to$&$ M(S=1)_L+q'   $& 1 4 3 2\\
73 &$ \gamma+ q     $&$\to$&$ M(S=1)_T+q'   $& 1 4 3 2\\
74 &$ \gamma+ \qbar $&$\to$&$ M(S=0)+\qbar' $& 1 4 3 2\\
75 &$ \gamma+ \qbar $&$\to$&$M(S=1)_L+\qbar'$& 1 4 3 2\\
76 &$ \gamma+ \qbar $&$\to$&$M(S=1)_T+\qbar'$& 1 4 3 2\\\hline}%
{Direct photon subprocesses.\label{tab.11}}}

\subsubsection[\vv{IPROC}=1900--1999: Higgs production by weak boson fusion]%
{\vv{IPROC}=1900--1999: Higgs boson production by weak boson fusion}

{\sloppy
The $q\qbar\to q^{(')}\qbar^{(')} VV\to q^{(')}\qbar ^{(')}\SMH$
subprocesses, for $VV=W^+W^-,Z^0Z^0$, summed over initial-
and final-state quarks can be invoked by setting
\vv{IPROC}=1900+\vv{ID}, with \vv{ID} used to identify the Higgs decay.

}

The formulae used are well known in the literature and can be found
e.g.\ in~\cite{Moretti:2002eu}. This process is administered by the
subroutine \vv{HWHIGW}, which also handles the similar cases initiated
by $e^+e^-$ and $e^\pm p$ collisions.

\subsubsection{\vv{IPROC}=2000--2008: single top production}

The process of single top quark production by $W$-boson exchange
includes so far only those processes initiated by a $b$-quark (or
antiquark) and a first- or second-generation quark or antiquark.  Note
that this requires $b$-quarks to be available in the parton
distribution functions.
\subsubsection[\vv{IPROC}=2100--2170: electroweak boson plus jet production]
{\vv{IPROC}=2100--2170: electro-\\weak boson plus jet production}
The electroweak boson decay correlations and width are now correctly
included in these processes.

\subsubsection{\vv{IPROC}=2200: direct photon pair production}

See section~\ref{direct} for \vv{IHPRO} codes (61--63).

\subsubsection{\vv{IPROC}=2300--2399: Higgs boson plus jet production}
\label{qcd_higgs}

High transverse momentum, scalar Higgs production via one-loop diagrams,
in association with
a jet, is available as \vv{IPROC}=2300, within the \SM. Only the top
quark is included in the loops with \vv{IAPHIG} controlling the
approximation used: \vv{IAPHIG}=0 gives the zero top mass limit, 1
(default) the exact result, 2 the infinite top mass limit.
The various subprocesses are illustrated in table~\ref{tab.12}.

\TABULAR{|c|rcl|c|}{
\hline
\vv{IHPRO} &$ 1 + 2         $&$\to$&$ 3 + 4         $& c/f conn. \\
\hline
81 &$ q    + \qbar  $&$\to$&$ g    +\SMH      $& 2 3 1 4 \\
82 &$ q    + g      $&$\to$&$ q    +\SMH      $& 3 1 2 4 \\
83 &$ \qbar + q     $&$\to$&$ g    +\SMH      $& 3 1 2 4 \\
84 &$ \qbar + g     $&$\to$&$ \qbar +\SMH     $& 2 3 1 4 \\
85 &$ g    + q      $&$\to$&$ q    +\SMH      $& 2 3 1 4 \\
86 &$ g    + \qbar  $&$\to$&$ \qbar +\SMH     $& 3 1 2 4 \\
87 &$ g    + g      $&$\to$&$ g    +\SMH      $& 2 3 1 4 \\
\hline
}{Higgs plus jet subprocesses.\label{tab.12}}

{\sloppy
Note that the Higgs boson is co\-lour/flavour connected to itself.
   
The relevant routines \vv{HWHGJ1, HWHGJA/B/C/D, HWUCI2} and \vv{HWULI2}
use (non-standard \Fos) \vv{DOUBLE COMPLEX} variables which may
not be accepted by some compilers and are called \vv{COMPLEX*16} by
others.  Users can change to \vv{COMPLEX} variables, but this involves
a risk of rounding errors spoiling numerical cancellations.

}

\subsubsection{\vv{IPROC}=2400--2450: colour singlet exchange}

\vv{IPROC}=2400: Two-to-two parton scattering via exchange of a colour
singlet, Mueller-Tang pomeron~\cite{MuTan}.  The fixed $\as$ and
$\omega_0$ are given by \vv{ASFIXD} (default 0.25) and \vv{OMEGA0} (0.3)
respectively.\\
\vv{IPROC}=2450: Photon exchange, for like-flavour $q\qbar$ pairs
including the $t$-channel component of the interference with
$q\qbar\to q\qbar$ via an $s$-channel photon or $Z^0$.

\subsubsection{\vv{IPROC}=2500--2599: Higgs boson plus top quark pair 
production}
\label{SMQQHiggs}

The \SM\ $2\to3$ Higgs production subprocesses of the type $gg\to
t\tbar\SMH$ and $q\qbar\to t\tbar\SMH$, for any flavour of initial
state quarks $q$, are new to \HW\ version~6 and are handled by the
subroutines \vv{HWHIGQ} and \vv{HWH2QH}. They are invoked by setting
\vv{IPROC}=2500+\ID\ (both $gg$ and $q\qbar$), with \ID\ administering
the Higgs decay channels as in \vv{IPROC}=300+\ID.  The initial state
quark flavours $q$ are always summed over. Notice that, given the size
of the Yukawa couplings of the Higgs boson to quarks, in practice only
the associated production with top quarks is of phenomenological
relevance in the SM.  The matrix elements used in the implementation
can be found in ref.~\cite{Moretti:2002eu}.  The treatment of the Higgs width
here is as described in section~\ref{sec:gaugebosons}.

\subsubsection{\vv{IPROC}=2600--2799: Higgs plus weak boson production}

The associated production of \SM\ Higgs scalars with $W^\pm$
(\vv{IPROC}=2600-2699) and $Z^0$ (\vv{IPROC}=2700-2799) gauge bosons
initiated by quark-antiquark fusion via the $2\to2$ processes
$q\qbar\to W^\pm\SMH$ and $q\qbar\to Z^0\SMH$ is now available. A
summation is as usual intended on the incoming quarks. The formulae
given in~\cite{Moretti:2002eu} are used.  The production processes are
generated by the two new subroutines \vv{HWHIGV} and \vv{HWH2VH} whereas
the Higgs decays are administered through the \ID\ increment, as in
\vv{IPROC}=300+\ID. Again, the treatment of the Higgs boson width here
is as discussed in section~\ref{sec:gaugebosons}.

\subsubsection{\vv{IPROC}=2800--2825: Gauge boson pair production}
  In version 6.3,
  the code already included in \HW\ for $e^+e^-\to$ WW/ZZ \cite{Gunion:1986mc} 
  was adapted for hadron-hadron collisions, including photons and the photon/Z
  interference for the resonant diagrams.

  All of these processes use a cut \vv{EMMIN} (default value 20 GeV)
  on the mass of the gauge bosons produced. The cut \vv{PTMIN} (default 10 GeV)
  on the transverse momentum of the bosons is also used.
  Both these cuts should not be taken to zero simultaneously if photon terms
  are included. The phase space for these processes contains a number of peaks
  and it was therefore necessary to use an adaptive multi-channel phase-space
  integration method which is described below.

  In versions 6.4 and above, the hard process scale \vv{EMSCA} for this
  process has been  changed to  the 
  parton-parton  centre-of-mass  energy  from  the  average  of  the 
  produced boson masses, which was used previously.

\subsubsection{\vv{IPROC}=2850--2880: Gauge boson pairs using MC@NLO}
These codes activate the interface to the program \MN\ for diboson production
at next-to-leading order (see sect.~\ref{sec:MCNLO}).

\subsubsection[\vv{IPROC}=2900--2916: $Q\Qbar Z$ production]
{\vv{IPROC}=2900--2916: \boldmath{$Q\Qbar Z$} production}

  The matrix elements of \cite{Kleiss:1985yh} were used for the massless
  case and an independent calculation,
  using the approach of \cite{vanEijk:1990zp},
  which was checked both for gauge invariance and against
  the massless case for the massive result.

  In both cases the decay of the $Z$ is fully included and is selected using
  \vv{MODBOS}. \vv{PTMIN} controls the minimum transverse momentum of the
  outgoing quarks.
  As with gauge boson pair production it was necessary to use an optimized
  multi-channel phase-space integrator which is described below. 

  The phase space for both gauge boson pair production and $Q\Qbar Z$
  is complicated, as these processes are both treated as $2\to4$ processes.
  In order to obtain a reasonable efficiency it was necessary to adopt a
  multi-channel approach based on that described in
  \cite{Berends:1995xn,Kleiss:1994qy}.

  In each case a number of different channels are included which attempt
  to map the phase space for the different processes.
  The default weights for these different channels 
  have been chosen to optimize the efficiency for the Tevatron
  and LHC; a choice of which to use is made based on the beam energy.
  However, the choice is affected by the phase space cuts applied.
  Therefore if these are significantly altered the weights for
  the different channels need re-optimizing. 

  This is controlled by the new variable \vv{OPTM} (default \vv{.FALSE.}).
  If \vv{OPTM}=\vv{.TRUE.}, before performing the initial search for
  the maximum weight \HW\ will attempt to optimize the
  efficiency using the procedure suggested in \cite{Kleiss:1994qy}. 

  This is done by generating \vv{IOPSTP} (default 10) iterations of
  \vv{IOPSH} (default 1000) events.
  The choice of \vv{IOPSTP} and \vv{IOPSH} is a compromise between run
  time and accuracy.
  The value of \vv{IOPSH} should not be significantly reduced because the
  procedure
  attempts to minimize the error on the Monte Carlo evaluation of the cross
  section and if \vv{IOPSH} is small the error on the error can be significant.
  If you need to re-optimize the weights we would recommend a
  long run just to optimize the weights which can then be used in all the
  runs to generate events. 
  The new subroutine \vv{HWIPHS} was added to initialise the phase space.

\subsection{Hadron-hadron supersymmetric processes (MSSM)}\label{susy}
\label{sec:SUSYproc}

The R-parity conserving \SY\ processes occupy the \vv{IPROC} entries of
the series 3000 (sparticle processes) and
3300--3600 (Higgs boson production),
while \RPV\ processes have $\vv{IPROC}=4000-4199$.

As with the lepton-antilepton SUSY processes the SUSY particle data
must be read in from an input file before event generation as
described in section~\ref{sec:SUSYdata}.  In particular, unlike those of
the SM Higgs boson $\SMH$, the widths and decay modes of the \SY\
Higgs bosons are not computed by \HW.

A large number of final states involving the production
of both neutral and charged Higgs bosons of the Minimal Supersymmetric
Standard Model
(MSSM) have been made available in version 6.3.  
They all proceed via $2\to3$ body hard scattering subprocesses.
They are listed below, with corresponding process numbers
(\IQ\ and \ISQ\ are as detailed in the following subsections).
Further details of their implementation can be found in \cite{Moretti:2002eu}

The new subroutines introduced to administer the following processes
are \vv{HWHIBQ} and \vv{HWH2BH} for
\vv{IPROC}=3500, plus \vv{HWHISQ} and \vv{HWH2SH} for 
\vv{IPROC}=3100, 3200. In addition, \vv{HWHIGQ} has been modified
to accommodate the \vv{IPROC}=3800 series.

\subsubsection{\vv{IPROC}=3000--3030: sparton, gaugino and/or slepton production}

With \vv{IPROC}=3000 one obtains the three following processes,
\vv{IPROC}=3010, 3020, 3030, in the correct proportions. The variable
\vv{IHPRO} gives the subprocess actually generated (table \ref{tab.13}).

The matrix elements have been derived independently and the cross
sections are in good agreement with those from \IS~\cite{ISA}.

The hard process scale \vv{EMSCA} has to be chosen globally for all
sparton processes, e.g.\ for the argument of the \QCD\ coupling.  This
is done using the kinematics appropriate to production of the lightest
supersymmetric particle (\sss{LSP}) and
$$
\vv{EMSCA}=\sqrt{2\:\shat\:\that'\:\uhat'\over\shat^2+\that'^2+\uhat'^2}\,.
$$
with $\that'=\that-m^2$, $\uhat'=\uhat-m^2$ where $m$ is the \sss{LSP} mass.

$\vv{IPROC}=3010$ gives 2-parton $\to$ 2-sparton processes.  All \QCD\
sparton, i.e.\ squark and gluino, pair production processes are
implemented. The matrix elements and the scheme for separating
different colour flow parts are as given in~\cite{ko_colour}.

$\vv{IPROC}=3020$ gives 2-parton $\to$ 2-gaugino or gaugino+sparton
processes.  All gaugino, i.e.\ chargino and neutralino, pair
production processes and gaugino-sparton associated production
processes are implemented.

The gaugino and sfermion mixing conventions of Haber and Kane
~\cite{HabKan} are used as described in section~\ref{sec_susy_lept}.

The various subprocesses, which include the $1\leftrightarrow2$ and
charge conjugate reactions omitted below for brevity, are shown in
table~\ref{tab.13}.

\TABULAR{|c|rcl|c|}{
\hline
\vv{IHPRO} &$  1 + 2  $&$\to$&$             3 + 4            $& c/f conn.\\
\hline
3021 &$ q+\qbar $&$\to$&$ \gaugino^\pm_a+\gaugino^\mp_b $& 2 1 3 4 \\
3022 &$ q+\qbar $&$\to$&$ \gaugino^0_i  +\gaugino^0_j   $& 2 1 3 4 \\
3023 &$ q+\qbar'$&$\to$&$ \gaugino^\pm_a+\gaugino^0_i   $& 2 1 3 4 \\
3024 &$ q+\qbar $&$\to$&$ \gaugino^0_i  +\gluino        $& 2 4 3 1 \\
3025 &$ q+\qbar'$&$\to$&$ \gaugino^\pm_a+\gluino        $& 2 4 3 1 \\
3026 &$ g+ q    $&$\to$&$ \gaugino^0_i  +\squark        $& 2 4 3 1 \\
3027 &$ g+ q    $&$\to$&$ \gaugino^\pm_a+\squark'       $& 2 4 3 1 \\
\hline
}{\SY\ subprocesses.\label{tab.13}}
Note that the gauginos connect to themselves. The indices $a,b,i,j$
label gauginos in the order of increasing mass and take values
$a,b=1-2$ and $i,j=1-4$.

{\sloppy
Gaugino mixing matrices are implemented in all subprocesses. The associated
production subprocesses \vv{IHPRO}=3026, 3027 include stop and sbottom
left-right mixings. CKM mixing is implemented in subprocesses
\vv{IHPRO}=3023, 3025, 3027 but neglected in subprocess \vv{IHPRO}=3021.

$\vv{IPROC}=3030$ gives 2-parton $\to$ 2-slepton processes.
All Drell-Yan slepton production processes are implemented.
The formulae agree with those of refs.~\cite{EHLQ,DEQ}
in the limit of no stau left-right mixing.

}

\subsubsection{\vv{IPROC}=3110--3178: charged Higgs plus squark pair
production}\label{sqsqc}

The production of charged Higgs bosons of the MSSM in association
with squark pairs, of bottom and top flavours only, is implemented
via the 3100 series of \vv{IPROC} numbers (table \ref{tab:sqch}).
Their phenomenological relevance has been discussed in \cite{sqsqh1}.

\TABULAR[h]{|c|rcl|c|}{
\hline
       \vv{IPROC} & partons         &$\to$& spartons         & Higgs \\
\hline
           3110  &$ gg+q\qbar $&$\to$&$ {\tilde q}_i{\tilde q}^{'*}_j$&$H^\pm$ \\
           3111  &$ gg+q\qbar $&$\to$&$ {\tilde b}_1{\tilde t}^*_1 $& $H^+$ \\
           3112  &$ gg+q\qbar $&$\to$&$ {\tilde b}_1{\tilde t}^*_2 $& $H^+$ \\
           3113  &$ gg+q\qbar $&$\to$&$ {\tilde b}_2{\tilde t}^*_1 $& $H^+$ \\
           3114  &$ gg+q\qbar $&$\to$&$ {\tilde b}_2{\tilde t}^*_2 $& $H^+$ \\
           3115  &$ gg+q\qbar $&$\to$&$ {\tilde t}_1{\tilde b}^*_1 $& $H^-$ \\
           3116  &$ gg+q\qbar $&$\to$&$ {\tilde t}_1{\tilde b}^*_2 $& $H^-$ \\
           3117  &$ gg+q\qbar $&$\to$&$ {\tilde t}_2{\tilde b}^*_1 $& $H^-$ \\
           3118  &$ gg+q\qbar $&$\to$&$ {\tilde t}_2{\tilde b}^*_2 $& $H^-$ \\
\hline}{Processes for \vv{IPROC}=3110--3178.\label{tab:sqch}}

One should add 30(60) to \vv{IPROC} for $gg$($q\qbar$)-only initiated processes.

\subsubsection[\vv{IPROC}=3210--3298: neutral Higgs plus squark pair
production]{\vv{IPROC}=3210--3298: neutral\\ Higgs plus squark pair
production}\label{sqsqn}

The production of neutral Higgs bosons of the MSSM in association
with squark pairs, of bottom and top flavours only, is implemented
via the 3200 series of \vv{IPROC} numbers (table \ref{tab:sqne}).
Their phenomenological relevance has been discussed in 
\cite{sqsqh1,sqsqh2}.

One should add 30(60) to \vv{IPROC} for $gg$($q\qbar$)-only initiated processes.

\subsubsection[\vv{IPROC}=3310--3375: Higgs pair and Higgs-W/Z production]
	{\vv{IPROC}=3310--3375: Higgs-Higgs and
	Higgs-gauge boson pair\\ production}\label{bosbos}

The production of Higgs-Higgs and Higgs-gauge boson pairs of the
\MS\ is implemented at tree level. We include gauge boson mediated
contributions but not Higgstrahlung from the initial state, the only
exception being $W^\pm H^\mp$ production (\vv{IPROC}=3350), which does
include diagrams where the Higgs boson couples to the initial partons
as well as those mediated by neutral Higgs states~\cite{BBK}.  Some of
these processes are the \MS\ equivalent of \vv{IPROC}=2600 and 2700
described earlier for the case of the \SM. Here, the cases
\vv{IPROC}=3310(3320) correspond to $W^\pm h^0(W^\pm H^0)$ and
\vv{IPROC}=3360(3370) to $Z^0 h^0(Z^0 H^0)$ final states.  (No similar
$A^0$ production can occur at leading order.) The array \vv{ENHANC} is
used to implement the \MS\ couplings of Higgs scalars to gauge
bosons.

\TABULAR[t]{|c|rcl|c|}{
\hline
       \vv{IPROC} & partons         &$\to$& spartons         & Higgs \\
\hline
3210(3220)[3230]&$gg+q\qbar $&$\to$&$ {\tilde q}_i{\tilde q}^*_j $& $h(H)[A]$ \\
3211(3221)[3231]&$gg+q\qbar $&$\to$&$ {\tilde b}_1{\tilde b}^*_1 $& $h(H)[A]$ \\
3212(3222)[3232]&$gg+q\qbar $&$\to$&$ {\tilde b}_1{\tilde b}^*_2 $& $h(H)[A]$ \\
3213(3223)[3233]&$gg+q\qbar $&$\to$&$ {\tilde b}_2{\tilde b}^*_1 $& $h(H)[A]$ \\
3214(3224)[3234]&$gg+q\qbar $&$\to$&$ {\tilde b}_2{\tilde b}^*_2 $& $h(H)[A]$ \\
3215(3225)[3235]&$gg+q\qbar $&$\to$&$ {\tilde t}_1{\tilde t}^*_1 $& $h(H)[A]$ \\
3216(3226)[3236]&$gg+q\qbar $&$\to$&$ {\tilde t}_1{\tilde t}^*_2 $& $h(H)[A]$ \\
3217(3227)[3237]&$gg+q\qbar $&$\to$&$ {\tilde t}_2{\tilde t}^*_1 $& $h(H)[A]$ \\
3218(3228)[3238]&$gg+q\qbar $&$\to$&$ {\tilde t}_2{\tilde t}^*_2 $& $h(H)[A]$ \\
\hline}{Processes for \vv{IPROC}=3210--3298.
\label{tab:sqne}}
\subsubsection{\vv{IPROC}=3410--3450: Higgs boson plus heavy quark
production}\label{Higghvy}

We have included so far only those processes initiated by a $b$-quark
(or antiquark) and a gluon. Note that this requires $b$-quarks
to be available in the parton distribution functions.

\subsubsection{\vv{IPROC}=3500: charged Higgs boson from
$bq$-initiated processes}\label{bq}

This process is relevant for charged Higgs scalar production at large 
$\tan\beta$ values, see \cite{SMKO}.  

\subsubsection[\vv{IPROC}=3610--3630: parton fusion to neutral Higgs]
{\vv{IPROC}=3610--3630: neutral Higgs production by parton fusion}
\label{Higgfus}

These processes are the \MS\ analogues of the \SM\ processes 1600
etc.  Recall however that the \MS\ Higgs decay modes are controlled
by the \SY\ input data (see section~\ref{sec:SUSYdata}) and not by the
value of \vv{IPROC}. The subroutines \vv{HWHIGS} and \vv{HWHIGT} have
been modified to take account of squark loop contributions and
parity-violating Higgs-fermion couplings in the \MS\ case.

\subsubsection[\vv{IPROC}=3710--3720: weak boson fusion to neutral Higgs]
	{\vv{IPROC}=3710--3720: neutral Higgs production via 
                                 weak boson fusion}

These processes are the \MS\ counterparts of the \SM\ process
of weak vector-vector fusion in hadronic collisions (\vv{IPROC}=1900). They
are computed using the same subroutines and setting
the $\Phi^0 VV$ couplings to $\sin(\beta-\alpha)$ for 
$\Phi^0=h^0$ (\vv{IPROC}=3710) and to $\cos(\beta-\alpha)$ 
for $\Phi^0=H^0$ (\vv{IPROC}=3720), respectively, where $V=W^\pm,Z^0$.
(There is no $A^0VV$ coupling at tree level.)

These reactions are two of the major direct production channels of neutral
CP-even Higgs bosons of the \MS\ at hadron colliders, such as the 
LHC (see e.g.\ \cite{Spira}).

\subsubsection[IPROC=3811--3899: Higgs  plus heavy quark 
                                 pair production]{IPROC=3811--3899: Higgs boson plus heavy quark 
                                 pair production}\label{MSSMQQHiggs}

For the case of neutral Higgs states,
\vv{IPROC}=3810+\IQ\ corresponds to $h^0$ production,
\vv{IPROC}=3820+\IQ\ to $H^0$ and \vv{IPROC}=3830+\IQ\ to $A^0$.
(For the last case, the variable \vv{PARITY} is set to $-1$.)
Note also the production of charged Higgs states, via
\vv{IPROC}=3839, 3869 and 3899, in association with pairs of top-bottom 
quarks. 

In the usage of the \vv{IPROC} numbers corresponding to neutral
Higgs states, when $b$-quarks are involved in $gg$-fusion modes
(\vv{IPROC}(+30)=3845, 3855 or 3865),
the user should take care to avoid double-counting the chosen 
process with the corresponding $2\to1$ and $2\to 2$ cases 
of \vv{IPROC}={3610}--3630 and \vv{IPROC}=3410--3430
initiated by quark-antiquark annihilation, i.e.\ $b \bar b\to$ Higgs,
and (anti)quark-gluon scattering, i.e. $bg\to b$ Higgs, respectively:
see \cite{Scott}.  
Similar arguments hold for the charged Higgs states, as
the $gg$-induced process (\vv{IPROC}(+30)=3869) is an alternative 
implementation of \vv{IPROC}=3450 \cite{Borz}.

The associate production of neutral Higgs bosons (both CP-even and
CP-odd) of the \MS\ with heavy $Q\Qbar$ pairs ($Q=b$ and $t$) is
of extreme phenomenological relevance as a discovery mode of Higgs
scalars, both at the Tevatron (Run 2) and the LHC 
(see e.g.\ \cite{Spira,TeV2}),
as is the case of the charged Higgs channel \cite{Borz,charged}.

\subsection{Other hadron-hadron non-Standard-Model processes}

\subsubsection{\vv{IPROC}=4000--4199: R-parity violating SUSY processes}
\label{sec:RPVhadronproc}

We include all the possible production processes of resonant sleptons
and squarks in hadron-hadron collisions, for arbitrary numbers of
non-zero \RPV\ couplings. These processes are implemented as
two-to-two processes, i.e.\ with the decay of the resonant particle
included.  This allows us to include the $t$-channel diagrams where
these occur. However we have not implemented those processes which can
only occur via a $t$-channel diagram, or where the resonance will
never be accessible. So for example while we include the process $u_i
d_j \rightarrow \ino{b}^*_1 Z^0$, which can occur via a resonant
$\ino{b}^*_2$, we do not include the process $u_i d_j \rightarrow
\ino{b}^*_2 Z^0$, which cannot occur via a resonant diagram. In all
cases both the processes listed and their charge conjugates are
included.

The scale choice is $\sqrt{\shat }$ rather than the conventional
transverse mass, due to the large number of different processes which
must be calculated. The colour connection structure of these processes
and their matrix elements can be found in ref.~\cite{Rtheory}.

\subsubsection{\vv{IPROC}=4200--4299 : graviton resonance production}

In some models with extra dimensions, Kaluza-Klein excitations of the
graviton can be produced with significant cross sections at TeV-scale
colliders.  When the scale of the extra dimensions is not large, the
excitations are manifest as discrete resonances.

The production of a resonant excitation of the graviton is implemented
as a $2\to2$ process including the decay of the resonance.  The
process is treated in a model-independent way, assuming only that
there is a universal coupling of the graviton resonance to the \SM\
fields. The effective lagrangian is given by
$$
  {\mathcal L}_I = -\frac{1}{\Lambda_\pi}h^{\mu\nu}T_{\mu\nu}\,,
$$
where $h^{\mu\nu}$ is the spin-2 field and $T_{\mu\nu}$ is the
energy-momentum tensor of the \SM\ fields. Although in these models
there are usually many resonances, we have implemented only
one. Others can be studied by making appropriate changes in the
parameters. Graviton resonance production is described in more detail
in~\cite{Allanach:2000nr}.

The process is controlled by the coupling $\vv{GRVLAM}=\Lambda_\pi$,
with dimensions of mass and default value 10 TeV, and by the mass
\vv{EMGRV} (default 1 TeV) and width \vv{GAMGRV} of the resonance.  If
the width is set to zero (the default), the subroutine \vv{HWHGRV}
which calculates the cross section also calculates the width.

The parton-level cross section for this process is non-unitary and is
proportional to $\shat/\vv{EMGRV}^4$ at high energies. The fall-off of
the parton distribution functions is not sufficient to suppress this
bad high energy behaviour. Hence the parameters \vv{EMMIN} and
\vv{EMMAX} controlling the minimum and maximum values of $\sqrt{\shat}$
must be set. The default is to set these to 90\% and 110\% of the
graviton resonance mass respectively. If the width of the resonance is
more than a few percent of its mass then these limits should be reset.

After event generation, \vv{IHPRO} is set to 50 for $q\qbar$ initiated
subprocesses and to 51 for $gg$ initiated subprocesses.

\subsection{Photon-hadron and photon-photon processes}

\subsubsection{\vv{IPROC}=5000--5520: pointlike photon-hadron processes}

Pointlike photon-hadron scattering to produce \QCD\ jets is available
as $\vv{IPROC}=5000-5206$.  This is suitable for fixed-target
photoproduction, provided events are generated in a frame in which the
target has high momentum, and then boosted back to the lab.  This is
done if $\vv{USECMF}=\vv{.TRUE.}$, the default, in which case the frame
for event generation is the beam-target c.m.\ frame.
$\vv{IPROC}=5100+\IQ$ gives flavour \IQ\ pair production, $\gamma+g\to
Q\Qbar$, and $\vv{IPROC}=5200+\IQ$ gives flavour \IQ\ single
excitation, $\gamma+Q\to g+Q$, both including quark masses.
$\vv{IPROC}=5000$ gives a sum over all processes and flavours, 5100 and
5200, with massless quark kinematics.  In all cases, after event
generation the code \vv{IHPRO} is set to 51, 52 or 53 according to the
hard subprocess, as specified in section~\ref{direct}.  $\vv{IPROC}=5300$
gives Compton scattering, $\gamma + q\to\gamma + q$.

The direct, higher twist, production of light ($u,d,s$) L=0 mesons by
point-like photons is also available: \vv{IPROC} = 5500 for all spin =0
and 1 mesons; 5510 for only S=0 mesons; and 5520 for only S=1
mesons. The vector mesons are produced with transverse or longitudinal
polarization and decayed accordingly.  The corresponding \vv{IHPRO}
codes (71--76) are also listed in section~\ref{direct}.

All these processes are available with lepton as well as hadron beams,
using the Equivalent Photon Approximation.  The phase-space variable
\vv{WHMIN} sets the minimum allowed hadronic mass and affects
photoproduction reactions ($\gamma$-hadron and $\gamma$-$\gamma$) and
\DIS.  In lepton-hadron \DIS\ it is largely irrelevant since there is
already a cut on Bjorken $y$ which at fixed $s$ is almost the same,
but for lepton-gamma \DIS\ it makes a big difference.  Direct
$\gamma+\gamma^*\to q+\qbar$ is included in the hard correction for
lepton-gamma \DIS.

\subsubsection{\vv{IPROC}=6000--6010: pointlike photon-photon processes}

Direct $\gamma\gamma\to$ charged particle pairs has been implemented
with $\vv{IPROC}=6000+\IQ$: if $\IQ=1-6$ then only quark flavour \IQ\
is produced, if $\IQ=7,8$ or 9 then only lepton flavour $e$, $\mu$ or
$\tau$ is produced and if $\IQ=10$ then only $W$ pairs are produced: in
these cases particle mass effects are included. If $\IQ=0$, the
natural mix of quark pairs is produced using massless matrix elements
but including a mass threshold cut. The range of allowed transverse
momenta is controlled by \vv{PTMIN} and \vv{PTMAX} as usual.

\subsection{Baryon number violating processes}

\subsubsection{\vv{IPROC}=7000--7999: generated by the HERBVI package}

  See section~\ref{sect:HERBVI} for details.

\subsection{Minimum bias soft hadron-hadron collisions}\label{minbias}

\subsubsection[\vv{IPROC}=8000: minimum bias soft hadron-hadron event]
	      {\vv{IPROC}=8000: minimum bias soft hadron-hadron event}

Non-diffractive, soft hadronic, minimum bias events (\vv{IPROC}=8000)
can be generated for the following combinations of beam and target:
$p,\pbar,\pi^\pm,K^\pm,e^\pm,\mu^\pm,\gamma$ on target $p$ (or vice
versa); $p,\pbar$ on target $n$ (or vice versa); or $\gamma$ on
$\gamma$.  The event weight is the estimated cross section based on
the parameterizations of Donnachie and Landshoff~\cite{DonLan}.  The
non-diffractive cross section is assumed to be 70\% of the total.  For
lepton beams a photon is first generated using the Effective Photon
Approximation (see section~\ref{beams}) and then the on-shell photon
cross section is used.
See section~\ref{minbiproc} for discussion of the model used and the relevant
parameters.

\subsection{Deep inelastic lepton scattering}\label{sec_DIS}

Deep inelastic (\DIS) processes are broadly divided into those that
start at ${\cal O}(\as^0)$ (\vv{IPROC}=90**) and those like heavy quark
and dijet production which start at ${\cal O}(\as^1)$
(\vv{IPROC}=91**).  Note that the \DIS\ ${\cal O}(\as)$ jet production
processes, \vv{IPROC}=92**, have been withdrawn since they are subsumed
within \vv{IPROC}=91**.

The default limits on $Q^2$ in \DIS\ processes (\vv{Q2MIN}, \vv{Q2MAX})
have been set very small/large (0, 10$^{10}$\,GeV) and are reset to
the kinematic limits unless changed by the user.  This means the
default \vv{Q2MIN} is not suitable for simple neutral current \DIS\
(\vv{IPROC}=9000 etc), but is appropriate for jet and heavy quark
photoproduction.  The range of the Bjorken-$y$ variable ($y=Q^2/xs$)
can be limited by \vv{YBMIN} and \vv{YBMAX}.

The kinematic reconstruction of \DIS\ processes can now take place in
the Breit frame, if \vv{BREIT}=\vv{.TRUE.} (the default value).
Previous versions used the lab frame. Although the reconstruction is
fully invariant under Lorentz boosts along the incoming hadron's
direction, it is not under transverse boosts, so there should be some
difference between the two frames. The boost is not performed for very
small $Q^2 (<10^{-4})$ to avoid numerical instabilities, but the two
frames are in any case equivalent for such small values.

The phase space for boson-gluon fusion is controlled by the parameters
\vv{EMMIN}, \vv{EMMAX} (see section~\ref{twophot}). The default values (0,
$\sqrt s$) correspond to the behaviour of version~5.1.

Lepton beam polarization effects are included in all \DIS\ processes
apart from $J/\psi$ production. The polarization is specified as in
lepton-antilepton processes, i.e.\ by setting the 3-vectors \vv{EPOLN}
and \vv{PPOLN}: component 3 is longitudinal and 1,2
transverse. Transverse only occurs in $\ee$ routines; recall that two
transverse `measurements' are needed to see an effect so it should not
arise elsewhere. Note that in \DIS\ processes one has to set either
\vv{EPOLN} if it is a lepton or (exclusive) \vv{PPOLN} if an antilepton.

All the \DIS\ processes \vv{IPROC}=9000--9599 are available in $\ee$ as
well as lepton-hadron collisions.  The program generates a photon from
the second beam (only) in the Equivalent Photon Approximation and the
default is to use Drees-Grassie~\cite{Drees,DK} structure functions
for \DIS\ on the photon.

The parameter \vv{WHMIN} sets the minimum allowed hadronic mass in
\DIS.  In lepton-hadron \DIS\ it is largely irrelevant since there is
already a cut on Bjorken $y$ which at fixed $s$ is almost the same but
for lepton-gamma \DIS\ it makes a big difference.

In addition to the processes listed here, a full simulation of \QCD\
instanton-in\-duced processes in \DIS~\cite{DISinst} is available
through an interface to the program\,{\sf QCDINS} \cite{QCDINS}. For
details, see the web page
{\small \href{http://www.desy.de/~t00fri/qcdins/qcdins.html}
{\tt http://www.desy.de/~t00fri/qcdins/qcdins.html}}

\subsubsection{\vv{IPROC}=9000--9006: neutral current}

Matrix-element corrections to \DIS\ are available, following the
general method described in section~\ref{mecorr}.  The hard correction
is implemented in \vv{HWBDIS} and the soft correction is included in
routines \vv{HWBRAN} and \vv{HWSBRN} for the final- and initial-state
radiation respectively.

\subsubsection{\vv{IPROC}=9010--9016: charged current}

These are the charged current processes corresponding to those above,
with the same treatment of hard and soft matrix element corrections.

\subsubsection{\vv{IPROC}=9100--9130: ${\cal O}(\as)$ neutral current processes}

The photoproduction processes have been extended from the original
heavy quark production program, to include all quark pair production
(\vv{IPROC}=9100-9106) and \QCD\ Compton (\vv{IPROC}=9110--9122), as well
as the sum of the two (\vv{IPROC}=9130).  The possible flavours for
processes 9100, 9110 and 9130 are limited by the input parameters
\vv{IFLMIN} and \vv{IFLMAX} (defaults are 1 and 3, i.e.\ only $u,d,s$
flavours).

A sign error has been corrected that led to the incorrect sign for the
lepton-jet azimuthal correlation in \QCD\ Compton processes in
versions before 5.7.

$J/\psi$ production (\vv{IPROC}=9107) now uses the Equivalent Photon
Approximation instead of Weizsacker-Williams, with the same
phase-space cuts as hadronic processes with lepton beams (see
section~\ref{beams}).

The argument of the running coupling is controlled by the parameter
\vv{BGSHAT} --- see section~\ref{control}.

\subsubsection{\vv{IPROC}=9140--9144: charged-current heavy quark production}
    At present only $W^\pm g$ fusion processes are implemented.

\subsubsection{\vv{IPROC}=9500--9599:  Higgs  production by weak boson fusion}

The process of $W^+ W^-/Z^0Z^0$ fusion into the \SM\ Higgs boson is
also available in $e^\pm p$ collisions, as $\vv{IPROC} = 9500+\vv{ID}$.
For details of the implementation, see section~\ref{sec:gaugebosons} and
the corresponding processes initiated by lepton-lepton and
hadron-hadron collisions, \vv{IPROC}=400+\ID\ and \vv{IPROC}=1900+\ID,
respectively.

\subsection{Including new subprocesses}
\label{including}

The procedure for including further subprocesses remains substantially
as described in ref.~\cite{hw51} but is repeated here for
completeness. However, we now recommend that users make use of the Les Houches
interface, see Section~\ref{sec:lesh}.

The parton and hard subprocess 4-momenta, masses and identity codes
need to be entered in \vv{COMMON/HEPEVT/} with the appropriate status
codes $\vv{ISTHEP(I)}=110$--114 to tell the program which is which (see
the table in section~\ref{status}).  The colour/flavour structure should
be specified by the second mother and daughter pointers as explained
in section~\ref{qcd}.

The \HW\ identity codes $\vv{IDHW(I)}$ in \vv{COMMON/HWEVNT/} also need
to be set correctly. The \vv{IDHW} codes can be listed in a run with
$\vv{IPRINT}=2$: the most important are the quarks 1--6 (as \vv{IDHEP}),
antiquarks 7--12, gluon 13, overall centre-of-mass 14, hard
centre-of-mass 15, soft centre-of-mass 16, photon 59, leptons
121--126, antileptons 127--132.

The utility subroutine \vv{HWUIDT(IOPT,IPDG,IHWG,NAME)} is provided to
translate between Particle Data Group code \vv{IPDG}, \HW\ code
\vv{IHWG}, and \HW\ \vv{CHARACTER*8 NAME}, with $\vv{IOPT}=1,2,3$
depending on which of \vv{IPDG}, \vv{IHWG} and \vv{NAME} is the input
argument.

Consider for example the process of virtual photon-gluon fusion to
make $b+\bbar$ in proton-electron collisions (in fact this process is
included as $\vv{IPROC}=9105$).  We assume the user provides a
\pagebreak[3]
subroutine to generate the momenta \vv{PHEP} for the hard subprocess
$e+g\to e b \bbar$.  The colour structure is
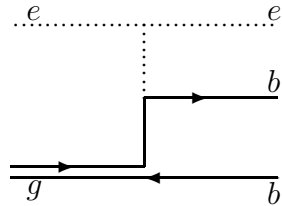
\begin{center}
{\begin{picture}(65,35)(0,-2)\thicklines
\put(57,0){\vector(-1,0){25}}
\put(32,0){\line(-1,0){25}}
\put( 7,2){\vector(1,0){12}}
\put(19,2){\line(1,0){13}}
\put(32,2){\line(0,1){13}}
\put(32,15){\vector(1,0){12}}
\put(44,15){\line(1,0){13}}
\multiput(32,15)(0,1.5){9}{\circle*{.3}}
\multiput(32,28.5)(1.5,0){17}{\circle*{.3}}
\multiput(32,28.5)(-1.5,0){17}{\circle*{.3}}
\put(10,-3){$g$}
\put(10,29.5){$e$}
\put(55,-5){$\bar{b}$}
\put(55,16){$b$}
\put(55,29.5){$e$}
\end{picture}}
\end{center}

Thus the momenta generated, together with those of the initial beams
and the overall centre of mass, could be entered in the 
sequence shown in table~\ref{tab.14}.

{\renewcommand{\belowcaptionskip}{-7pt}
\TABULAR[t]{|c|c|c|c|c|c|c|}{\hline
\vv{IHEP}&Entry&\vv{ISTHEP}&\vv{IDHEP}&\vv{JMOHEP}&\vv{JDAHEP}&\vv{IDHW}\\
\hline
1 & $e$ beam &  101 &   11&  0  0&  0  0& 121\\
2 & $p$ beam &  102 & 2212&  0  0&  0  0&  73\\
3 & $ep$ c.m.&  103 &    0&  0  0&  0  0&  14\\
\hline
4 & $e$ in   &  111 &   11&  6  7&  0  7& 121\\
5 & gluon    &  112 &   21&  6  9&  0  8&  13\\
6 & hard cm  &  110 &    0&  4  5&  7  9&  15\\
7 & $e$ out  &  113 &   11&  6  4&  0  4& 121\\
8 & $b$      &  114 &    5&  6  5&  0  9&   5\\
9 & $\bbar$  &  114 &  $-5$&  6  8&  0  5&  11\\
\hline
}{Event record entries for $eg\to eb\bar b$.\label{tab.14}}}

Note that if there are more than two outgoing partons, the first has
status 113 and all the others 114. Each parton has $\vv{JMOHEP(1,I)}=6$
to indicate the location of the hard c.m.\  for this subprocess, while
$\vv{JMOHEP(2,I)}$ gives the location of the colour mother (treating
the incoming gluon as outgoing) or the connected
electron. $\vv{JDAHEP(1,I)}$ will be set by the jet generator
\vv{HWBGEN}, while $\vv{JDAHEP(2,I)}$ points to the anticolour mother
(or connected electron). Finally the \HW\ identifiers $\vv{IDHW(I)}$
could be set to the indicated values by means of the translation
subroutine \vv{HWUIDT} as follows:\vspace{-1em}
\small\begin{verbatim}
                CHARACTER*8 NAME
                .....
                NHEP=9
                IDHEP(1)=11
                IDHEP(2)=2212
                .....
                IDHEP(9)=-5
                DO 10 I=1,NHEP
             10 CALL HWUIDT(1,IDHEP(I),IDHW(I),NAME)
                IDHW(6)=15
\end{verbatim}\normalsize
The last statement is needed because $\vv{IDPDG(I)}=0$ returns
$\vv{IDHW(I)}=14$.  If subroutine \vv{HWBGEN} is now called, it will
find the coloured partons and generate \QCD\ jets from
them. Subsequent calls to \vv{HWCFOR} etc.\ can then be used to form
clusters and hadronize them.

\pagebreak[3]

If the hard subprocess routine is called from \vv{HWEPRO}, like those
already provided, it must have two options controlled by the logical
variable \vv{GENEV} in \vv{COMMON/HWHARD/}. For $\vv{GENEV}=\vv{.FALSE.}$,
an event weight (normally the cross section in nanobarns) is generated
and stored as \vv{EVWGT} in \vv{COMMON/HWEVNT/}. If this weight is
accepted by \vv{HWEPRO}, the subroutine is called a second time with
$\vv{GENEV}=\vv{.TRUE.}$ and the corresponding event data should then be
generated and stored as explained above. On certain computers it will
be necessary to \vv{SAVE} those variables that determine event
characteristics between the two subroutine calls.

The parameter \vv{NMXJET} sets the maximum number of outgoing partons
in a hard subprocess (default 200).

\section{Parameters}\label{sec:params}
\label{input}

The quantities that may be regarded as adjustable parameters are
indicated in table~\ref{tab.15}. Notes on parameters are given below.

\TABULAR[b]{|c|l|c|}{
\hline
Name   &     Description                  & Default\\
\hline
\vv{QCDLAM}  & $\Lambda_{QCD}$ (see below)      & 0.18  \\
\hline
$\vv{RMASS(1)}$& Down    quark mass               & 0.32  \\
$\vv{RMASS(2)}$& Up      quark mass               & 0.32  \\
$\vv{RMASS(3)}$& Strange quark mass               & 0.50  \\
$\vv{RMASS(4)}$& Charmed quark mass               & 1.55  \\
$\vv{RMASS(5)}$& Bottom  quark mass               & 4.95  \\
$\vv{RMASS(6)}$& Top     quark mass               & 174.3 \\
\hline
$\vv{RMASS(13)}$& Gluon effective mass             & 0.75  \\
\hline
\vv{VQCUT}   & Quark virtuality cutoff (added to& 0.48  \\
            & quark masses in parton showers)  &       \\
\vv{VGCUT}   & Gluon virtuality cutoff (added to& 0.10  \\
            & effective mass in parton showers)&       \\
\vv{VPCUT}   & Photon virtuality cutoff         & 0.40  \\
\hline
\vv{CLMAX}   & Maximum cluster mass parameter   & 3.35  \\
\vv{CLPOW}   & Power in maximum cluster mass    & 2.00  \\
\vv{PSPLT(1)}   & Split cluster spectrum parameter & 1.00  \\
\vv{PSPLT(2)}   & 1: light cluster, 2 heavy $b$-cluster & \vv{PSPLT(1)} \\ 
\hline
\vv{QDIQK}   & Maximum scale for gluon$\to$diquarks& 0.00  \\
\vv{PDIQK}   & Gluon$\to$diquarks rate parameter& 5.00  \\
\hline
\vv{QSPAC}   & Cutoff for spacelike evolution   & 2.50  \\
\vv{PTRMS}   &Intrinsic $p_T$ in incoming hadrons& 0.00  \\
\hline
}{Adjustable parameters.\label{tab.15}}

\begin{itemize}
\item
\vv{QCDLAM} can be identified at high momentum fractions ($x$ or $z$) with
the fundamental 5-flavour \QCD\ scale $\lms^{(5)}$.  However, this
relation does not necessarily hold in other regions of phase space,
since higher order corrections are not treated  precisely enough to
remove renormalisation scheme ambiguities~\cite{Catani1}.
\item
$\vv{RMASS(1,2,3,13)}$ are effective light quark and gluon masses used in
the  hadron\-ization phase  of the program.  They can be  set to zero
provided the parton shower cutoffs \vv{VQCUT} and \vv{VGCUT} are large
enough to prevent divergences (see below).
\item
For cluster hadronization, it must be possible to split gluons into
$q\qbar$, i.e.\ $\vv{RMASS(13)}$ must be at least twice the lightest quark
mass.  Similarly it may be impossible for heavy-flavoured clusters
to decay if $\vv{RMASS(4,5)}$ are too low.
\item
\vv{VQCUT} and \vv{VGCUT} are needed if the  quark and gluon effective
masses become small. The condition to avoid divergences in  parton showers is
$$ 
\frac{1}{Q_i} + \frac{1}{Q_j} < \frac{1}{\vv{QCDL3} }
$$
for either $i$ or $j$ or both gluons, where
$Q_i=\vv{RMASS}(i)+\vv{VQCUT}$ for quarks, $\vv{RMASS(13)}+\vv{VGCUT}$
for gluons, and \vv{QCDL3} is the three-flavour \QCD\ scale used
internally by \HW. \vv{QCDL3} is obtained by matching at the $b$- and
$c$-quark mass scales from the internal five-flavour scale
$$
\vv{QCDL5} = \vv{QCDLAM} \times \exp\left(\frac{151 - 9
\pi^2}{138}\right)/\sqrt{2} = 1.109 \times \vv{QCDLAM}\,.
$$
Note that, in the notation of ref.~\cite{Catani1} and
section~\ref{sec:showers}, $\vv{QCDL5}=\Lambda_{\rm phys}/\sqrt{2}$
for five flavours.
\item 
\vv{VPCUT} is the analogous quantity for photon emission. It now
defaults to 0.4\,GeV.  Previous versions defaulted
to $\sqrt s$, switching off such emission. Results after  experimental
cuts are insensitive to its exact value in the range 0.1 to 1.0\,GeV.
\item
\vv{CLMAX} and \vv{CLPOW} determine the maximum allowed mass of a
cluster made from quarks $i$ and $j$ as follows
$$
M^{\vv{CLPOW}} < \vv{CLMAX}^{\vv{CLPOW}} +
(\vv{RMASS}(i)+\vv{RMASS}(j))^{\vv{CLPOW}}\,.
$$
Since the cluster mass spectrum falls rapidly at high mass, results
become insensitive to \vv{CLMAX} and \vv{CLPOW} at large values of \vv{CLMAX}.
Smaller values of \vv{CLPOW} will increase the yield of heavier clusters
(and hence of baryons)  for heavy quarks,  without affecting  light
quarks  much.  For example,  the default  value gives  no $b$-baryons
whereas $\vv{CLPOW}=1.0$ makes $b$-baryons/$b$-hadrons about 1/4.
\item
\vv{PSPLT}  determines the  mass  distribution in the  cluster splitting
$C\!\l_1 \to C\!\l_2 + C\!\l_3$ when $C\!\l_1$ is above the maximum
allowed mass.
The masses of $C\!\l_2$ and $C\!\l_3$ are generated uniformly in
$M^{\vv{PSPLT}}$. As long as the number of split clusters is small,
dependence on \vv{PSPLT} is weak.\vspace{-2pt}
\item
\vv{QDIQK} greater than twice the lightest diquark mass enables
non-perturbative gluon splitting into diquarks as well as quarks. The
probability of this is $\vv{PDIQK}\times dQ/Q$ for scales $Q$ below
\vv{QDIQK}. The
diquark masses are taken to be the sum of constituent quark masses. Thus
the default value $\vv{QDIQK}=0$ suppresses gluon $\to$ diquark splitting.\vspace{-2pt}
\item
\vv{QSPAC} is the scale below which the structure functions of incoming
hadrons are frozen and non-valence constituent partons are forced to
evolve to valence partons, if $\vv{ISPAC}=0$.  For $\vv{ISPAC}=2$,
structure functions are frozen at scale \vv{QSPAC}, but evolution
continues down to the infrared cutoff.\vspace{-2pt}
\item
\vv{PTRMS} is the width of the (gaussian) intrinsic transverse momentum
distribution of valence partons in incoming hadrons at scale
\vv{QSPAC}.\vspace{-2pt}

\end{itemize}

In practice, the parameters that have been found most effective in
fitting data are \vv{QCDLAM}, the gluon effective mass $\vv{RMASS(13)}$,
and the cluster mass parameter \vv{CLMAX}.  Note that \vv{QSPAC},
\vv{PTRMS} and \vv{ENSOF} do not affect lepton-lepton collisions.

The default parameter values are based on those that were found to
give good agreement when comparing earlier versions with event shape
distributions at \LEP.  However, the substantial changes in this
version mean that a re-tuning of parameters would be very worthwhile.

Up-to-date details of \HW\ parameter tunes can be found via the
official web page cited in section~\ref{using}.

\vspace{-5pt}

\section{Control switches, constants and options}\label{control}
\vspace{-2pt}
A number of quantities can be reset to control the program and
various options:
\setlongtables
\setlength{\LTcapwidth}{\textwidth}
\begin{longtable}[t]{|@{\,}c@{\,}|@{\,}l@{\,}|@{\,}c@{\,}|}
\hline
Name & Description & Default \\
\hline
\endhead
\caption*{\small {\bf Table~\ref{tab.16}:} Control switches, constants and options. (Continues)}
\endfoot
\caption*{\small {\bf Table~\ref{tab.16}:} Control switches, constants and options.}
\endlastfoot
\label{tab.16}
\vv{NEVHEP}    & Current number of events                     & 0 \\
\vv{NHEP}      & Current number of entries in \vv{/HEPEVT/}    & 0 \\
\hline
\vv{IPRINT}    & Information to include in print out          & 1 \\
\vv{MAXPR}     & Number of events to print out                & 1 \\
\vv{PRVTX}     & Include vertex information in print out      & \vv{.TRUE.}\\
\vv{NPRFMT}    & Controls number of sig. figs. in  print out  & 1 \\
\vv{PRNDEC}    & Use decimal/hexadecimal in print out         & \vv{.TRUE.} \\
\vv{PRNDEF}    & Produce ASCII (stout) version of print out   & \vv{.TRUE.} \\
\vv{PRNTEX}    & Produce \LaTeX\ version of print out          & \vv{.FALSE.} \\
\vv{PRNWEB}    & Produce html version of print out            & \vv{.FALSE.} \\
\hline
\vv{MAXER}     & Maximum number of errors to tolerate         & 10 \\
\hline
\vv{LWEVT}     & Unit for writing output events               & 0 \\
\hline
\vv{LRSUD}     & Unit for reading Sudakov table               & 0 \\
\vv{LWSUD}     & Unit for writing Sudakov table               & 77 \\
\vv{SUDORD}    & $\as$ order in Sudakov table                 & 1 \\
\vv{INTER}     & Order of interpolation in Sudakov tables     & 3 \\
\hline
\vv{NRN(1)}    & Random number seed 1                         & 17673 \\
\vv{NRN(2)}    & Random number seed 2                         & 63565 \\
\vv{WGTMAX}    & Max. weight (0 to search for it)             & 0.0 \\
\vv{NOWGT}     & Generate unweighted events with \vv{EVWGT=AVWGT}& \vv{.TRUE.}\\
\vv{AVWGT}     & Mean event weight                            & 1.0 \\
\vv{EFFMIN}    & Min. acceptable Monte Carlo efficiency       & 0.001 \\
\vv{NEGWTS}   & Whether or not to allow negative weight events & \vv{.FALSE.} \\
\hline
\vv{AZSOFT}    & Include soft gluon azimuthal correlations    & \vv{.TRUE.} \\
\vv{AZSPIN}    & Include gluon spin azimuthal correlations    & \vv{.TRUE.} \\
\hline
\vv{HARDME}    & Use hard matrix-element corrections          & \vv{.TRUE.} \\
\vv{SOFTME}    & Use soft matrix-element corrections          & \vv{.TRUE.} \\
\vv{GCUTME}    & Gluon energy cut in top M.E.\ correction     & 2.0 \\
\hline
\vv{NCOLO}     & Number of colours                            & 3 \\
\vv{NFLAV}     & Number of (producible) flavours              & 6 \\
\hline
\vv{MODPDF(I)} & \PDF\ parton set and author group for beam   & $-1$ \\
\vv{AUTPDF(I)} & \vv{I} (=1,2) (if $\vv{MODPDF}<0$ do not use \PDF)
                                                             & \vv{'MRS'} \\
\vv{NSTRU}     & Input parton set (1,2 = Duke-Owens sets 1,2; &   \\
              & 3,4 = \sss{EHLQ } sets 1,2; 5 = Owens set 1.1, & 8 \\
              & 6,7,8 = \sss{MRST}, see table~\ref{tab:MRST}) & \\
\hline
\vv{PRSOF}    & Probability of soft underlying event          & 1.0 \\
\hline
\vv{ENSOF}    & Multiplicity enhancement for SUE:           &  \\ 
	     & \hfill $n=\langle n_{p\pbar}\rangle(\vv{ENSOF}\sqrt s)$	  & 1.0 \\
\vv{PMBN1}    & Mean multiplicity in SUE/Min.\ bias event     & +9.110 \\
\vv{PMBN2}    & $\langle n_{p\pbar}\rangle(\sqrt s)=
                \vv{PMBN1}s^{\vv{PMBN2}}+\vv{PMBN3}$            & +0.115 \\
\vv{PMBN3}    &                                               & $-9.500$ \\
\vv{PMBK1}    & \hbox{Negative binomial param.}               &      \\
             & \hfill  $k^{-1}=\vv{PMBK1}
               \log_e(s)+\vv{PMBK2}$                          & +0.029 \\
\vv{PMBK2}    &                                               & $-0.014$ \\
\hline
\vv{PMBM1}    & Soft cluster mass spectrum:                   &   \\
	     & \hfill $(M-M_1-M_2
               -\vv{PMBM1})e^{-\vv{PMBM2}M}$                   &  0.2 \\
\vv{PMBM2}    &                                               & 2.0 \\
\vv{PMBP1}    & Soft cluster $P_T$ spectrum:  
               $p_Te^{-\vv{PMBPi}
               \sqrt{p^2_T+M^2}}$,                            &   \\ 
             &  \hfill$d,u$ \hbox{quarks}                     & 5.2 \\ 
\vv{PMBP2}    &  \hfill $s,c$ quarks                           & 3.0 \\ 
\vv{PMBP3}    &  \hfill  diquarks                               & 5.2 \\
\hline
\vv{IOPREM}   & Options for treatment of remnant clusters     & 1 \\
\hline
\vv{BTCLM}    & Mass parameter in remnant fragmentation       & 1.0 \\
\hline
\vv{VMIN2}    & Min.\ parton virtuality$^2$ in distance calcs. & 0.1 \\
\hline
\vv{CLRECO}   & Include colour rearrangement                  & \vv{.FALSE.} \\
\vv{PRECO}    & Probability for rearrangement                 & 1/9 \\
\vv{EXAG}     & Lifetime scaling for weak bosons              & 1.0 \\
\hline
\vv{ETAMIX}   & $\eta/\eta'$ \hfill mixing angle in degrees   & $-23$ \\
\vv{PHIMIX}   & $\phi/\omega$ \hfill mixing angle in degrees  & +36 \\
\vv{H1MIX}    & $h_1(1380)/h_1(1170)$ \hfill mixing angle in degrees &
                                                      $\tan^{-1}(1/\sqrt 2)$ \\
\vv{F0MIX}    & $-/f_0(1370)$ \hfill mixing angle in degrees  &
                                                      $\tan^{-1}(1/\sqrt 2)$ \\
\vv{F1MIX}    & $f_1(1420)/f_1(1285)$ \hfill mixing angle in degrees & 
                                                      $\tan^{-1}(1/\sqrt 2)$ \\
\vv{F2MIX}    & $f'_2/f_2$ \hfill mixing angle in degrees     & +26 \\
\vv{ET2MIX}   & $\eta_2(1645)/\eta_2(1870)$\hfill  mixing angle in degrees &
                                                      $\tan^{-1}(1/\sqrt 2)$ \\
\vv{OMHMIX}   & $-/\omega(1600)$ \hfill mixing angle in degrees &
                                                      $\tan^{-1}(1/\sqrt 2)$ \\
\vv{PH3MIX}   & $\phi_3/\omega_3$ \hfill mixing angle in degrees & +28 \\
\hline
\vv{B1LIM}    & B cluster $\to$ 1 hadron parameter            &  0.0 \\
\vv{CLDIR(I)} & Decay orientation of perturbative clusters,   & 1,1\\
              & 0: isotropic, 1: along quark direction        &  \\
\vv{CLSMR(I)} & Width of gaussian angle smearing,             & 0.0,0.0 \\
              & (\vv{I}=1: light cluster, \vv{I}=2: heavy $b$-cluster) & \\
\hline
\vv{PWT(I)}   & A priori weights for $f\bar f$-pairs in cluster decay, & 1.0 \\
             & \vv{I}=1-6: $f=d,u,s,c,b,t $ \vv{I}=7: $f=qq'$ & \\
\vv{REPWT(L,J,N)}& A priori weight for $n^{(2S+1)}L_J$ mesons & 1.0 \\
\vv{SNGWT}    & A priori weight for singlet baryons           & 1.0 \\
\vv{DECWT}    & A priori weight for decuplet baryons          & 1.0 \\
\hline
\vv{PLTCUT}   & Minimum lifetime for particle to be set stable 
                                                        & $1.0\times10^{-8}$ \\
\hline
\vv{VTOCDK(I)}& Veto decay of clusters to hadron \vv{I}        & \vv{.FALSE.} \\
\vv{VTOCDK(I)}& Veto decay of resonances to hadron \vv{I}      & \vv{.FALSE.} \\
             & \vv{I}=290-293, $f_0(980),a_0(980)$            & \vv{.TRUE.} \\
\hline
\vv{PIPSMR}   & Smear the primary vertex                      & \vv{.FALSE.} \\
\vv{VIPWID(1)}& $x$ width (mm)                                & 0.25 \\
\vv{VIPWID(2)}& $y$ width (mm)                                & 0.015 \\
\vv{VIPWID(3)}& $z$ width (mm)                                & 1.8 \\
\hline
\vv{MAXDKL}   & Veto decays outside given volume              & \vv{.FALSE.} \\
\vv{IOPDKL}   & Option for volume: 1=cylinder, 2=sphere       & 1 \\
\vv{DXRCYL}   & Radius for cylindrical option (mm)            & 20 \\
\vv{DXZMAX}   & Length for cylindrical option (mm)            & 500 \\
\vv{DXRSPH}   & Radius for spherical option (mm)              & 100 \\
\hline
\vv{BDECAY}   & Controls which B Decay package is used.    & \vv{'HERW'} \\
              & Allowed values are: \vv{'HERW'}, \vv{'EURO'} or \vv{'CLEO'} & \\
\vv{MIXING}   & Include neutral B meson mixing                & \vv{.TRUE.} \\
\vv{XMIX(1)}  & $\Delta M/\Gamma$ for $B^0_s$                 & 10.0  \\
\vv{XMIX(2)}  & $\Delta M/\Gamma$ for $B^0_d$                 & 0.7  \\
\vv{YMIX(1)}  & $\Delta\Gamma/2\Gamma$ for $B^0_s$            & 0.2  \\
\vv{YMIX(2)}  & $\Delta\Gamma/2\Gamma$ for $B^0_d$            & 0.0  \\
\hline
\vv{RMASS(198)}& $W^+$ mass                                   & 80.42 \\
\vv{RMASS(199)}& $W^-$ mass                                   & \vv{RMASS(198)}\\
\vv{GAMW}      & $W^\pm$ width                                & 2.12 \\
\vv{RMASS(200)}& $Z^0$ mass                                   & 91.188 \\
\vv{GAMZ}      & $Z^0$ width                                  & 2.495 \\
\vv{WZRFR}     & Use $W/Z$ rest frame for decay parton showers & \vv{.TRUE.} \\
\vv{MODBOS(I)} & Force decay modes for weak bosons,           & \\
              &    see sect.~\ref{sec:gaugebosons}            & 0 \\
\hline
\vv{RMASS(201)}& SM Higgs mass                                & 115 \\
\vv{IOPHIG}    & Options for large Higgs mass distribution    & 3 \\
\vv{GAMMAX}    & Limit on range of Higgs mass distribution    & 10. \\
\vv{ENHANC(I)} & Enhancement factor for SM Higgs decay  mode \vv{I} & 1.0 \\
\hline
\vv{RMASS(209)}& Hypothetical 4th generation `bottom' quark mass & 200. \\
\vv{RMASS(215)}& corresponding    antiquark mass      & \vv{RMASS(209)}\\
\hline
\vv{ALPHEM}   & Thompson limit value of $\alpha_{em}(0)$      & 0.0072993 \\
\vv{SWEIN}    & Value of $\sin^2\theta_W$                     & 0.2319 \\
\vv{QFCH(I)}  & Fermion electric charge \hfill \vv{I}=1-6: $d,..,t$ & \\
\vv{AFCH(I,J)}& Fermion weak axial charge \hfill \vv{I}=10-16: $e,..,\nu_\tau$ &
                                                    see sect.~\ref{couples} \\
\vv{VFCH(I,J)}& Fermion weak vector charge \hfill \vv{J}=1: $Z$, \vv{J}=2: $Z'$&\\
\vv{ZPRIME}   & Include a $Z'$ in $\gamma^\star/Z^0$ processes& \vv{.FALSE} \\
\vv{RMASS(202)}& Mass of the $Z'$                             & 500. \\
\vv{GAMZP}    & Width of the $Z'$                             & 5.0 \\
\vv{VCKM(I,J)}& Cabibbo-Kobayashi-Maskawa matrix elements: &
\raisebox{-6pt}[0pt][0pt]{\parbox[t]{3.5cm}
{\centering\scriptsize$\left(\begin{array}{lll}
0.9512&0.0488&0\\0.0488&0.9492&0.002\\0&0.002&0.998\end{array}\right)$}} \\
             & $V^2_{KL}$,\vv{K}=1-3: $u,c,t\>\>\>$ \vv{L}=1-3: $d,s,b$   & \\
\vv{SCABI}    & Value of $\sin^2\theta_{\mbox{\scriptsize Cabibbo}}$ & 0.0488 \\
\hline
\vv{EPOLN(1)} & Electron and positron beam                    & 0.0  \\
\vv{EPOLN(2)} & polarizations in \DIS\ and $\ee$              & 0.0  \\
\vv{EPOLN(3)} & annihilation. First two cmpts are             & 0.0  \\
\vv{PPOLN(1)} & transverse and only used in $\ee$,            & 0.0  \\
\vv{PPOLN(2)} & 3rd cmpt is longitudinal, and is              & 0.0  \\
\vv{PPOLN(3)} & $+/-1$ for fully rh/lh polarized                & 0.0  \\
\hline
\vv{QLIM}     & Upper limit on hard process scale             & $10^8$ \\
\hline
\vv{THMAX}    & Max.\ value of thrust in \vv{IPROC}=110--116   & 0.9 \\
\vv{Y4JT}     & Min.\ jet separation in \vv{IPROC}=600--656    & 0.01 \\
\vv{DURHAM}   & Use Durham/JADE algorithm in \vv{IPROC}=600--656 &\vv{.TRUE.} \\
             & Colour interferences in \vv{IPROC}=600--656:                  & \\
\vv{IOP4JT(1)}& $q\qbar gg$ 0: neglect, 1: extreme 3142, 2: extreme 4123  & 0 \\
\vv{IOP4JT(2)}& $q\qbar q\qbar$ 0: neglect, 1: extreme 4123, 2: extreme 2143    & 0 \\
\hline
\vv{BGSHAT}   & Boson-gluon fusion scale (see below)          & \vv{.TRUE.} \\
\hline
\vv{BREIT}    & Use Breit frame for \DIS\ kinematics          & \vv{.TRUE.} \\
\vv{USECMF}   & Use hadron-hadron c.m.\frame                  & \vv{.TRUE.} \\
\hline
\vv{NOSPAC}   & Switch off spacelike showers                 & \vv{.FALSE.} \\
\vv{ISPAC}    & Changes meaning of \vv{QSPAC}                  & 0 \\
             & (see the earlier notes on \vv{QSPAC})          & \\
\hline
\vv{TMNISR}   & Min.\ value of $\shat/S$ for photon ISR        & $10^{-4}$ \\
\vv{ZMXISR}   & Max.\ momentum fraction for photon ISR         & $1-10^{-6}$ \\
\hline
\vv{ASFIXD}   & Values of fixed $\alpha_s$ and $\omega=12\log_e(2)\alpha_s/\pi$
                                                              & 0.25 \\
\vv{OMEGA}    & for Mueller-Tang cross section                 & 0.3 \\
\hline
\vv{IAPHIG}   & Approx.\ used in Higgs+jet production 
                                                              & 1 \\
             & \vv{IPROC}=2300-2312                            & \\
\hline
\vv{PHOMAS}   & Damp structure functions for off mass-shell    & 0.0 \\
             & photons (0 for no damping)                     &  \\
\hline
\vv{PRESPL} & Preserve longitudinal momentum of hard c.m. & \vv{.TRUE.} \\
\hline
\vv{PTMIN}    & Min.\ $p_T$ in hadronic jet production         & 10.0  \\
\vv{PTMAX}    & Max.\ $p_T$ in hadronic jet production         & $10^8$\\
\vv{PTPOW}    & $1/p_T^{\vv{PTPOW}}$ for jet sampling           & 4.0   \\
\vv{YJMIN}    & Min.\ jet rapidity                             & $-8.0$ \\
\vv{YJMAX}    & Max.\ jet rapidity                             & +8.0   \\
\hline
\vv{EMMIN}    & Min.\ dilepton mass in Drell-Yan               & 10.0  \\
\vv{EMMAX}    & Max.\ dilepton mass in Drell-Yan               & $10^8$\\
\vv{EMPOW}    & $1/m^{\vv{EMPOW}}$ for Drell-Yan sampling       & 4.0   \\
\hline

\vv{Q2MIN}    & Min.\ $Q^2$ in deep inelastic scattering       & 0 \\
\vv{Q2MAX}    & Max.\ $Q^2$ in deep inelastic scattering       & $10^{10}$ \\
\vv{Q2POW}    & $1/Q^{2\vv{Q2POW}}$ for \DIS\ sampling          & 2.5   \\
\hline
\vv{YBMIN}    & Min.\ and max.\ Bjorken-$y$ in \DIS            & 0.0 \\
\vv{YBMAX}    &                                                & 1.0 \\
\hline
\vv{WHMIN}    & Min.\ hadronic mass in                          & 0.0 \\
             &   $\gamma$-induced  processes (inc.\ \DIS)     &   \\
\hline
\vv{ZJMAX}    & Max.\ $z$ in $J/\psi$ production               & 0.9 \\
\hline
\vv{Q2WWMN}   & Min.\ and max.\ $Q^2$ in                       & 0.0  \\
\vv{Q2WWMX}   & Equivalent Photon Approximation                & 4.0  \\
\vv{YWWMIN}   & Min.\ and max.\ photon light-cone fraction     & 0.0  \\
\vv{YWWMAX}   & in Equiv.\ Photon Approx.\                     & 1.0  \\
\hline
\vv{CSPEED}   & Speed of light in vacuum (mm/s)                  &
                                                      $2.99792\times10^{11}$ \\
\vv{GEV2NB}   & Value of $(\hbar c/e)^2$                         & 389 379 \\
\hline
\vv{IBSH}     & Number of shots for initial max.\ weight search  & 10 000 \\
\vv{IBRN(1)}  & 1st random number seed for  max.\ weight search & 1246579 \\
\vv{IBRN(2)}  & 2nd random number seed for  max.\ weight search & 8447766 \\
\hline
\vv{NQEV}     & Number of entries in Sudakov FF        & \\
             & look-up table    & 1024 \\
\vv{ZBINM}    & Max.\ bin size for $z$ in spacelike branching   & 0.05 \\
\vv{NZBIN}    & Max.\ number of $z$ bins in spacelike branching & 100 \\
\hline
\vv{NBTRY}    & Max.\ number of attempts to branch a parton      & 200 \\
\vv{NCTRY}    & Max.\ number of attempts to decay a cluster      & 200 \\
\vv{NETRY}    & Max.\ number of attempts to generate a mass & 200 \\
\vv{NSTRY}    & Max.\ number of attempts at soft subprocess      & 200 \\
\hline
\vv{ACCUR}    & Precision for soft gaussian integration          & $10^{-6}$ \\ 
\hline
\vv{RPARTY}   &  R-parity conservation in \SY                  & \vv{.TRUE.} \\
\hline
\vv{SUSYIN}   & Check to see if SUSY data are already loaded   & \vv{.FALSE.}\\
\vv{LRSUSY}   & Unit for reading SUSY data (if needed)         & 66 \\
\hline 
\vv{SYSPIN}   & Spin correlations in decays & \vv{.TRUE.} \\
\vv{THREEB}   & SUSY three body decays      & \vv{.TRUE.} \\
\vv{FOURB}    & SUSY four body decays       & \vv{.FALSE.}\\
\hline
\vv{TAUDEC}   & Tau decay package (\HW\ or \TA)                  & \vv{HERWIG}\\
\hline
\vv{LHSOFT}   & Generation of soft event for Les Houches interface & \vv{.TRUE.}\\
\vv{LHGLSF}   & Self-connected gluons for Les Houches interface & \vv{.FALSE.}\\
\hline
\vv{OPTM}     & Optimisation of phase space & \vv{.FALSE.}\\
\vv{IOPSTP}   & Number of steps for phase space optimisation&10\\
\vv{IOPSH}    & Number of weights for phase space optimisation&1000\\
\hline
\end{longtable}

Printout options are listed in table~\ref{tab.17}.

{\renewcommand{\belowcaptionskip}{-12pt}
\TABLE[h]{\centerline{\begin{tabular}{|rl|}\hline
$\vv{IPRINT} = 0$& Print program title only                   \\
               1 & Print selected input parameters            \\
               2 & 1 + table of particle codes and properties \\
               3 & 2 + tables of Sudakov form factors         \\\hline
	  \end{tabular}}%
\caption{Printout options.\label{tab.17}}}}

The contents of \vv{/HEPEVT/} can by printed by calling \vv{HWUEPR},
those of \vv{/HWPART/} (the last parton shower) by calling \vv{HWUBPR}.
The logical variable \vv{PRNDEC} (default \vv{.TRUE.} unless
$\vv{NMXHEP}>9999$) causes track numbers in event listings to be
printed in decimal, or hexadecimal if false.  The latter is necessary
for very large events such as those generated by the \vv{HERBVI}
package (see above).

The maximum number of errors \vv{MAXER} refers to errors from which the
program cannot recover without killing an event and starting a new
one.  Such errors are not necessarily a cause for grave concern
because the phase space for backward evolution of initial-state
showers is complicated and the program may occasionally step outside
it (in which case the event weight should be zero anyway).  When
generating large numbers of events, it is advisable to increase
\vv{MAXER} in proportion, e.g.\ to \vv{MAXEV}/100.

See section~\ref{form} on form factors for details of \vv{LRSUD},
\vv{LWSUD} and \vv{SUDORD}.

The parameter \vv{EFFMIN} sets the minimum allowed efficiency for the
generation of unweighted events. A warning is printed once in every
10/\vv{EFFMIN} weights if the efficiency is below
10$\times$\vv{EFFMIN}, and running is stopped if the efficiency is
below \vv{EFFMIN}. See sect.~\ref{sect:negwgts} for details of \vv{NEGWTS}.

Variables \vv{HARDME} and \vv{SOFTME} invoke hard and soft
matrix-element corrections respectively, as described in subsection
\ref{mecorr}.

If \vv{BGSHAT} is \vv{.TRUE.}, the scale used for heavy quark production
via boson-gluon fusion in lepton-hadron collisions will be the hard
subprocess c.m.\ energy $\shat$. If it is \vv{.FALSE.}, the scale used
will be
$$
     {2\:\shat\:\that\:\uhat\over\shat^2+\that^2+\uhat^2}\,,
$$
except in the case of $J/\psi+g$ production, where $\uhat$ is used.

If \vv{BREIT} is true, the kinematic reconstruction of deep inelastic
events takes place in the Breit frame (i.e.\ the frame where the
exchanged boson is purely spacelike, and collinear with the incoming
hadron).  In fact the reconstruction procedure is invariant under
longitudinal boosts, so any frame in which the boson and hadron are
collinear would be equivalent, and it is only the transverse part of
the boost that has an effect.  The \vv{BREIT} frame option becomes very
inaccurate for very small $Q^2$.  It is therefore only used if $Q^2 >
10^{-4}$ (the lab and Breit frames are anyway equivalent for such
small $Q^2$).  If \vv{BREIT} is false, reconstruction takes place in
the lab frame.

If \vv{USECMF} is true, the entire event record is boosted to the
hadron-hadron c.m.\ frame before event processing, and boosted back
afterwards.  This means that fixed-target simulation can be done in
the lab frame, i.e.\  with $\vv{PBEAM2}=0$.  For hadronic processes with
lepton beams, this boosting is always done, regardless of the value of
\vv{USECMF}.

    In version 6.5,
    a new logical input variable, \vv{PRESPL} [\vv{.TRUE.}],  has  been
    introduced to control whether  the  longitudinal  momentum 
    (\vv{PRESPL} = \vv{.TRUE.}),  or  rapidity (\vv{PRESPL} = \vv{.FALSE.}),
    of the hard process centre-of-mass is preserved in
    hadron collisions after initial-state parton showering.   At present
    the only function of this variable is to allow users to study the
    effects of momentum reshuffling,  which is necessary after showering
    to compensate for jet masses. In future, it is anticipated that setting
    \vv{PRESPL}=\vv{.FALSE.} will simplify the treatment of other processes
    in \MN\ (see sect.~\ref{sec:MCNLO}).

The quantities from \vv{PTMIN} to \vv{ZJMAX} control the region of phase
space in which events are generated and importance sampling inside
those regions.  See section~\ref{weights} on event weights for further
details on these quantities and the use of \vv{WGTMAX} and \vv{NOWGT}.

If hadronic processes with lepton beams are requested, the photon
emission vertex includes the full transverse-momentum-dependent
kinematics (the Equivalent Photon Approximation).  The variables
\vv{Q2WWMN} and \vv{Q2WWMX} set the minimum and maximum virtualities
generated respectively. For normal simulation, \vv{Q2WWMN} should be
zero, and \vv{Q2WWMX} should be the largest $Q^2$ through which the
lepton can be scattered without being detected.  The variables
\vv{YWWMIN} and \vv{YWWMAX} control the range of lightcone momentum
fraction generated.

In addition there are options to give different weights to the various
flavours of quarks and diquarks, and to resonances of different
spins. So far, these options have not been used. See the comments in
the initialisation routine \vv{HWIGIN} for details.

\subsection{Negative weights option}
\label{sect:negwgts}
In a number of new applications such as \MN\ (see sect.~\ref{sec:MCNLO}),
parton configurations with negative weight are used to produce more
accurate predictions, and therefore the possibility of negative event
weights has to be considered.  In general,
a Monte Carlo program generates $N_w$ weights $\{w_i\}$ such that
the estimated cross section is
\begin{equation}
\sigma =  \frac{1}{N_w}\sum_{i=1}^{N_w} w_i \equiv \overline w\;.
\end{equation}
The corresponding error depends on the width of the weight distribution:
\begin{equation}
\frac{\delta\sigma}{\sigma} = \frac{1}{\sqrt{N_w}}
\frac{\delta w_{\mbox{\scriptsize rms}}}{\overline w}\;.
\end{equation}

If only {\it positive weights} are generated, and there exists a
maximum weight $w_{\mbox{\scriptsize max}}$, then
{\it unweighted} events can be generated by `hit and miss':
$w'_i = 0$ or $1$ with probability
$P(w'_i = 1) = w_i/w_{\mbox{\scriptsize max}}$.
Then
\begin{equation}
\frac{\delta\sigma}{\sigma} = \sqrt{
\frac{w_{\mbox{\scriptsize max}}-\overline w}{N_w\overline w}}
= \sqrt{
\frac{w_{\mbox{\scriptsize max}}-\overline w}
{N_e w_{\mbox{\scriptsize max}}}}\sim\frac{1}{\sqrt{N_e}}
\end{equation}
where $N_e=N_w \overline w/w_{\mbox{\scriptsize max}}$ is the number of
{\it events} generated. The time needed (especially for detector simulation)
depends mainly on the number of events. Hence the inefficiency of
`hit and miss' is not necessarily a disaster.  This is the usual approach
adopted in Monte Carlo event generators.

{\it Negative weights} can be generated by subtraction procedures for
matrix element corrections. These are not a problem of principle but
prevent naive `hit and miss'.
To generalize `hit and miss', one can generate unweighted
events ($ w'_i = 1$) and `antievents' ($ w'_i = -1$)
with
\begin{eqnarray}\label{sign_w}
\mbox{sign}(w'_i) &=&\mbox{sign}(w_i)\;,\\ \label{prob_w}
P(w'_i = \pm 1) & =&|w_i|/|w|_{\mbox{\scriptsize max}}\;.
\end{eqnarray}
Then
\begin{equation}
\frac{\delta\sigma}{\sigma} = \sqrt{
\frac{\overline{|w|}|w|_{\mbox{\scriptsize max}}
-{\overline w}^2}{N_w{\overline w}^2}}
\sim\frac{1}{\sqrt{N_e}}\frac{\overline{|w|}}{\overline w}
\end{equation}
where $ N_e=N_w \overline{|w|}/|w|_{\mbox{\scriptsize max}}$ is
now the total number of {\it events}+{\it antievents} generated.
Again, the time needed is almost proportional to $N_e$, so this
is tolerable as long as $\overline{|w|}\sim\overline w$.
The cross section after any cuts that may be applied is
\begin{equation}\label{sigma_c}
\sigma_c = \frac{\overline{|w|}}{N_e}(N_+ - N_-)
\end{equation}
where $ N_+$ events and $ N_-$ antievents pass the cuts.

To allow for the possibility of negative weights, a new
logical parameter \vv{NEGWTS} has been introduced. 
The default (\vv{NEGWTS}=\vv{.FALSE.}) is as before: negative weights are
forbidden. If one is detected, a non-fatal warning is issued and the
event weight is set to zero.

If \vv{NEGWTS}=\vv{.TRUE.}, negative weights are allowed. Statistics are
computed and printed accordingly. If unweighted events are requested
(\vv{NOWGT}=\vv{.TRUE.}), the initial search stores the maximum and mean
absolute weights, $|w|_{\mbox{\scriptsize max}}$ and $\overline{|w|}$.
Events and antievents are selected according to
Eqs.~(\ref{sign_w},\ref{prob_w}) and \vv{EVWGT} is reset to
$\overline{|w|}\mbox{sign}(w_i)$, so that the numerator in 
Eq.~(\ref{sigma_c}) is the sum of \vv{EVWGT}s for contributing
(anti)events.

\section{Particle data}\label{pdata}

From \HW\ version~5.9 onwards, new 8-character particle names have
been introduced and the revised 7 digit \PD\ numbering scheme, as
advocated in the LEP2 report~\cite{LEP2}, has been adopted. All hadron
and lepton masses are given to five significant figures whenever
possible.

Unstable hadrons from clusters produced in both the hard and soft
components of the event decay according to simplified decay schemes,
which can be tabulated by specifying the print option $\vv{IPRINT}=2$.
Decays modes are `invented' where necessary to make the branching
ratios add up to 100\%. Phase space distributions are assumed except
where stated otherwise. See section~\ref{heavy} for the treatment of
heavy quark decays. After a $t$, partonic $b$ or quarkonium decay,
secondary parton showers are produced by outgoing partons as discussed
in ref.~\cite{March4}; these are hadronized in the same way as primary
jets.

There have been a number of additions/changes to the default hadrons
included via \vv{HWUDAT}.  Here the identification of hadrons follows
the \PD\ ~\cite[table 13.2]{PDG}, numbered according to their
section~30.

All S and P wave mesons are present including the $^1$P$_0$ and
$^3$P$_1$ states and many new, excited B$^{**}$, B$_c$ and quarkonium
states. Also all D wave kaons and some `light' I=3 states [$\pi_2,
\rho(1700)$ and $\rho_3$].  All the baryons (singlet/octet/decuplet)
containing up to one heavy ($c,b$) quark are included.

New isoscalars states have been added to try to complete the $1^3D_3$,
$1^1D_2$ and $1^3D_1$ multiplets (table~\ref{tab.18}).

\TABULAR[t]{|rlr|rlr|}{
\hline
\vv{IDHW} & \vv{RNAME}  &  \vv{IDPDG} &
\vv{IDHW} & \vv{RNAME}  &  \vv{IDPDG} \\
\hline
395 & \vv{OMEGA\_3}  &     227 &   396  & \vv{PHI\_3}    &   337 \\
397 & \vv{ETA\_2(L)} &   10225 &   398  & \vv{ETA\_2(H)} & 10335 \\
399 & \vv{OMEGA(H)}  &   30223 &        &               & \\
\hline
}{New isoscalar states.\label{tab.18}}

{\renewcommand\belowcaptionskip{-1em}
\TABULAR[t]{|rlr|rlr|}{
\hline
\vv{IDHW} & \vv{RNAME}  &  \vv{IDPDG} &
\vv{IDHW} & \vv{RNAME}  &  \vv{IDPDG} \\
\hline
57 & \vv{FH\_1}    &   20333  &      &            &          \\
293 & \vv{F0P0}    & 9010221  &  294 & \vv{FH\_00}  &    10221 \\
62 & \vv{A\_0(H)0} &   10111  &  290 & \vv{A\_00}  &  9000111 \\
63 & \vv{A\_0(H)+} &   10211  &  291 & \vv{A\_0+}  &  9000211 \\
64 & \vv{A\_0(H)-} & $-10211$  &  292 & \vv{A\_0-}  &$-9000211$ \\
\hline
}{Re-identified/replaced states.\label{tab.19}}}

\TABULAR{|r|l|}{\hline
\vv{ETAMIX}& $\eta/\eta'$,\\
\vv{PHIMIX}& $\omega/\phi$, \\
\vv{H1MIX}&  $h_1(1170)/h_1(1380)$,\\
\vv{F0MIX}&  $f_0(1300)/f_0(980)$,\\
\vv{F1MIX}&  $f_1(1285)/f_1(1510)$,\\
\vv{F2MIX}&  $f_2/f_2'$.\\\hline}{Mixing angles.\label{tab.20}}

Also the states in table~\ref{tab.19} have been redefined.
The $f_1(1420)$ state completely replaces the $f_1(1520)$ in the
$1^3P_0$ multiplet, taking over 57. The $f_0(1370)$ (294) replaces the
$f_0(980)$ (293) in the $1^3P_0$ multiplet; the latter is retained as
it appears in the decays of several other states.  The new $a_0(1450)$
states (62--64) replace the three old $a_0(980)$ states (290--292)
in the $1^3P_0$ multiplet; the latter are kept as they appear in
$f_1(1285)$ decays.

By default production of the $f_0(980)$ and $a_0(980)$ states in
cluster decays is vetoed.

The mixing angles (in degrees) of all the light, I=0 mesons can now be
set using table~\ref{tab.20}.

There were previously some inconsistencies and ambiguities in our
conventions for the mixing of flavour `octet' and `singlet' mesons.
They are now as in table~\ref{tab.21}.

\TABULAR{|cccc|}{
\hline
Multiplet   &  Octet     &   Singlet    &     Mixing Angle \\
\hline
$1^1$S$_0$  & $\eta$     & $\eta'$      &  \vv{ETAMIX}=$-23$. \\
$1^3$S$_1$  & $\phi$     & $\omega$     &  \vv{PHIMIX}=+36. \\
$1^1$P$_1$  & $h_1$(1380)& $h_1$(1170)  &  \vv{H1MIX}=\vv{ANGLE} \\
$1^3$P$_0$  & missing    & $f_0$(1370)  &  \vv{F0MIX}=\vv{ANGLE} \\
$1^3$P$_1$  & $f_1$(1420)& $f_1$(1285)  &  \vv{F1MIX}=\vv{ANGLE} \\
$1^3$P$_2$  & $f'_2$     & $f_2$        &  \vv{F2MIX}=+26. \\
$1^1$D$_2$  & $\eta_2$(1645)& $\eta_2$(1870) & \vv{ET2MIX}=\vv{ANGLE} \\
$1^3$D$_1$  &   missing  & $\omega$(1600)& \vv{OMHMIX}=\vv{ANGLE} \\
$1^3$D$_3$  & $\phi_3$    & $\omega_3$   &  \vv{PH3MIX}=+28. \\
\hline
}{Mixed meson states.\label{tab.21}}

After mixing, the quark content of the physical states is given in
terms of the mixing angle, $\theta$, by table~\ref{tab.22} where
$\tan\theta_0=\sqrt 2$.

\TABULAR{|ccc|}{
\hline
State & $(d\dbar+u\ubar)/\sqrt 2$ & $s\sbar$ \\
\hline
Octet & $\cos(\theta+\theta_0)$     & $-\sin(\theta+\theta_0)$ \\
Singlet & $\sin(\theta+\theta_0)$     & $\cos(\theta+\theta_0)$ \\
\hline
}{Quark content of mixed states.\label{tab.22}}

{\sloppy
Hence, using the default value of $\vv{ANGLE}=\arctan(1/\sqrt 2)=+35.3$
for $\theta$ gives ideal mixing, that is, the `octet' state$=s\sbar$
and the `singlet'$=(d\dbar+u\ubar)/\sqrt 2$.  This choice is important
to avoid large isospin violations in the $1^3P_0$ and $1^3D_1$
multiplets in which the octet member is unknown.

}

Since version~6 contains a large number of supersymmetry processes
many new hypothetical particles have been added --- see
section~\ref{sec:SUSY}.  These new states do not interfere with the
user's ability to add new particles as described below.
\pagebreak[3]

The resonance decay tables supplied in the program have also been
largely revised. Measured/expected modes with branching fraction at or
above 1 per mille are given, including 4 and 5 body decays. To print
the new tables call \vv{HWUDPR}.

The layout of \vv{HWUDAT} has been altered to make it easier to
identify and modify particle properties. Three new arrays have been
introduced \vv{RLTIM}, \vv{RSPIN} and \vv{IFLAV}. These are: the
particle's lifetime (s), spin, and a code which specifies the flavour
content of each hadron --- used (in \vv{HWURES}) to create sets of
iso-flavour hadrons for cluster decay.  Using the standard numbering
of quark flavours the convention is:

\begin{itemize}
\item Mesons: $n_q n_{\qbar}$, e.g.\ $\pi^+$:  21, $\pi^-$: 12;
\item Baryons: $\pm n_{q1} n_{q2} n_{q3}$, e.g.\ $\Xi^0$: 332, $\bar\Xi^0$:
 $-332$  etc.\ ($<0$ for antibaryons; digits in decreasing order);
\item Light, neutral mesons are identified as: 11 if isovector
($\pi^0,\rho^0,\dots$), 33 if isoscalar ($\eta,\eta',\dots$).
\end{itemize}

Some parts of the program have been automated so that it is possible
for the user to add new particles by specifying their properties via
the arrays in \vv{/HWPROP/} and \vv{/HWUNAM/} and increasing \vv{NRES}
appropriately: this should be done before a call to \vv{HWUINC}.

As an example, the following lines add an isoscalar, spin-2 state
\vv{'STAN'} and a (very light) stable toponium state \vv{'BEER'} with
the decay mode: $\vv{STAN}\to\vv{BEER}+\vv{BEER}+\vv{BEER}$.
\small\begin{verbatim}
      NRES=NRES+1
      RNAME(NRES)='STAN    '
      IDPDG(NRES)=666
      IFLAV(NRES)=11
      ICHRG(NRES)=0.
      RMASS(NRES)=0.5
      RLTIM(NRES)=1.000D-10
      RSPIN(NRES)=2.0
      NRES=NRES+1
      RNAME(NRES)='BEER    '
      IDPDG(NRES)=66
      IFLAV(NRES)=66
      ICHRG(NRES)=0.
      RMASS(NRES)=0.1
      RLTIM(NRES)=1.000D+30
      RSPIN(NRES)=0.0
      CALL HWMODK(666,1.D0,0,66,66,66,0,0)
\end{verbatim}\normalsize

Using the logical arrays \vv{VTOCDK} and \vv{VTORDK} the production of
specified particles can be stopped in both cluster decays and via the
decay of other unstable resonances.

A priori weights for the relative production rates in cluster decays
of mesons and baryons differing only via their S and L quantum numbers
can be supplied using \vv{SNGWT} and \vv{DECWT} for singlet
(i.e.\ $\Lambda$-like) and decuplet baryons and \vv{REPWT} for
mesons. The old \vv{VECWT} now corresponds to \vv{REPWT}(0,1,0) and
\vv{TENWT} to \vv{REPWT}(0,2,0).

The arrays \vv{FBTM}, \vv{FTOP} and \vv{FHVY} which stored the branching
fractions of the bottom, top and heavier quarks' `partonic' decays are
now no longer used.  Such decays are specified in the same way as all
other decay modes: this permits different decays to be given to
individual heavy hadrons.  Partonic decays of charm hadrons and
quarkonium states are also now supported.  As already mentioned, the
products' order in a partonic decay mode is significant: see
discussion in section~\ref{heavy}.

The structure of the program has been altered so that the secondary
hard subprocess and subsequent fragmentation associated with each
partonic heavy hadron decay appears separately. Thus pre-hadronization
top quark decays are treated individually, as are any subsequent
bottom hadron partonic decays.

Additionally decays of heavy hadrons to exclusive non-partonic final
states are supported. No check against double counting from partonic
modes is included. However this is not expected to be a major problem
for the semi-leptonic and 2-body hadronic modes supplied.

B decays can also be performed by the {\sf EURODEC} or {\sf CLEO} Monte
Carlo packages.  The new variable \vv{BDECAY} controls which package is
used: \vv{'HERW'} for \HW; \vv{'EURO'} for {\sf EURODEC}; \vv{'CLEO'} for
{\sf CLEO}. The {\sf EURODEC} package can be obtained from the \CERN\
library. The {\sf CLEO} package is available by kind permission of the
{\sf CLEO} collaboration.

An array \vv{NME} has been introduced to enable a possible matrix element
to be specified for each decay mode.
\begin{itemize}
\item $\vv{NME} = 0$:   Isotropic decay.
\item $\vv{NME} = 100$: Free particle $(V-A)*(V-A)$, $(p_0\cdot p_2)(p_1\cdot p_3)$.
\item $\vv{NME} = 101$: Bound quark $(V-A)*(V-A)$,   $(p_0\cdot p_2)(p_1\cdot
[p_3 - x_s p_0])$, $x_s = m_Q/M_0 =$ spectator quark momentum fraction.
\item $\vv{NME} = 130$: Ore and Powell ortho-positronium matrix element for:
onium$\to gg+g/\gamma$.
\item $\vv{NME} = 200$: Free-particle $t\to b$ quark decay
through a scalar-fermion-fermion current.
\item \vv{NME} = 300: Gaugino and gluino three-body \RPV\ decays.
This also implements the angular ordering procedure in the \RPV\ gluino decays.
\end{itemize}

The list of matrix elements currently supported is modest; users are
urged to contact an author to have others implemented. It should be noted however
that a number of additional matrix elements are now available via the spin
correlation algorithm.

A $Z'$ has been introduced with \PD\ code 32, \HW\ identifier 202,
default mass 500\,GeV, width \vv{GAMZP} (default 5\,GeV) and name
\vv{'Z0PR '}.  It is invoked by setting \vv{ZPRIME=.TRUE.} (default
\vv{.FALSE.}).

The decay tables can be written to/read from a file by using
\vv{HWIODK}, adopting the format advocated in the LEP2
report~\cite{LEP2}.  In addition to the \PD\ numbering of particles
the \HW\ numbers or character names can be used. This permits easy
alteration of the decay tables. In \vv{HWUINC} a call is made to
\vv{HWUDKS} which sets up \HW\ internal pointers and performs some
basic checks of the decay tables. Each decay mode must conserve charge
and be kinematically allowed and not contain vetoed decay
products. The sum of all branching ratios is set to 1 for all
particles.  Also a warning is printed if an antiparticle does not have
all the charge conjugate decays modes of the particle.

\vv{HWMODK} enables changes to the decay tables to be made by
altering or adding single decay modes including on an event-by-event
basis. This can be done before calling \vv{HWUINC}, in which case when
altering the branching ratio and/or matrix element code of an existing
mode a warning is given of a duplicate second mode which superseeds
the first. Branching ratios set below 10$^{-6}$ are eliminated, whilst
if one mode is within 10$^{-6}$ of unity all other modes are removed.
Note that some forethought is required if the branching ratios of two
modes of the same particle are changed since the operation of
rescaling the branching ratio sum to unity causes a non-commutativity
in the order of the calls.

It is possible to create particle property and event listings in any
combination of 3 formats~--- standard ASCII, \LaTeX\ or html.  These
options are controlled by the logical variables \vv{PRNDEF} (default
\vv{.TRUE.})  \vv{PRNTEX} (default \vv{.FALSE.}) and \vv{PRNWEB} (default
\vv{.FALSE.}).  The ASCII output is directed to stout (screen/log file)
as in previous versions. When a listing of particle properties is
requested (\vv{IPRINT.GE.2} or \vv{HWUDPR} is called explicitly) then
the following files are produced:
\begin{tabbing}
That is,m\=If (\vv{PRNTEX}): \=  \vv{HW\_decays.tex} \kill
         \>If (\vv{PRNTEX}): \>  \vv{HW\_decays.tex} \\
         \>If (\vv{PRNWEB}): \>  \vv{HW\_decays/index.html} \\
         \>                 \>   \vv{/PART0000001.html} etc.
\end{tabbing}
The \vv{HW\_decays.tex} file is written to the working directory whilst
the many \vv{**.html} files appear in the sub-directory \vv{HW\_decays/}
which must have been created previously.  Paper sizes and offsets for
the \LaTeX\ output are stored at the top of the block data file
\vv{HWUDAT}: they may need modifying to suit a particular printer. When
event listings are requested (\vv{NEVHEP.LE.MAXPR} or \vv{HWUEPR} is
called explicitly) the following files are created in the current
working directory:
\begin{tabbing}
That is,m\=If (\vv{PRNTEX}): \=  \vv{HW\_decays.tex} \kill
         \>If (\vv{PRNTEX}): \>  \vv{HWEV\_*******.tex} \\
         \>If (\vv{PRNWEB}): \>  \vv{HWEV\_*******.html} \\
\end{tabbing}
where \vv{*******}=0000001 etc.\ is the event number.

Note that the html file automatically makes links to the
\vv{index.html} file of particle properties, assumed to be in the
\vv{HW\_decays} sub-directory.

A new integer variable \vv{NPRFMT} (default 1) has been introduced to
control how many significant figures are shown in each of the 3 event
outputs.  Basically \vv{NPRFMT=1} gives short compact outputs whilst
\vv{NPRFMT=2} gives long formats.

Note that all the \LaTeX\ files use the package \vv{longtable.sty} to
format the tables.  Also if \vv{NPRFMT=2} or \vv{PRVTX=.TRUE.} then the
\LaTeX\ files are designed to be printed in landscape mode.

\section{Structure and output}

\subsection{Main program}
\label{program}

The main program \vv{HWIGPR} has the following form:
\small\begin{verbatim}
      PROGRAM HWIGPR
C---COMMON BLOCKS ARE INCLUDED AS FILE HERWIG65.INC
      INCLUDE 'HERWIG65.INC'
      INTEGER N
      EXTERNAL HWUDAT
C---MAX NUMBER OF EVENTS THIS RUN
      MAXEV=100
C---BEAM PARTICLES
      PART1='P'
      PART2='P'
C---BEAM MOMENTA
      PBEAM1=7000.
      PBEAM2=PBEAM1
C---PROCESS
      IPROC=3000
C---INITIALISE OTHER COMMON BLOCKS
      CALL HWIGIN
C---USER CAN RESET PARAMETERS AT
C   THIS POINT, OTHERWISE DEFAULT
C   VALUES IN HWIGIN WILL BE USED.
      PRVTX=.FALSE.
      MAXER=MAXEV/100
      MAXPR=0
      PTMIN=100.
C   N.B. TO READ SUDAKOV FORM FACTOR FILE ON UNIT 77
C   INSERT THE FOLLOWING TWO LINES IN SUBSEQUENT RUNS
C      LRSUD=77
C      LWSUD=0
C---READ IN SUSY INPUT FILE, IN THIS CASE LHC SUGRA POINT 2
      OPEN(UNIT=LRSUSY,FORM='FORMATTED',STATUS='OLD',ERR=999,
     &     FILE='sugra_pt2.1200.in')
      CALL HWISSP
      CLOSE(UNIT=LRSUSY)
C---COMPUTE PARAMETER-DEPENDENT CONSTANTS
      CALL HWUINC
C---CALL HWUSTA TO MAKE ANY PARTICLE STABLE
      CALL HWUSTA('PI0     ')
C---USER'S INITIAL CALCULATIONS
      CALL HWABEG
C---INITIALISE ELEMENTARY PROCESS
      CALL HWEINI
C---LOOP OVER EVENTS
      DO 100 N=1,MAXEV
C---INITIALISE EVENT
      CALL HWUINE
C---GENERATE HARD SUBPROCESS
      CALL HWEPRO
C---GENERATE PARTON CASCADES
      CALL HWBGEN
C---DO HEAVY OBJECT DECAYS
      CALL HWDHOB
C---DO CLUSTER FORMATION
      CALL HWCFOR
C---DO CLUSTER DECAYS
      CALL HWCDEC
C---DO UNSTABLE PARTICLE DECAYS
      CALL HWDHAD
C---DO HEAVY FLAVOUR HADRON DECAYS
      CALL HWDHVY
C---ADD SOFT UNDERLYING EVENT IF NEEDED
      CALL HWMEVT
C---FINISH EVENT
      CALL HWUFNE
C---USER'S EVENT ANALYSIS
      CALL HWANAL
  100 CONTINUE
C---TERMINATE ELEMENTARY PROCESS
      CALL HWEFIN
C---USER'S TERMINAL CALCULATIONS
      CALL HWAEND
      STOP
 999  WRITE (6,*)
      WRITE (6,*) 'SUSY input file did not open correctly.'
      WRITE (6,*) 'Please check that it is in the right place.'
      WRITE (6,*) 'Examples can be obtained from the ISAWIG web page.'
      WRITE (6,*)
      END
\end{verbatim}\normalsize

The declaration \vv{EXTERNAL HWUDAT} is recommended to help the linker
with finding the block data on some systems.

Various phases of the simulation can be suppressed by deleting the
corresponding subroutine calls, or different subroutines may be
substituted.  For example, in non-\SY\ studies the call to \vv{HWISSP}
should be omitted, and in studies at the parton level everything from
\vv{CALL HWCFOR} to \vv{CALL HWMEVT} can be omitted.

Note that the functionality of the routine \vv{HWUINE} in earlier
versions has now been split between it and a new routine,
\vv{HWUFNE}. A call to the latter {\em must} be made between the calls
to \vv{HWMEVT} and \vv{HWANAL}. A check is built in to prevent execution
if this is not done.

The analysis routine \vv{HWANAL} should always begin with the line
\small\begin{verbatim}
      IF (IERROR.NE.0) RETURN
\end{verbatim}\normalsize
since if an event is cancelled, each of the routines is still called
in turn until reaching the end of the main loop.

\subsection{Form factor file}\label{form}

\HW\ uses look-up tables of Sudakov form factors for the evolution of
initial- and final-state parton showers.  These can be read from an
input file rather than being recomputed each time.  The reading,
writing and computing of form factor tables is controlled by integer
parameters \vv{LRSUD} and \vv{LWSUD} (table~\ref{tab.23}).

\TABULAR{|ll|}{\hline
$\vv{LRSUD}=\vv{N}>0$ & Read form factors for this run from unit \vv{N}\\
$\vv{LRSUD} = 0$   & Compute new form factor tables for this run     \\
$\vv{LRSUD} < 0$   & Form factor tables are already loaded           \\
$\vv{LWSUD}=\vv{N}>0$ & Write form factors on unit \vv{N} for future use\\
$\vv{LWSUD} = 0$   & Do not write new form factor tables             \\\hline
}{Form factor read/write options. \label{tab.23}}

The option $\vv{LRSUD}<0$ allows the program to be initialised several
times in the same run (e.g.\ to generate various event types) without
recomputing or rereading form factors.

Note that the Sudakov form factors depend on the parameters
\vv{QCDLAM}, \vv{VQCUT}, \vv{VGCUT}, \vv{NCOLO}, \vv{NFLAV},
$\vv{RMASS(13)}$ and $\vv{RMASS}(i)$ for $i=1,\ldots,\vv{NFLAV}$.
Consequently form factor tables {\em must} be recomputed every time
any of these parameters is changed.  These parameters are written/read
with the form factor tables and checks are performed to ensure
consistency.

The parton showering algorithm uses the two-loop running coupling,
with matching at each flavour threshold. However, the Sudakov table
can be computed with either the one-loop or two-loop form, according
to the variable \vv{SUDORD} (= 1 or 2 respectively, default=1). If
$\vv{SUDORD}=1$ the two-loop value is recovered using the veto
algorithm in the shower, whereas if $\vv{SUDORD}=2$ no vetoes are used
in the final-state evolution.  This means that the relative weight of
any shower configuration can be calculated in a closed form, hence
that showers can be `forced'.

To next-to-leading order the two possibilities should be identical,
but they differ at beyond-NLO, so some results may change a little.
The most noticeable difference is that the form factor table takes a
factor of about five times longer to compute with $\vv{SUDORD}=2$ than
with 1.

When \vv{SUDORD}=2, no veto is needed for gluon splitting to
quarks. This means that no vetoes are needed for final state
showering, except for the previously-mentioned transverse momentum
cut. The removal of vetoes allows preselection of the flavours that a
jet will contain, giving a huge increase in the efficiency of rare
process simulation ~\cite{Seymour:1994ca}.

\subsection{Event data}
\label{event}

\vv{/HEPEVT/} is the \LEP\ standard common block containing
current event data (table~\ref{tab.24}).

\TABULAR{|ll|}{\hline
 \vv{NEVHEP}      & event number                                              \\
 \vv{NHEP}        & number of entries for this event                          \\
$\vv{ISTHEP(I)}$  & status of entry \vv{I} (see below)                         \\
$\vv{IDHEP(I)}$   & identity of entry \vv{I} (Particle Data Group code)        \\
$\vv{JMOHEP(1,I)}$& pointer to  first mother of entry \vv{I} (see below)       \\
$\vv{JMOHEP(2,I)}$& pointer to second mother of entry \vv{I} (see below)       \\
$\vv{JDAHEP(1,I)}$& pointer to first daughter of entry \vv{I} (see below)      \\
$\vv{JDAHEP(2,I)}$& pointer to  last daughter of entry \vv{I} (see below)      \\
$\vv{PHEP(*,I)}$  & $(p_x,p_y,p_z,E,M)$ of entry \vv{I}:
                         $M={\rm sign}\left(\sqrt{{\rm abs}(m^2)},m^2\right)$\\
$\vv{VHEP(*,I)}$  & $(x,y,z,t)$ of production vertex of entry \vv{I}
                         (see sect.~\ref{spacetime})\\\hline
}{Current event data. \label{tab.24}}

All momenta are given in GeV/$c$ in the laboratory frame, in which the
input beam momenta are \vv{PBEAM1} and \vv{PBEAM2} as specified by the
user and point along the $+z$ and $-z$ directions respectively.  Final
state particles have $\vv{ISTHEP(I)} = 1$.  See section~\ref{status} for
a complete list of the special status codes used by \HW.

\looseness=1 The identity codes \vv{IDHEP} are almost as recommended by the \LEP\
Working Group~\cite{LEP2}, i.e.\ the revised \PD~\cite{PDG} numbers
where defined, \vv{IDHEP} = 91 for clusters, 94 for jets, and 0 for
objects with no \PD\ code.  The only exception is our use of \vv{IDHEP}
= 26 for the lightest \MS\ Higgs boson, to distinguish it from the
\SM\ Higgs boson (\PD\ code 25).  In addition, the
`generator-specific' (\vv{IDHEP}=81-100) codes 98 and 99 are used for
remnant photons and nucleons, respectively (see section~\ref{sec:sue}).

\HW\ also has its own internal identity codes $\vv{IDHW(I)}$, stored in
\vv{/HWEVNT/}. The utility subroutine \vv{HWUIDT} translates between
\HW\ and \PD\ identity codes.  See section~\ref{including} for further
details.

The mother and daughter pointers are standard, except that
$\vv{JMOHEP(2,I)}$ and $\vv{JDAHEP(2,I)}$ for a \emph{parton} are its
\emph{colour mother} and \emph{colour daughter}, i.e.\ the partons to
which its colour and anticolour are connected, respectively.  For this
purpose the primary partons from a hard subprocess are all regarded as
outgoing (see examples in section~\ref{qcd}, \ref{including}
and~\ref{guide}).  Since a quark has no anticolour, $\vv{JDAHEP(2,I)}$
is used to point to its \emph{flavour} partner. Similarly for
$\vv{JMOHEP(2,I)}$ in the case of an antiquark.

In addition to entries representing partons, particles, clusters etc,
\vv{/HEPEVT/} contains purely informational entries representing the
total centre-of-mass momentum, hard and soft subprocess momenta, etc.
See section~\ref{status} for the corresponding status codes.

Information from all stages of event processing is retained in
\vv{/HEPEVT/} so the same particle may appear several times with
different status codes. For example, an outgoing parton from a hard
scattering (entered initially with status 113 or 114) will appear
after processing as an on-mass-shell parton before \QCD\ branching
(status 123,124), an off-mass-shell entry representing the flavour and
momentum of the outgoing jet (status 143,144), and a jet constituent
(157). It might also appear again in other contexts, e.g.\ as a
spectator in a heavy flavour decay (status 154,160).

Incoming partons (entered with status 111 or 112, changed to 121 or
122 after branching) give rise to spacelike jets (status 141 or 142)
with $m^2<0$, indicated by $\vv{PHEP(5,IHEP)}<0$, due to the loss of
momentum via initial-state bremsstrahlung. The same applies in
principle to incoming leptons, but in that case emission of at most a
single photon is permitted from each initial-state lepton and the
off-shell lepton is given status code 3 (see section~\ref{status}).

\looseness=1 Each parton jet begins with a status 141--144 jet entry giving the
total flavour and momentum of the jet.  The first mother pointer of
this entry gives the location of the parent hard parton, while the
second gives that of the subprocess centre-of-mass momentum.  If \QCD\
branching has occurred, this is followed by a lightlike \vv{CONE}
entry, which fixes the angular extent of the jet and its azimuthal
orientation relative to the parton with which it interferes. The
interfering parton is listed as the second mother of the cone.  Next
come the actual constituents of the jet.  If no branching has
occurred, there is no cone and the single jet constituent is the same
as the jet.

Since version~5.1, the event record has been modified to retain
entries for all partons before hadronization (with status
\vv{ISTHEP}=2).  During hadronization, the gluons are split into
quark-antiquark, while other partons are copied to a location
(indicated by \vv{JDAHEP}(1,*)) where their momenta may be shifted
slightly, to conserve momentum, during heavy cluster
splitting. Previously the original momenta were shifted, so momentum
appeared not to be conserved at the parton level.

Since version~6.3, to take account of the increased energy and
complexity of interactions at
LHC and future colliders, the default value of the parameter \vv{NMXHEP},
which sets the array sizes in the standard \vv{/HEPEVT/} common block,
has been increased to 4000.

\subsubsection{Status codes}
\label{status}

A complete list of currently-used \HW\ status codes is given below
(table~\ref{tab.25}). Many are used only in intermediate stages of event
processing.  The most important for users are probably 1 (final-state
particle), 101--3 (initial state), 141--4 (jets), and 199 (decayed
$b$-flavoured hadrons).

For technical reasons, some \HW\ status codes \vv{ISTHEP} between 153
and 165 have changed their meanings since version~5.1.

The event status \vv{ISTAT} in common \vv{/HWEVNT/} is roughly
$\vv{ISTHEP}-100$ where
\vv{ISTHEP} is the status of entries being processed.
For completed events, $\vv{ISTAT}=100$.

\setlongtables
\begin{longtable}[t]{|c|l|}
\hline
\vv{ISTHEP}   &  Description \\         
\hline
\endhead
\caption*{\small {\bf Table~\ref{tab.25}:} Status codes. (Continues)}
\endfoot
\caption*{\small {\bf Table~\ref{tab.25}:} Status codes.}
\endlastfoot
\label{tab.25}
                    1  &  final state particle        \\
                    2  &  parton before hadronization \\
                    3  &  documentation line          \\
\hline
                  100  &  cone limiting jet evolution \\
                  101  & `beam'   (beam 1)            \\
                  102  & `target' (beam 2)            \\
                  103  &  overall centre of mass      \\
\hline
                  110  & unprocessed hard process c.m.\\
                  111  & unprocessed beam parton      \\
                  112  & unprocessed target parton    \\
                  113  & unproc.\ first outgoing parton\\
                  114  & unproc.\ other outgoing parton\\
                  115  & unprocessed spectator parton \\
\hline
                120--25& as 110--15, after processing \\
\hline
                  130  & lepton in jet (unboosted)    \\
                131--34& as 141--44, unboosted to c.m.\\
                  135  & spacelike parton (beam, unboosted)\\
                  136  & spacelike parton (target,unboosted)\\
                  137  & spectator (beam, unboosted)  \\
                  138  & spectator (target, unboosted)\\
                  139  & parton from branching (unboosted)\\
                  140  & parton from gluon splitting (unboosted)\\
\hline
                141--44& jet from parton type 111--14 \\
                145--50& as 135--40 boosted, unclustered\\
\hline
                  151  & as 159, not yet clustered    \\
                  152  & as 160, not yet clustered    \\
                  153  & spectator from beam          \\
                  154  & spectator from target        \\
                  155  & heavy quark before decay     \\
                  156  & spectator before heavy decay \\
                  157  & parton from \QCD\ branching  \\
                  158  & parton from gluon splitting  \\
                  159  & parton from cluster splitting\\
                  160  & spectator after heavy decay  \\
\hline
                  161  & beam spectator after gluon splitting\\
                  162  & target spectator after gluon splitting\\
                  163  & other cluster before soft process\\
                  164  & beam cluster before soft process\\
                  165  & target cluster before soft process\\
                  167  & unhadronized beam cluster    \\
                  168  & unhadronized target cluster  \\
\hline
                  170  & soft process centre of mass  \\
                  171  & soft cluster (beam, unhadronized)\\
                  172  & soft cluster (target, unhadronized)\\
                  173  & soft cluster (other, unhadronized)\\
\hline
                  181  & beam cluster  (no soft process)\\
                  182  & target cluster (no soft process)\\
                  183  & hard process cluster (hadronized)\\
                  184  & soft cluster (beam, hadronized)\\
                  185  & soft cluster (target, hadronized)\\
                  186  & soft cluster (other, hadronized)\\
\hline
                190--93& as 195--98, before decays     \\
                  195  & direct unstable non-hadron   \\
                  196  & direct unstable hadron (1-body clus.)\\
                  197  & direct unstable hadron (2-body clus.)\\
                  198  & indirect unstable hadron or lepton \\
                  199  & decayed heavy flavour hadron \\
\hline
                  200  & neutral B meson, flavour at prod'n \\
\hline
\end{longtable}

\subsubsection{Event weights}
\label{weights}

The default is to generate unweighted events
($\vv{EVWGT}=\vv{AVWGT}$). Then event distributions are generated by
computing a weight proportional to the cross section and comparing it
with a random number times the maximum weight.  Set \vv{WGTMAX} to the
maximum weight, or to zero for the program to compute it.  If a weight
greater than \vv{WGTMAX} is generated during execution, a warning is
printed and \vv{WGTMAX} is reset. If this occur too often, output event
distributions could be distorted.

To generate weighted events, set $\vv{NOWGT}=\vv{.FALSE.}$ in common
\vv{/HWEVNT/}.

In \QCD\ hard scattering and heavy flavour and direct photon
production (\vv{IPROC}= 1500--1800) the transverse energy distribution
of weighted events (or the efficiency for unweighted events) can be
varied using the parameters \vv{PTMIN}, \vv{PTMAX} and \vv{PTPOW}.

Similarly in Drell-Yan processes ($\vv{IPROC} = 1350$ etc.) the lepton
pair mass distribution is controlled by the parameters \vv{EMMIN},
\vv{EMMAX} and \vv{EMPOW}, and in deep inelastic scattering the $Q^2$
distribution depends on \vv{Q2MIN}, \vv{Q2MAX} and \vv{Q2POW}.

Data on weights generated are output at the end of the run. The mean
weight is an estimate of the cross section (in nanobarns) integrated
over the region used for event generation.  Note that the mean weight
is the sum of weights divided by the total number of {\em weights}
generated, not the total number of {\em events}.

The maximum weight is now always printed in full precision. This is
needed to be sure of generating the same events in repeated runs.

  In versions 6.3 and higher, the option of generating negative weights has
  been available. This is controlled by the input parameter
  \vv{NEGWTS}~[\vv{.FALSE.}], see sect.~\ref{sect:negwgts} for details.

\subsection{Error conditions}
\label{error}

Certain combinations of input parameters may lead to problems in
execution.  \HW\ tries to detect these and print a warning.  Errors
during execution are dealt with by \vv{HWWARN}, which prints the
calling subprogram and a code and takes appropriate action.  In
general, the larger the code the more serious the problem.  Refer to
the source code to find out why \vv{HWWARN} was called. It is important
to note the subprogram from which the call was issued: many different
subprograms use the same error code, but each code is unique within a
given subprogram.

Events can be rerun by setting the random number seeds \vv{NRN(1)} and
\vv{NRN(2)} to the values given in the error message or event dump, and
\vv{MAXWGT} to the maximum weight encountered in the run. The contents
of \vv{/HEPEVT/} can by printed by calling \vv{HWUEPR}, those of
\vv{/HWPART/} (the last parton shower) by calling \vv{HWUBPR}.

Note that if \vv{WGTMAX} is increased during event generation, so that
this type of message is printed:
\small\begin{verbatim}
   HWWARN CALLED FROM SUBPROGRAM HWEPRO: CODE =   1
   EVENT      21:   SEEDS =  836291635 & 1823648329  WEIGHT = 0.3893E-08
   EVENT SURVIVES. EXECUTION CONTINUES
            NEW MAXIMUM WEIGHT = 0.428217360829367E-08
\end{verbatim}\normalsize
then to regenerate any later events, \vv{WGTMAX} must be set to the
printed value, as well as setting \vv{NRN} to the appropriate seeds.

Examples of error messages are:
\small\begin{verbatim}
   HWWARN CALLED FROM SUBPROGRAM HWSBRN: CODE = 101
   EVENT      31:   SEEDS =  422399901 &  771980111  WEIGHT = 0.3893E-08
   EVENT KILLED.   EXECUTION CONTINUES
\end{verbatim}\normalsize
Spacelike (initial-state) parton branching had no phase space. This
can happen due to cutoffs which are slightly different in the hard
subprocess and the parton shower. \\ 
Action taken: program throws away this event and starts a new one.
\small\begin{verbatim}
   HWWARN CALLED FROM SUBPROGRAM HWCHAD: CODE = 102
   EVENT      51:   SEEDS = 1033784787 & 1428957533  WEIGHT = 0.3893E-08
   EVENT KILLED.   EXECUTION CONTINUES
\end{verbatim}\normalsize
A cluster has been formed with too low a mass to represent any hadron
of the correct flavour, and there is no colour-connected cluster from
which the necessary additional mass could be transferred. \\ Action
taken: program throws away this event and starts a new one.
\small\begin{verbatim}
   HWWARN CALLED FROM SUBPROGRAM HWUINE: CODE= 200
   EVENT SURVIVES.  RUN ENDS GRACEFULLY
\end{verbatim}\normalsize
\sss{CPU} time limit liable to be reached before generating \vv{MAXEV}
events.\\ 
Action taken: skips to terminal calculations using existing events.
\small\begin{verbatim}
   HWWARN CALLED FROM SUBPROGRAM HWBSUD: CODE= 500
   RUN CANNOT CONTINUE
\end{verbatim}\normalsize
The table of Sudakov form factors read on unit \vv{LRSUD} does not
extend to the maximum momentum scale \vv{QLIM} specified for this
run. \\
Action taken: run aborted.  The user must either reduce \vv{QLIM} or
set \vv{LRSUD} to zero to make a bigger table (set \vv{LWSUD} non-zero
to write it).
\small\begin{verbatim}
   HWWARN CALLED FROM SUBPROGRAM HWBSUD: CODE= 515
   RUN CANNOT CONTINUE
\end{verbatim}\normalsize
The table of Sudakov form factors read on unit \vv{LRSUD} is for a
different value of a relevant parameter (in this case the $b$ quark
mass). \\
Action taken: run aborted. The user must make a new table (set
\vv{LWSUD} non-zero to write it).

\subsection{Sample output}\label{sample}

This is the output from the main program listed in
section~\ref{program}, with no event printout or user analysis.
 
\small\begin{verbatim}
          HERWIG 6.500    16 Oct 2002 

          Please reference:  G. Marchesini, B.R. Webber,
          G.Abbiendi, I.G.Knowles, M.H.Seymour & L.Stanco
          Computer Physics Communications 67 (1992) 465
                             and
          G.Corcella, I.G.Knowles, G.Marchesini, S.Moretti,
          K.Odagiri, P.Richardson, M.H.Seymour & B.R.Webber,
          JHEP 0101 (2001) 010


                                      

          Since SUSY processes are called,
          please also reference: S.Moretti, K.Odagiri,
          P.Richardson, M.H.Seymour & B.R.Webber,
          JHEP 0204 (2002) 028

          Reading in SUSY data from unit 66

          INPUT CONDITIONS FOR THIS RUN

          BEAM 1 (P       ) MOM. =   7000.00
          BEAM 2 (P       ) MOM. =   7000.00
          PROCESS CODE (IPROC)   =    3000
          NUMBER OF FLAVOURS     =    6
          STRUCTURE FUNCTION SET =    8
          AZIM SPIN CORRELATIONS =    T
          AZIM SOFT CORRELATIONS =    T
          QCD LAMBDA (GEV)       =    0.1800
          DOWN     QUARK  MASS   =    0.3200
          UP       QUARK  MASS   =    0.3200
          STRANGE  QUARK  MASS   =    0.5000
          CHARMED  QUARK  MASS   =    1.5500
          BOTTOM   QUARK  MASS   =    4.9500
          TOP      QUARK  MASS   =  175.0000
          GLUON EFFECTIVE MASS   =    0.7500
          EXTRA SHOWER CUTOFF (Q)=    0.4800
          EXTRA SHOWER CUTOFF (G)=    0.1000
          PHOTON SHOWER CUTOFF   =    0.4000
          CLUSTER MASS PARAMETER =    3.3500
          SPACELIKE EVOLN CUTOFF =    2.5000
          INTRINSIC P-TRAN (RMS) =    0.0000
          DECAY SPIN CORRELATIONS=    T
          SUSY THREE BODY ME     =    T
          SUSY FOUR  BODY ME     =    F

          NO EVENTS WILL BE WRITTEN TO DISK

          B_d: Delt-M/Gam =0.7000 Delt-Gam/2*Gam =0.0000
          B_s: Delt-M/Gam = 10.00 Delt-Gam/2*Gam =0.2000

          PDFLIB NOT USED FOR BEAM 1
          PDFLIB NOT USED FOR BEAM 2


          Checking consistency of particle properties


          Checking consistency of decay tables

 Line    1 is the same as line 2634
 Take BR 0.333 and ME code 100 from second entry
 Line    2 is the same as line 2635
 Take BR 0.333 and ME code 100 from second entry
 Line    3 is the same as line 2636
 Take BR 0.111 and ME code 100 from second entry
 Line    4 is the same as line 2637
 Take BR 0.111 and ME code 100 from second entry
 Line    5 is the same as line 2638
 Take BR 0.111 and ME code 100 from second entry
 Line    6 is the same as line 2639
 Take BR 0.333 and ME code 100 from second entry
 Line    7 is the same as line 2640
 Take BR 0.333 and ME code 100 from second entry
 Line    8 is the same as line 2641
 Take BR 0.111 and ME code 100 from second entry
 Line    9 is the same as line 2642
 Take BR 0.111 and ME code 100 from second entry
 Line   10 is the same as line 2643
 Take BR 0.111 and ME code 100 from second entry

          WRITING SUDAKOV TABLE ON UNIT  77

          WRITING MATRIX ELEMENT TABLE ON UNIT  88

          CHECKING SUSY DECAY MATRIX ELEMENTS

          PARTICLE TYPE  21=PI0      SET STABLE

          INITIAL SEARCH FOR MAX WEIGHT

          PROCESS CODE IPROC =        3000
          RANDOM NO. SEED 1  =     1246579
                     SEED 2  =     8447766
          NUMBER OF SHOTS    =       10000
          NEW MAXIMUM WEIGHT =  2.4359288434851706E-03
          NEW MAXIMUM WEIGHT =  3.7017893759328114E-03
          NEW MAXIMUM WEIGHT =  1.9578185739344733E-02

 HWWARN CALLED FROM SUBPROGRAM HWSMRS: CODE =   5
 EVENT       0:   SEEDS =      17673 &      63565  WEIGHT = 0.0000E+00
 EVENT SURVIVES. EXECUTION CONTINUES
 WARNING: MRST98 CALLED WITH Q OUTSIDE ALLOWED RANGE!
 Q VALUE=3.869E+03, MINIMUM=1.118E+00, MAXIMUM=3.162E+03
 NO FURTHER WARNINGS WILL BE ISSUED
          NEW MAXIMUM WEIGHT =  2.2550328399448680E-02
          NEW MAXIMUM WEIGHT =  5.5633613959185035E-02
          NEW MAXIMUM WEIGHT =  6.9166986913034176E-02
          NEW MAXIMUM WEIGHT =  0.1197405440791255    
          NEW MAXIMUM WEIGHT =  0.1488818465259009    

          INITIAL SEARCH FINISHED

          OUTPUT ON ELEMENTARY PROCESS

          N.B. NEGATIVE WEIGHTS NOT ALLOWED

          NUMBER OF EVENTS   =           0
          NUMBER OF WEIGHTS  =       10000
          MEAN VALUE OF WGT  =  4.1213E-03
          RMS SPREAD IN WGT  =  1.2862E-02
          ACTUAL MAX WEIGHT  =  1.3755E-01
          ASSUMED MAX WEIGHT =  1.4888E-01

          PROCESS CODE IPROC =        3000
          CROSS SECTION (PB) =   4.121    
          ERROR IN C-S  (PB) =  0.1286    
          EFFICIENCY PERCENT =   2.768    

          SUBROUTINE TIMEL CALLED BUT NOT LINKED.
          DUMMY TIMEL WILL BE USED. DELETE DUMMY
          AND LINK CERNLIB FOR CPU TIME REMAINING.

          OUTPUT ON ELEMENTARY PROCESS

          N.B. NEGATIVE WEIGHTS NOT ALLOWED

          NUMBER OF EVENTS   =         100
          NUMBER OF WEIGHTS  =        3699
          MEAN VALUE OF WGT  =  3.7473E-03
          RMS SPREAD IN WGT  =  1.2205E-02
          ACTUAL MAX WEIGHT  =  1.3900E-01
          ASSUMED MAX WEIGHT =  1.4888E-01

          PROCESS CODE IPROC =        3000
          CROSS SECTION (PB) =   3.747    
          ERROR IN C-S  (PB) =  0.2007    
          EFFICIENCY PERCENT =   2.517    
\end{verbatim}\normalsize

\subsubsection{Guide to sample output}\label{guide}

See ref.~\cite{hw51} for a full discussion of the basic features of
\HW\ output, including a listing of a sample event.  Here we point out
only some new features in comparison to version~5.1.

\pagebreak[3]

The beam particles, their energies and the process code \vv{IPROC}=3000
indicate the \SY\ process of squark and/or gluino production at \LHC.
The call to \vv{HWISSP} triggers a request for a \SY\ particle data
file, in the format specified in section~\ref{sec:SUSY}, which is read
in from the default unit.  In this case the file corresponds to the
second \LHC\ \sss{SUGRA} point discussed in section~\ref{sec:SUSYdata}.

First the program lists the main relevant input parameters, including
$B_d$ and $B_s$ mixing parameters.  Parton distributions were not
requested from the \PDF\ library and therefore the default MRST set
\cite{Martin:1998np} is used~(see sect.~\ref{sec_pdfs}.)

Next the program performs some basic checks on the particle data
provided. Here it finds that the input file read by \vv{HWISSP}
contains top quark decay modes which duplicate the default modes
stored in \vv{HWUDAT}. The branching ratios and matrix element codes
are accordingly updated to those in the input file.

After an initial search for the maximum weight, the program prints its
estimate of the relevant cross section and the expected Monte Carlo
efficiency for event generation. During the search for the maximum weight
the default parton distributions are called with a parameter, in this case
the scale, outside the allowed range and a warning is issued. 
The maximum or minimum value of the parameter is used instead, depending
on whether the value requested is too large or too small.

In the course of event generation a new cross section estimate is
generated, which is printed together with the actual Monte Carlo
efficiency at the end of the run.

\subsection{Subroutine descriptions}
\label{subroutine}

We give in table~\ref{tab.26} a list of all subroutines with their
functions.  Note that the third letter of the name usually follows a
rough classification scheme.

\setlongtables
\begin{longtable}[t]{|c|l|}
\hline
          Name & Description  \\
\hline
\endhead
\caption*{\small {\bf Table~\ref{tab.26}:} Subroutines. (Continues)}
\endfoot
\caption*{\small {\bf Table~\ref{tab.26}:} Subroutines.}
\endlastfoot
\label{tab.26}
& Main program and initialisation \\
\hline
    \vv{HWIGPR} & Main program                                \\
    \vv{HWIGIN} & Default initialisations                     \\
\hline
& Reading/writing/altering decay modes \\
\hline
    \vv{HWIGUP} & Initialisation for Les Houches interface\\
    \vv{HWIMDE} & Adds modes for SUSY four body decays\\
    \vv{HWIODK} & Inputs/outputs formatted decay tables \\
    \vv{HWISPC} & Initialisation of couplings for spin correlations and SUSY decays\\
    \vv{HWISPN} & Initialisation for spin correlations and SUSY decays\\
    \vv{HWISP2} & Initialisation of two body decays for spin correlations and SUSY decays \\
    \vv{HWISP3} & Initialisation of three body decays for spin correlations and SUSY decays\\
    \vv{HWISP4} & Initialisation of four body decays for spin correlations and SUSY decays\\
    \vv{HWIPHS} & Initialises optimized phase space \\
    \vv{HWISSP} & Inputs supersymmetric particle data \\
    \vv{HWMODK} & Modifies or adds an individual decay mode \\
\hline
& User-provided analysis routines  \\
\hline
    \vv{HWABEG} & Initialises user's analysis                 \\
    \vv{HWAEND} & Terminates user's analysis                  \\
    \vv{HWANAL} & Performs user's analysis on event           \\
\hline
& Parton branching with interfering gluons  \\
\hline
    \vv{HWBAZF} & Computes azimuthal correlation functions    \\
    \vv{HWBCON} & Makes colour connections between jets       \\
    \vv{HWBDED} & Correction to the `dead zone' in $\ee$      \\
    \vv{HWBDIS} & Correction to the `dead zone' in DIS        \\
    \vv{HWBDYP} & Correction to the `dead zone' in Drell-Yan  \\
    \vv{HWBFIN} & Transfers external lines of jet to \vv{/HEPEVT/} \\
    \vv{HWBGEN} & Finds unevolved partons and generates jets  \\
    \vv{HWBGUP} & Makes colour and flavour connections for Les Houches events\\
    \vv{HWBJCO} & Combines jets with correct kinematics       \\
    \vv{HWBMAS} & Computes masses and trans. momenta in jet   \\
    \vv{HWBRAN} & Generates a timelike parton branching       \\
    \vv{HWBRCN} & Replaces \vv{HWBCON} if R-parity is violated \\
    \vv{HWBRC1} & Finds colour partner in gluino decay \\
    \vv{HWBRC2} & Finds colour partner in jet \\
    \vv{HWBSPA} & Computes momenta in spacelike jet           \\
    \vv{HWBSPN} & Computes spin density/decay matrices        \\
    \vv{HWBSU1} & First  term in quark Sudakov form factor    \\
    \vv{HWBSU2} & Second term in quark Sudakov form factor    \\
    \vv{HWBSUD} & Computes (or reads) Sudakov form factors    \\
    \vv{HWBSUG} & Integrand in gluon Sudakov form factor      \\
    \vv{HWBSUL} & Logarithmic part of Sudakov form factor     \\
    \vv{HWBTIM} & Computes momenta in timelike jet            \\
    \vv{HWBTOP} & Correction to the `dead zone' in top decay  \\
    \vv{HWBVMC} & Virtual mass cutoff for parton type \vv{ID}  \\
\hline
& Cluster hadronization model \\
\hline
    \vv{HWCBCT} & Cuts a massive baryon cluster in two \\
    \vv{HWCBVI} & Clusters quarks from a \BNV\ interaction \\ 
    \vv{HWCBVT} & Finds which \BNV\ interaction partons came from\\
    \vv{HWCCCC} & Finds colour connections after gluon splitting if \BNV\\
    \vv{HWCCUT} & Cuts a massive cluster in two               \\
    \vv{HWCDEC} & Decays clusters into primary hadrons        \\
    \vv{HWCFLA} & Sets up flavours for \vv{HWCHAD}             \\
    \vv{HWCFOR} & Forms clusters                              \\
    \vv{HWCGSP} & Splits gluons                               \\
    \vv{HWCHAD} & Decays a cluster into one or two hadrons    \\
\hline
& Particle and heavy quark decays   \\
\hline
    \vv{HWD2ME} & Finds maximum weight for a two body decay for spin correlations\\
    \vv{HWD3ME} & SUSY three body decays and spin correlations master routine\\
    \vv{HWD3M0} & Calculation of SUSY three body matrix element\\
    \vv{HWD3M1} & Helicity amplitude for $f\ra ff\bar{f}$ via vector boson exchange\\
    \vv{HWD3M2} & Helicity amplitude for $f\ra ff\bar{f}$ via Higgs exchange \\
    \vv{HWD3M3} & Helicity amplitude for $f\ra ff\bar{f}$ via antisfermion exchange\\
    \vv{HWD3M4} & Helicity amplitude for $f\ra ff\bar{f}$ via sfermion exchange\\
    \vv{HWD3M5} & Helicity amplitude for $\bar{f}\ra\bar{f}f\bar{f}$
			 via vector  boson exchange\\
    \vv{HWD3M6} & Helicity amplitude for $\phi\ra\phi f\bar{f}$ 
			via vector boson exchange \\
    \vv{HWD3M7} & Helicity amplitude for $f\ra\tilde{G}f\bar{f}$
			via vector boson exchange\\
    \vv{HWD3M8} & Helicity amplitude for $f\ra\bar{f}\bar{f}f$
	 via scalar exchange (1st diagram)  \\
    \vv{HWD3M9} &  Helicity amplitude for $f\ra\bar{f}\bar{f}f$
	 via scalar exchange (2nd diagram)  \\
    \vv{HWD3MA} &  Helicity amplitude for $f\ra\bar{f}\bar{f}f$
	 via scalar exchange (3rd diagram)  \\
    \vv{HWD3MB} & Helicity amplitude for $f\ra fff$
	 via scalar exchange (1st diagram)  \\
    \vv{HWD3MC} & Helicity amplitude for $f\ra fff$
	 via scalar exchange (2nd diagram)  \\
    \vv{HWD3MD} & Helicity amplitude for $f\ra fff$
	 via scalar exchange (3rd diagram)  \\
    \vv{HWD3MF} &  Helicity amplitude for $f\ra\bar{f}\bar{f}\bar{f}$
	 via scalar exchange (1st diagram)  \\
    \vv{HWD3MG} & Helicity amplitude for $f\ra\bar{f}\bar{f}\bar{f}$
	 via scalar exchange (2nd diagram)  \\
    \vv{HWD3MH} & Helicity amplitude for $f\ra\bar{f}\bar{f}\bar{f}$
	 via scalar exchange (3rd diagram)  \\
    \vv{HWD3MI} & Helicity amplitude for $\bar{f}\ra\bar{f}f\bar{f}$
			 via scalar exchange\\
    \vv{HWD4ME} & Master routine for SUSY four body decays\\
    \vv{HWD4M0} & Matrix Element for $\phi\ra V^*V^*\ra f\bar{f}f\bar{f}$\\
    \vv{HWDBOS} & Finds and decays W and Z bosons             \\
    \vv{HWDBOZ} & Chooses decay mode of W and Z bosons        \\ 
    \vv{HWDBZ2} & Copy of \vv{HWDBOZ} used by hadronic WW, WZ and ZZ production \\
    \vv{HWDCHK} & Checks given decay mode is self-consistent  \\
    \vv{HWDCLE} & Interface to {\sf CLEO} package for B decays      \\
    \vv{HWDEUR} & Interface to {\sf EURODEC} package for B decays \\
    \vv{HWDFIV} & Generates a five-body decay                 \\
    \vv{HWDFOR} & Generates a four-body decay                 \\
    \vv{HWDHAD} & Generates decays of unstable hadrons        \\
    \vv{HWDHGC} & Higgs $\to\gamma\gamma$ decay          \\
    \vv{HWDHGF} & Higgs $\to W^+ W^-$ decay              \\
    \vv{HWDHIG} & Finds and decays Higgs bosons \\
    \vv{HWDHOB} & Finds and decays heavy objects  \\
    \vv{HWDHO1} & Selects decay mode for heavy object\\
    \vv{HWDHO2} & Calculates momenta of heavy object's decay products\\
    \vv{HWDHO3} & Makes colour connects for heavy object decay\\
    \vv{HWDHO4} & Performs parton shower for heavy object decay\\
    \vv{HWDHO5} & Checks for colour disconnections in heavy object decay\\
    \vv{HWDHO6} & Perform \RPV\ colour connections in heavy object decay\\
    \vv{HWDHVY} & Finds and decays heavy flavours             \\
    \vv{HWDHWT} & Subroutine for top decay via a virtual $H^\pm$\\
    \vv{HWDRCL} & Colour connections for a \BNV\ \SY\ decay\\
    \vv{HWDRME} & Main \RPV\ 3-body decay matrix element subroutine\\
    \vv{HWDRM1} & \RPV\ 3-body decay matrix element  subroutine \\
    \vv{HWDRM2} & \RPV\ 3-body decay matrix element  subroutine \\
    \vv{HWDRM3} & \RPV\ 3-body decay matrix element  subroutine \\
    \vv{HWDRM4} & \RPV\ 3-body decay matrix element  subroutine \\
    \vv{HWDRM5} & \RPV\ 3-body decay matrix element  subroutine \\
    \vv{HWDPWT} & Phase space three-body decay weight         \\
    \vv{HWDSIN} & Performs decays with spin correlations\\
    \vv{HWDSI1} & Picks next particle to decay for spin correlations\\
    \vv{HWDSI2} & Pick moment of decay products with spin correlations\\
    \vv{HWDSI3} & Selects $\tau$ polarization and passes it to the decay routine\\
    \vv{HWDSM2} & Two body decay with spin correlations\\
    \vv{HWDSM3} & Three body decay with spin correlations\\
    \vv{HWDSM4} & Four body decay with spin correlations\\
    \vv{HWDTAU} & Interface to \TA\ for $\tau$ decays\\
    \vv{HWDTHR} & Generates a three-body decay                \\
    \vv{HWDTOP} & Decides whether to decay top quark          \\
    \vv{HWDTWO} & Generates a two-body decay                  \\
    \vv{HWDWWT} & Weak ($V-A$) three-body decay weight        \\
    \vv{HWDXLM} & Tests if decay vertex lies in given volume  \\
\hline
& Elementary subprocess generation \\
\hline
    \vv{HWECIR} & Interface to {\sf CIRCE} \\
    \vv{HWEFIN} & Final calculations on elementary subprocess \\
    \vv{HWEGAM} & Generates incoming photon                   \\
    \vv{HWEINI} & Initialises elementary subprocess           \\
    \vv{HWEISR} & Generates photon emission from $e$ or $\mu$ \\
    \vv{HWEONE} & Sets up a $2\to1$ hard subprocess           \\
    \vv{HWEPRO} & Generates elementary subprocess             \\
    \vv{HWETWO} & Sets up a $2\to2$ hard subprocess           \\
\hline
& Individual hard subprocesses \\
\hline
    \vv{HWH2BK} & Matrix element for $b\bbar\to W^\pm H^\mp$ \\ 
    \vv{HWH2BH} & Matrix element for $H^\pm$ production via $bq$-fusion \\
    \vv{HWH2DD} & Function   to return the $D$ function of \cite{Kleiss:1985yh} \\
    \vv{HWH2F1} & Subroutine to return the $F$ function of \cite{vanEijk:1990zp}
for a fixed first momentum\\
    \vv{HWH2F2} & Subroutine to return the $F$ function of \cite{vanEijk:1990zp}
for a fixed second momentum\\
    \vv{HWH2F3} & Subroutine to return the $F$ function of \cite{vanEijk:1990zp}
for all first and second momenta\\
    \vv{HWH2HE} & Matrix element for Higgs associated production \\
    \vv{HWH2M0} & Subroutine to compute the massless matrix element for $Q\Qbar Z$\\
    \vv{HWH2MQ} & Subroutine to compute the massive matrix element for $Q\Qbar Z$\\
    \vv{HWH2PS} & Subroutine to perform the phase-space for $Z$+two jets\\
    \vv{HWH2P1} & Subroutine to select quark masses for     \vv{HWH2PS}\\
    \vv{HWH2P2} & Subroutine to select quark masses for     \vv{HWH2PS}\\
    \vv{HWH2QH} & Matrix element for $q\qbar,gg\to Q\Qbar^{(')}$ Higgs \\ 
    \vv{HWH2SH} & Matrix element for squark pair plus Higgs production \\
    \vv{HWH2SS} & Subroutine to return the $S$ function of \cite{Kleiss:1985yh}\\
    \vv{HWH2T1} & Function   to return the $T_1$ function of \cite{Kleiss:1985yh}\\
    \vv{HWH2T2} & Function   to return the $T_2$ function of \cite{Kleiss:1985yh}\\
    \vv{HWH2T3} & Function   to return the $T_3$ function of \cite{Kleiss:1985yh}\\
    \vv{HWH2T4} & Function   to return the $T_4$ function of \cite{Kleiss:1985yh}\\
    \vv{HWH2T5} & Function   to return the $T_5$ function of \cite{Kleiss:1985yh}\\
    \vv{HWH2T6} & Function   to return the $T_6$ function of \cite{Kleiss:1985yh}\\
    \vv{HWH2T7} & Function   to return the $T_7$ function of \cite{Kleiss:1985yh}\\
    \vv{HWH2T8} & Function   to return the $T_8$ function of \cite{Kleiss:1985yh}\\
    \vv{HWH2T9} & Function   to return the $T_9$ function of \cite{Kleiss:1985yh}\\
    \vv{HWH2T0} & Function   to return the $T_{10}$ function of
\cite{Kleiss:1985yh}\\
    \vv{HWH2VH} & Matrix element for $q\qbar^{(')} \to V$ Higgs, 
                                    $V=W^\pm,Z^0$ \\
    \vv{HWH4JT} & Hard subprocess: $\ee\to$ 4 jets \\
    \vv{HWH4J1} & Matrix element for $\ee\to$ 4 jets \\
    \vv{HWHBGF} & Hard subprocess: boson-gluon fusion (\sss{BGF})\\
    \vv{HWHBKI} & Computes kinematics for \sss{BGF}             \\
    \vv{HWHBRN} & Returns a phase-space point for \sss{BGF}     \\
    \vv{HWHBSG} & Computes cross section for \sss{BGF}          \\
    \vv{HWHDIS} & Hard subprocess: deep inelastic $e/\mu$ quark\\
    \vv{HWHDYP} & Hard subprocess: Drell-Yan $Z^0/\gamma$ production\\
    \vv{HWHDYQ} & Subroutine for $Q\Qbar Z$\\
    \vv{HWHEGG} & Hard subprocess: two-photon processes in $\ee$\\
    \vv{HWHEGW} & Hard subprocess: $\gamma W$ processes in $\ee$\\
    \vv{HWHEGX} & Calculates cross section for \vv{HWHEGW}  \\
    \vv{HWHEPA} & Hard subprocess: $\ee \to q \qbar$       \\
    \vv{HWHEPG} & Hard subprocess: $\ee \to q \qbar g$     \\
    \vv{HWHESG} & Gaugino pair production in $\ee$ collisions \\
    \vv{HWHESL} & Slepton pair production in $\ee$ collisions \\
    \vv{HWHESP} & Sparticle pair production in $\ee$ collisions \\
    \vv{HWHESQ} & Squark pair production in $\ee$ collisions \\
    \vv{HWHEW0} & $\ee \to W^+ W^-$ subroutine             \\
    \vv{HWHEW1} & $\ee \to W^+ W^-$ subroutine             \\
    \vv{HWHEW2} & $\ee \to W^+ W^-$ subroutine             \\
    \vv{HWHEW3} & $\ee \to W^+ W^-$ subroutine             \\
    \vv{HWHEW4} & $\ee \to W^+ W^-$ subroutine             \\
    \vv{HWHEW5} & $\ee \to W^+ W^-$ subroutine             \\
    \vv{HWHEWW} & Hard subprocess: $\ee \to W^+ W^-$       \\
    \vv{HWHGBP} & Main routine for gauge boson pair production in hadron-hadron\\
    \vv{HWHGBS} & Phase space for gauge boson pair production in hadron-hadron\\
    \vv{HWHGB1} & Selects gauge boson mass for \vv{HWHGBS}\\
    \vv{HWHGB2} & Matrix element for $WW$ in hadron-hadron\\
    \vv{HWHGB3} & Matrix element for $ZZ$ in hadron-hadron\\
    \vv{HWHGB4} & Matrix element for $WZ$ in hadron-hadron\\
    \vv{HWHGB5} & Selects $t$ and $u$ for \vv{HWHGBS}\\
    \vv{HWHGRV} & Graviton resonance production             \\
    \vv{HWHGUP} & External hard process using Les Houches interface \\
    \vv{HWHHVY} & Hard subprocess: heavy quark production  \\
    \vv{HWHIBG} & Hard subprocess: for $bg \to Q~\rm{Higgs}$, with $Q=t,b$ \\
    \vv{HWHIBK} & Hard subprocess: $b\bbar\to W^\pm H^\mp$ \\ 
    \vv{HWHIBQ} & Subroutine for $H^\pm$ production via $bq$-fusion \\
    \vv{HWHIG1} & Matrix elements for Higgs + jet production\\
    \vv{HWHIGA} & Amplitudes squared for Higgs + jet       \\
    \vv{HWHIGB} & Loop integrals for Higgs + jet           \\
    \vv{HWHIGE} & Hard subprocess: Higgs associated production \\
    \vv{HWHIGH} & Hard subprocess: $q\qbar\to$ Higgs$_1$ + Higgs$_2$ \\
    \vv{HWHIGJ} & \QCD\ Higgs + jet production              \\
    \vv{HWHIGM} & Choose any Higgs mass for production routines\\
    \vv{HWHIGQ} & Hard subprocess: $q\qbar,gg\to Q\Qbar^{(')}$ Higgs \\
    \vv{HWHIGS} & Hard subprocess: $g g/q\qbar \to$ Higgs  \\
    \vv{HWHIGT} & Computes $gg \to$ Higgs cross section    \\
    \vv{HWHIGV} & Hard subprocess: $q\qbar^{(')} \to V$ Higgs, 
                                  $V=W^\pm,Z^0$ \\
    \vv{HWHIGW} & Hard subprocess: $W^+W^-/Z^0Z^0\to$ Higgs        \\
    \vv{HWHIGY} & Computes $\ee\to Z^0 \to Z^0 \SMH$ cross section \\
    \vv{HWHIGZ} & Hard subprocess: $\ell^+ \ell^- \to Z^0 \to Z^0 \SMH$ \\
    \vv{HWHIHH} & Hard subprocess: $\ell^+\ell^-$ \SY\ Higgs pair production\\
    \vv{HWHISQ} & Subroutine for squark pair plus Higgs production \\
    \vv{HWHPH2} & Hard subprocess: direct photon pairs        \\
    \vv{HWHPHO} & Hard subprocess: direct photon production   \\
    \vv{HWHPPB} & Box contribution to $gg\to\gamma\gamma$     \\
    \vv{HWHPPE} & Pointlike photon-parton (fixed flavour)    \\
    \vv{HWHPPH} & Pointlike photon-parton (fixed pair flavour)\\
    \vv{HWHPPM} & Pointlike photon-parton direct light meson  \\
    \vv{HWHPPT} & Pointlike photon-parton (all flavours)      \\
    \vv{HWHPQS} & Pointlike photon-quark (Compton) scattering \\
    \vv{HWHQCD} & \QCD\ $2\to 2$ hard subprocesses            \\
    \vv{HWHQCP} & Identifies \QCD\ $2\to 2$ hard subprocess   \\
    \vv{HWHQCM} & Hard subprocess: $\gamma\gamma\to q\bar{q},\ \ell^+\ell^-,\ W^+W^-$\\
    \vv{HWHRBB} & \RPV\ resonant squark to \SM\ particles \\
    \vv{HWHRBS} & \RPV\ resonant squark to \SY\ particles \\
    \vv{HWHREE} & \RPV\  SM particle production in $\ee$ collisions \\
    \vv{HWHREM} & Treats hard scattering remnants             \\
    \vv{HWHREP} & Decides which \RPV\  subroutine to use in $\ee$ \\
    \vv{HWHRES} & \RPV\ single sparticle production in $\ee$ collisions\\
    \vv{HWHRLL} & \RPV\ resonant slepton to \SM\ particles \\
    \vv{HWHRLS} & \RPV\ resonant slepton to \SY\ particles \\
    \vv{HWHRSP} & Decides which \RPV\ subroutine to use in hadron-hadron\\
    \vv{HWHRSS} & Identifies \RPV\ process \\
    \vv{HWHSCT} & Process extra hard scatterings              \\
    \vv{HWHSNG} & Colour singlet parton scattering            \\
    \vv{HWHSNM} & Colour singlet parton scattering matrix element \\
    \vv{HWHSPN} & Main routine for spin correlations in the hard process\\
    \vv{HWHS01} & Helicity amplitude for $f\bar{f}\ra ff$
                  via $s$-channel vector boson exchange\\
    \vv{HWHS02} &  Helicity amplitude for $f\bar{f}\ra ff$
                  via $t$-channel scalar exchange\\
    \vv{HWHS03} &  Helicity amplitude for $f\bar{f}\ra ff$
                  via $u$-channel scalar exchange\\
    \vv{HWHS04} &  Helicity amplitude for $f\bar{f}\ra f\bar{f}$
                  via $s$-channel vector boson exchange\\
    \vv{HWHS05} &  Helicity amplitude for $gg\ra f\bar{f}$
                  via $t$-channel fermion exchange\\
    \vv{HWHS06} &  Helicity amplitude for $gg\ra f\bar{f}$
                  via $u$-channel fermion exchange\\
    \vv{HWHS07} &  Helicity amplitude for $gg\ra f\bar{f}$
                  via $s$-channel gluon exchange\\
    \vv{HWHS08} & Helicity amplitude for $gf\ra f\phi$
                  via $t$-channel scalar exchange\\
    \vv{HWHS09} & Helicity amplitude for  $g\bar{f}\ra f \phi$
                  via $t$-channel scalar exchange\\
    \vv{HWHS10} & Helicity amplitude for $gf\ra f\phi$ 
                  via $s$-channel fermion exchange\\
    \vv{HWHS11} & Helicity amplitude for  $g\bar{f}\ra f \phi$
                  via $s$-channel antifermion exchange\\
    \vv{HWHS12} & Helicity amplitude for $gf\ra f\phi$ 
                  via $u$-channel fermion exchange \\
    \vv{HWHS13} & Helicity amplitude for $g\bar{f}\ra f \phi$
                  via $u$-channel fermion exchange\\
    \vv{HWHS14} & Helicity amplitude for $gg\ra ff$ 
                  via $t$-channel fermion exchange \\
    \vv{HWHS15} & Helicity amplitude for $gg\ra ff$ 
                  via  $u$-channel fermion exchange\\
    \vv{HWHS16} & Helicity amplitude for $gg\ra ff$ 
                  via $s$-channel gluon exchange \\
    \vv{HWHS17} & Helicity amplitude for $ff\ra ff$ 
                  via $t$-channel gauge boson exchange \\
    \vv{HWHS18} & Helicity amplitude for $f\bar{f}\ra f\bar{f}$ 
                  via $t$-channel gauge boson exchange\\
    \vv{HWHS19} & Helicity amplitude for $\bar{f}f\ra\bar{f}f$ 
                  via $t$-channel gauge boson exchange\\
    \vv{HWHS20} & Helicity amplitude for $\bar{f}\bar{f}\ra\bar{f}\bar{f}$ 
                  via $t$-channel gauge boson exchange \\
    \vv{HWHS21} & Helicity amplitude for $f\bar{f}\ra f\bar{f}$
		  via $s$-channel scalar exchange \\
    \vv{HWHS22} & Helicity amplitude for $f\bar{f}\ra f\bar{f}$
		  via $t$-channel scalar exchange \\
    \vv{HWHS23} & Helicity amplitude for $f\bar{f}\ra f\bar{f}$
		  via $u$-channel scalar exchange \\
    \vv{HWHS24} & Helicity amplitude for $\bar{f} f \ra ff$
		  via $s$-channel scalar exchange \\
    \vv{HWHS25} & Helicity amplitude for $\bar{f} f \ra ff$
		  via $t$-channel scalar exchange \\
    \vv{HWHS26} & Helicity amplitude for $\bar{f} f \ra ff$
		  via $u$-channel scalar exchange \\
    \vv{HWHS27} & Helicity amplitude for $ff\ra f\bar{f}$
	          via $s$-channel scalar exchange \\
    \vv{HWHS28} & Helicity amplitude for $ff\ra f\bar{f}$
                  via $t$-channel scalar exchange\\
    \vv{HWHS29} & Helicity amplitude for $ff\ra f\bar{f}$ 
                  via $u$-channel scalar exchange \\
    \vv{HWHS30} & Helicity amplitude for $\bar{f}\bar{f}\ra ff$ 
                 via $s$-channel scalar exchange \\
    \vv{HWHS31} & Helicity amplitude for $\bar{f}\bar{f}\ra ff$ 
                 via $t$-channel scalar exchange \\
    \vv{HWHS32} & Helicity amplitude for $\bar{f}\bar{f}\ra ff$ 
                 via $u$-channel scalar exchange\\
    \vv{HWHS33} & Helicity amplitude for $ff\ra ff$ 
                 via $s$-channel scalar exchange \\
    \vv{HWHS34} & Helicity amplitude for $\bar{f}\bar{f}\ra\bar{f}\bar{f}$
                 via $s$-channel scalar exchange \\
    \vv{HWHSS1} & Gaugino-gaugino production matrix element   \\
    \vv{HWHSS2} & Gaugino-gaugino production matrix element with polarizations   \\
    \vv{HWHSSG} & Gaugino-gaugino/gaugino-sparton production  \\
    \vv{HWHSSL} & Slepton pair production                     \\
    \vv{HWHSSP} & Combines \MS\ subprocesses                \\
    \vv{HWHSSQ} & \sss{SQCD} $2\to 2$  hard subprocesses      \\
    \vv{HWHSSS} & Identifies \MS\ hard subprocess           \\
    \vv{HWHV1J} & Hard subprocess $W$/$Z$ + jet production        \\
    \vv{HWHV2J} & Master subroutine for all gauge boson + two jet processes\\
    \vv{HWHVVJ} & Dummy $WW$,$WZ$,$ZZ$ production subroutine
     (see sect.~\ref{sec:MCNLO})          \\
    \vv{HWHWEX} & Top production by $W$ exchange                \\
    \vv{HWHWPR} & Hard subprocess: $W$ production               \\
\hline
& Soft minimum-bias or underlying event  \\
\hline
    \vv{HWMEVT} & Generates min bias or soft underlying event \\
    \vv{HWMLPS} & Generates longitudinal phase space          \\
    \vv{HWMNBI} & Computes negative binomial probability      \\
    \vv{HWMULT} & Chooses min bias charged multiplicity       \\
    \vv{HWMWGT} & Calculates weight for minimum bias events   \\
\hline
& Random number generators \\
\hline
    \vv{HWRAZM} & Randomly rotated azimuth                    \\
    \vv{HWREXP} & Random number: exponential distribution     \\
    \vv{HWREXQ} & Random number: exp. dist. with cutoff       \\
    \vv{HWREXT} & Random number: exponential transverse mass  \\
    \vv{HWRGAU} & Random number: gaussian                     \\
    \vv{HWRGEN} & Random number generator (l'Ecuyer's method) \\
    \vv{HWRINT} & Random integer                              \\
    \vv{HWRLOG} & Random logical                              \\
    \vv{HWRPIP} & Random primary interaction point            \\
    \vv{HWRPOW} & Random number: power distribution           \\
    \vv{HWRUNG} & Random number: uniform + gaussian tails     \\
    \vv{HWRUNI} & Random number: uniform                      \\
\hline
& Spacelike branching of incoming partons \\
\hline
    \vv{HWSBRN} & Generates spacelike parton branching        \\
    \vv{HWSDGG} & Drees-Grassie gluon distribution in photon  \\
    \vv{HWSDGQ} & Drees-Grassie quark distribution in photon  \\
    \vv{HWSFBR} & Chooses a spacelike branching               \\
    \vv{HWSFUN} & Hadron structure functions                  \\
    \vv{HWSGAM} & Gamma function (for structure functions)    \\
    \vv{HWSGEN} & Generates $x$ values for spacelike partons  \\
    \vv{HWSGQQ} & Inserts $g\to q\qbar$ part of gluon form factor \\
    \vv{HWSMRS} & Subroutine for MRST PDFs \\
    \vv{HWSSPC} & Replaces spacelike partons by spectators    \\
    \vv{HWSSUD} & Sudakov form factor/structure function      \\
    \vv{HWSTAB} & Interpolates in function table (for \vv{HWSSUD})\\
    \vv{HWSVAL} & Checks for valence parton                   \\
\hline
& Miscellaneous utilities \\
\hline
    \vv{HWUAEM} & Running electromagnetic coupling constant  \\
    \vv{HWUAER} & Real part of photon self-energy            \\
    \vv{HWUALF} & Two-loop \QCD\ running coupling constant   \\
    \vv{HWUANT} & Finds a particle's antiparticle            \\
    \vv{HWUATS} & Replaces \& with \tt{\ }\\
    \vv{HWUBPR} & Prints branching data for last parton shower\\
    \vv{HWUBST} & Boost event record to/from hadron-hadron c.m.f.\\
    \vv{HWUCFF} & Coefficients for $\ee$ and \sss{DIS} cross sections\\
    \vv{HWUCI2} & Logarithmic integral Ci$_2$                 \\
    \vv{HWUDAT} & Particle properties (N.B. \vv{BLOCK DATA})   \\
    \vv{HWUDKL} & Generates decay vertex of unstable particle \\
    \vv{HWUDKS} & Converts decay modes into internal format   \\
    \vv{HWUDPR} & Prints particle properties and decay modes  \\
    \vv{HWUECM} & Centre-of-mass energy                       \\
    \vv{HWUEDT} & Insert or delete entries in the event record\\
    \vv{HWUEEC} & Computes coefficients for $\ee$ cross section\\
    \vv{HWUEMV} & Moves entries within the event record       \\
    \vv{HWUEPR} & Prints event data                           \\
    \vv{HWUFNE} & Finishes an event                           \\
    \vv{HWUGAU} & Adaptive gaussian integration               \\
    \vv{HWUGUP} & Run termination for Les Houches Accord \\
    \vv{HWUIDT} & Translates particle identity codes          \\
    \vv{HWUINC} & Initial parameter-dependent calculations    \\
    \vv{HWUINE} & Initialises an event                        \\
    \vv{HWULB4} & Boost: rest frame to lab, no masses assumed \\
    \vv{HWULDO} & Lorentz 4-vector dot product                \\
    \vv{HWULF4} & Boost: lab frame to rest, no masses assumed \\
    \vv{HWULI2} & Logarithmic integral Li$_2$ (Spence function)\\
    \vv{HWULOB} & Lorentz transformation: rest frame $\to$ lab\\
    \vv{HWULOF} & Lorentz transformation: lab $\to$ rest frame\\
    \vv{HWULOR} & Multiplies by Lorentz matrix                \\
    \vv{HWUMAS} & Puts mass in 5th component of vector        \\
    \vv{HWUMBW} & Generates mass (Breit-Wigner distribution)  \\
    \vv{HWUMPO} & Spinor routine\\
    \vv{HWUMPP} & Spinor routine\\
    \vv{HWUNST} & Converts integer to character \\
    \vv{HWUPCM} & Centre-of-mass momentum                     \\
    \vv{HWUPUP} & Prints contents of Les Houches common block \\
    \vv{HWURAP} & Rapidity                                    \\
    \vv{HWURES} & Computes/prints resonance data              \\
    \vv{HWUROB} & Rotation by inverse of matrix $R$           \\
    \vv{HWUROF} & Rotation by matrix $R$                      \\
    \vv{HWUROT} & Computes rotation $R$ from vector to $z$-axis\\
    \vv{HWUSOR} & Sorts an array in ascending order           \\
    \vv{HWUSPR} & Print contents of spin correlations common block \\
    \vv{HWUSQR} & Square root with sign retention             \\
    \vv{HWUSTA} & Makes a particle type stable                \\
    \vv{HWUTAB} & Interpolates in a table                     \\
    \vv{HWUTIM} & Checks time remaining \\
\hline
& Vector manipulation  \\
\hline
    \vv{HWVDIF} & Vector difference                           \\
    \vv{HWVDOT} & Vector dot product                          \\
    \vv{HWVEQU} & Vector equality                             \\
    \vv{HWVSCA} & Vector times scalar                         \\
    \vv{HWVSUM} & Vector sum                                  \\
    \vv{HWVZRI} & Vector zero (integer)                       \\
    \vv{HWVZRO} & Vector zero                                 \\
\hline
    \vv{HWWARN} & Issues warnings and deals with errors       \\
\hline
\end{longtable}

In addition there are the routines for generating the
Schuler-Sj\"ostrand parton distributions of the photon:
\small\begin{verbatim}
   SASANO  SASBEH  SASDIR  SASGAM  SASVMD
\end{verbatim}\normalsize
  and for putting the output of \TA\ \cite{Jadach:1993hs}
  into the  \vv{/HEPEVT/} common block:
\small\begin{verbatim}
   FILHEP
\end{verbatim}\normalsize

Finally, there are dummy versions of external routines, which should
be deleted if the relevant packages are used:
\begin{itemize}
\item \CERN\ Library routine giving \vv{TRES}= CPU time (in seconds)
remaining (dummy returns \vv{TRES}=10$^{10}$):
\small\begin{verbatim}
   TIMEL
\end{verbatim}\normalsize

\item \CERN\ \PDF\ structure function package:
\small\begin{verbatim}
   PDFSET  STRUCTM
\end{verbatim}\normalsize

\item {\sf EURODEC} B decay package:
\small\begin{verbatim}
   EUDINI  FRAGMT  IEUPDG  IPDGEU
\end{verbatim}\normalsize

\item {\sf CLEO} B decay package:
\small\begin{verbatim}
   DECADD  QQINIT  QQLMAT
\end{verbatim}\normalsize

\item {\sf HERBVI} baryon number violation package:
\small\begin{verbatim}
   HVCBVI  HVHBVI
\end{verbatim}\normalsize

\item {\sf CIRCE} beamstrahlung package (see below):
\small\begin{verbatim}
   CIRCEE  CIRCES  CIRCGG
\end{verbatim}\normalsize

\item {\sf TAUOLA} $\tau$ lepton decay package (see below):
\small\begin{verbatim}
    DEXAY  INIETC  INIMAS  INIPHX  INITDK  PHOINI  PHOTOS
\end{verbatim}\normalsize

\item Les Houches interface (see below):
\small\begin{verbatim}
   UPINIT  UPEVNT
\end{verbatim}\normalsize
\end{itemize}

\section{Interfaces}

\subsection{Les Houches}\label{sec:lesh}
    Version 6.5 includes  support  for the  interface  between  parton  level
    generators and \HW\ using the Les Houches  Accord as  described in
    \cite{Boos:2001cv}. In  general we have  tried to code  the interface in
    such a way that from the user point of  view it behaves in  the same
    way as that already included in \PY\ \cite{Sjostrand:2000wi}.

    The interface  operates in the following  way. If the \vv{IPROC} code is
    set  negative  then \HW\  assumes  that the user wants an external
    hard process  using the Les Houches  accord.

    If this  option is  used the  initialisation will  call the  routine
    \vv{UPINIT} to  initialise  the external hard  process; this  name is the
    same  as that used by \PY. This routine  should set the values of
    the run parameters in the Les Houches run common block.

    After  the  initialisation  during the  event  generation  phase the 
    routine  \vv{UPEVNT} (again this name is the same as that used by \PY)
    is  called.  This  routine  should  fill the  event  common block as
    described in \cite{Boos:2001cv}.

    Dummy  copies of both these  routines are  supplied with  \HW\ and 
    should be deleted and  replaced if you are using this option. Due to
    the internal  structure of  \HW\ two  new parameters are needed to
    control  the  interface, in  addition  to those  in the  Les Houches
    common block. The logical input variable \vv{LHSOFT} [\vv{.TRUE}]
    controls the  generation of  the soft underlying event; the default
    is to generate a soft underlying event and so \vv{LHSOFT} must be set
    \vv{.FALSE.} for lepton-lepton  processes.  The second  variable
    \vv{LHGLSF} [\vv{.FALSE.}] controls the  treatment of colour
    self-connected  gluons, which  may occur  with some methods of colour
    decomposition. The  default is to kill events with self-connected
    gluons  whereas if \vv{LHGLSF} = \vv{.TRUE.} an  information-only 
    warning is issued instead.

    We have not included full support for the interface. In  particular
\begin{itemize}
\item It is assumed that all events have two particles of status \vv{IDUP}=--1;
\item We do not support  the status codes --2 and --9;
\item The treatment of the code \vv{IDUP}=3 is the same as \vv{IDUP}=2,
      i.e.\ all intermediate resonances have their masses preserved.
\end{itemize}
    These restrictions are the same as those 
    imposed by \PY. If you have any  problems with the  interface or 
    need the options we have not yet supported please let us know.

    It  should  also be noted  that while  the interface has been tested
    with  Standard  Model  processes, for example  all the  processes in
    \AL\ \cite{Mangano:2002ea}, it is  less well tested
    for \SY\ processes.

\subsection{CIRCE}

Simulation of beamstrahlung is now available, via an interface to the
\CI\ program \cite{Ohl:1997fi}.  It is implemented as a
modification of the standard bremsstrahlung implementation, so is only
available for processes for which that is available.  The produced
radiation is treated in the collinear limit, i.e.~\vv{COLISR} and
the option \vv{WWA} in routine \vv{HWEGAM} are forcibly set \vv{.TRUE.}.

For both types of distribution function, $f_{e/e}$ and $f_{\gamma/e}$,
we use a simplifying approximation: where the correct result should use
the convolution of beamstrahlung and bremsstrahlung, we actually use the
sum of the two.  For the photon distribution this is quite reasonable,
but for the electron it is perhaps more questionable.  It boils down to
the replacement:
\begin{equation}
   f_{e/e}(x) = \int_x^1 \frac{dz}z
   f_{e/e,\mathrm{brem}}\Bigl(z\Bigr)
   f_{e/e,\mathrm{beam}}\Bigl(\frac xz\Bigr)
   \to
   f_{e/e,\mathrm{brem}}(x) + f_{e/e,\mathrm{beam}}(x) - \delta(1-x).
\end{equation}
This is good to the extent that both distributions are dominated by the
$x\to1$ region.  A small utility program can be obtained from the
\HW\ web page to test this approximation.  For example, for TESLA running
at 500~GeV, the integral over $f_{e/e}(x)$ for large $x$, relevant for
threshold shapes, is accurate to 1\% for $x>0.95$ and to 10\% for
$x>0.99$, and the value of $f_{e/e}((\frac{90}{500})^2)$, relevant for
radiative return to the Z, is accurate to 2\%.

In order for the distribution to be probabilistic, the coefficient of
the delta function has to remain positive.  If it is not, \HW\ will
terminate.  This can be prevented by adjusting the \vv{ZMXISR}
parameter: the default value, \vv{ZMXISR}=0.999999 is too large and
\HW\ will fail, but reducing it to \vv{ZMXISR}=0.99999 works fine
for all the parameter sets currently available.  It is straightforward
to check that \vv{ZMXISR} values in this range do not have an
influence on any physical observables, by rerunning with it further
reduced.

The interface is controlled by five new variables, which are given in
the following table together with their default values:
\begin{center}
  \begin{tabular}{|l|l|}
    \hline
    \vv{CIRCOP} & 0 \\
    \vv{CIRCAC} & 2 \\
    \vv{CIRCVR} & 7 \\
    \vv{CIRCRV} & 9999 12 31 \\
    \vv{CIRCCH} & 0 \\
    \hline
  \end{tabular}
\end{center}
\vv{CIRCOP} is the main control option: \vv{CIRCOP}=0 switches
off beamstrahlung and uses standard \HW; \vv{CIRCOP}=1 switches
to collinear kinematics, but still leaves beamstrahlung switched off;
\vv{CIRCOP}=2 uses only beamstrahlung; and \vv{CIRCOP}=3 uses
both beamstrahlung and bremsstrahlung.  \vv{CIRCOP}=0 and
\vv{CIRCOP}=3 should therefore be regarded as `off' and `on',
respectively, with the other two options mainly for cross-checking
purposes.  The variables \vv{CIRCAC}, \vv{CIRCVR},
\vv{CIRCRV} and \vv{CIRCCH} are simply passed to \CI\ as
its input variables \vv{acc}, \vv{ver}, \vv{rev} and
\vv{chat}, as described in its documentation.  The default values
correspond to the most up-to-date revision of version~7 of the TESLA
parametrization, with minimal output.

\CI\ can be obtained from
\begin{center}
\href{http://heplix.ikp.physik.tu-darmstadt.de/nlc/beam.html}{\tt
http://heplix.ikp.physik.tu-darmstadt.de/nlc/beam.html}
\end{center}
  
\subsection{TAUOLA}
    An interface to the \TA\ decay package~\cite{Jadach:1993hs}
    has been added. To use this
    interface, the dummy  subroutines  \vv{DEXAY},  \vv{INIETC},  \vv{INIMAS},
  \vv{INIPHX}, 
    \vv{INITDK}, \vv{PHOINI}, \vv{PHOTOS} must be deleted. You should then link
 to both
    \TA\  and  {\sf PHOTOS}. The  easiest  way to do  this is  to obtain the
    {\sf TAUOLA/PHOTOS} versioning system from 
\begin{center}
\href{ http://wasm.home.cern.ch/wasm/test.html}{\tt
http://wasm.home.cern.ch/wasm/test.html}
\end{center}
    and produce a version with the correct  size  of the  \vv{HEPEVT}  common 
    block. This system  will provide a  \JS\ demo.  If you link \HW\ with
    the code produced by this  apart from the \JS\ interface and
    main program everything should work.

    This interface uses the  information from the spin density matrices
    to select the helicity of the decaying $\tau$ leptons if the spin
    correlation algorithm is being used;
    otherwise the  helicity of the decaying $\tau$ is averaged over.

\subsection{ISAWIG}
\label{sect:ISAWIG}
\IW\ is an interface which allows \SY\ spectra and decay tables
generated by \IS\ \cite{ISA} to be read into \HW.  The web address is
\begin{center}
\href{http://www.hep.phy.cam.ac.uk/~richardn/HERWIG/ISAWIG/}{\tt
http://www.hep.phy.cam.ac.uk/}$\sim${\tt richardn/HERWIG/ISAWIG/}
\end{center}

To  coincide with  the release of  \HW\ 6.5 we have  produced a new
version  of \IW.
    Since the original version of \IW\ there have been a number of new
    versions of \IS, often with changes in the  sizes of  the common
    blocks from which we extract the information we need. This has
    necessitated periodical updating of the code to run with the most
    recent version  of  \IS, with the result that the code could
    no longer be run with older versions. However, the Snowmass points and
    slopes \cite{Allanach:2002nj} are  defined  with  \IS\ version
    7.58 and so we  can no  longer
    continue  in  this way and  still be able to generate  these  points.
    Therefore, starting  with  the  new  \IW\ version  1.2, we are using C
    preprocessing so that users can define the version of \IS\ they
    are using at  compile time.  When compiling the main \IW\ code and 
    the modified {\small SUGRUN} and {\small SSRUN}
    programs the following compiler options should 
    be specified:
\begin{itemize}
\item \vv{-DISAJET758}  to use \IS7.58, 
\item \vv{-DISAJET763}  to use \IS7.63.
\end{itemize}
    The default at the moment is to compile code to run with \IS7.64.
    In  the future  this will change  so that the most recent version of 
    \IS\ becomes the default but we will continue to support the older
    versions.

\subsection{HDECAY}
    There has also been a new release of the \HD\ package \cite{Djouadi:1997yw},
    version 3.
    In order to support this  version, and the  previous version 2.0, we 
    are also using C  preprocessing  to control  the version  of  \HD\
    used. In order to achieve  this the \HD\  interface code has  been
    merged with the main \IW\ program. The following options should be
    used when compiling \IW\ if you are using \HD:
\begin{itemize}
\item \vv{-DHDECAY2} if you are using version 2 of \HD
\item \vv{-DHDECAY3} if you are using version 3 of \HD.
\end{itemize}
    As  before  the default  is to  compile a  dummy routine. If you are
    using \HD\ you must  link with a version of the \HD\  code which 
    has the main \HD\ program removed.

\subsection{MC@NLO}\label{sec:MCNLO}
  The program \MN\ \cite{Frixione:2002ik} generates events with
\begin{itemize}
\item  exclusive rates and distributions accurate to next-to-leading order
    when expanded in $\alpha_s$;
\item multiple  soft/collinear  parton emission generated by \HW\ parton
    showering;
\item hadronization according to the \HW\ cluster model.
\end{itemize}
    At present only the processes of gauge boson pair production in hadron
    ($h=p$ or $\bar p$) collisions are implemented in \MN, making use
    of the \HW\ process codes \vv{IPROC}=2850--2880 (see table~\ref{tab.8}).

To use these codes, one must delete the dummy \HW\ subroutine \vv{HWHVVJ}
and replace it by the one in the \MN\ package.  This will then read and
process an input file of parton configurations prepared by \MN.
For further details see the \MN\ web page:
\begin{center}
\href{http://www.hep.phy.cam.ac.uk/theory/webber/MCatNLO/}{\tt    
http://www.hep.phy.cam.ac.uk/theory/webber/MCatNLO/}
\end{center}

Implementation of other processes, such as heavy quark production, in \MN\ is
in progress \cite{Frixione:inprep}.  These future development will make use
of the Les Houches interface discussed in sect.~\ref{sec:lesh}, rather than
special process codes.
\subsection{JIMMY}
\label{sect:JIMMY}
Contrary to our original plans, the {\sf JIMMY} generator~\cite{jimmy}
for multiple parton-parton interactions in hadron-hadron, hadron-photon and
photon-photon collisions is still not integrated into \HW, and is still a separate add-on package. It can be obtained from the JIMMY web
page:
\begin{center}
\href{http://jetweb.hep.ucl.ac.uk/JIMMY/index.html}{\tt http://jetweb.hep.ucl.ac.uk/JIMMY/index.html}
\end{center}
where references to the relevant papers can also be found.

\subsection{HERBVI}
\label{sect:HERBVI}

An interface is provided to the \vv{HERBVI} package for electroweak
baryon number violation (\BNV), and other multi-$W^\pm$ production
processes~\cite{Gibbs}.  For full details, see

\noindent{\small
\href{http://www.hep.phy.cam.ac.uk/~richardn/HERWIG/HERBVI/}{\tt
http://www.hep.phy.cam.ac.uk/}$\sim${\tt richardn/HERWIG/HERBVI/}
}

\acknowledgments

We thank G.~Abbiendi, J.~Ch\'yla and L.~Stanco for their collaboration
on earlier versions of \HW\ and for their continuing interest.  We
also thank U.~Baur, M.J.~Gibbs, E.W.N.~Glover, Z.~Kunszt and D.R.~Ward
for contributing code to be incorporated into the program.
 
We are grateful to the many experimental colleagues who have made
valuable criticisms and suggestions over the years, especially
K.~Hamacher, R.~Hemingway and G.~Rudolph.

Since this is expected to be the last (Fortran) \HW\ version, we would
like to take this opportunity to thank again all those colleagues and users
who have contributed to the development of the program, whether by providing
code, suggesting improvements or reporting problems.  Thanks also to those
who worked so hard to establish the Les Houches accord, which should make
it possible to expand the application of Fortran \HW\ to new processes
without (much) further intervention by the authors.  Meanwhile, many of us
will be transferring our main efforts to \HP, in preparation for the new
era of object-oriented event generation.

\end{document}